\documentclass[acmsmall]{acmart}
\usepackage{amsmath,amsfonts}
\usepackage{algorithmic}
\usepackage{graphicx}
\usepackage{textcomp}
\usepackage{xcolor,colortbl}
\usepackage{booktabs}
\usepackage{tikz,pgfplots}
\usepackage{pgfplots}
\usepackage{pgfplotstable}
\usepackage{standalone}
\usepackage{multirow}
\usepackage[tikz]{bclogo}
\usepackage[listings,skins,breakable]{tcolorbox}
\usepackage{csquotes}
\usepackage{amsmath}
\usepackage{url}
\usepackage[tikz]{bclogo}
\usetikzlibrary{arrows.meta}
\usepackage{fancyhdr}
\usepackage{threeparttable}
\usepackage{subcaption}
\usepackage{makecell}
\usepackage{pbox}
\usepackage[ruled, linesnumbered, vlined, commentsnumbered]{algorithm2e}
\usepackage{tcolorbox}
\usepackage{xcolor}
\usepackage{pifont}
\usepackage{enumitem}
\usepackage{bm}
\usepackage[svgnames]{xcolor}

\newtcolorbox{quotebox}{colback=steel!10,boxrule=0.4pt,colframe=black,fonttitle=\bfseries,top=2pt,bottom=2pt}
\usepackage{xfp}

\definecolor{result}{rgb}{0.1, 0.3, 0.5} 
\definecolor{steel}{rgb}{0, 0.2, 0.9} 
\definecolor{red-}{HTML}{FCE4D6}
\definecolor{gold}{HTML}{FFF2CC}

\mathchardef\mhyphen="2D

\DeclareMathAlphabet\mathbfcal{OMS}{cmsy}{b}{n}

\newcolumntype{P}[1]{>{\centering\arraybackslash}m{#1}}
\newcolumntype{Y}{>{\centering\arraybackslash}X}

\newtcbox{\mytag}{nobeforeafter,colframe=mycolor,colback=mycolor!30!white,boxrule=0.7pt,arc=0pt,
 boxsep=-3pt,left=6pt,right=6pt,top=4pt,bottom=5pt,tcbox raise base}
 
\newcommand{\rev}[1]{\textcolor{black}{#1}}

\newcommand{\revminor}[1]{\textcolor{black}{#1}}

\usetikzlibrary{pgfplots.statistics}

\newcommand{\squart}[4]{\begin{adjustbox}{max width=.1\textwidth}\begin{picture}(100,5)
{\color{black}\put(0,5){\line(1,0){100}}\color{black}\put(0,5){\line(0,1){10}}\put(50,5){\line(0,1){10}}\put(100,5){\line(0,1){10}}\put(25,5){\line(0,1){5}}\put(75,5){\line(0,1){5}}\put(-2,-8){\LARGE$0$}\put(42,-8){\LARGE$0.5$}\put(96,-8){\LARGE$1$}}\end{picture}\end{adjustbox}}

\newtcolorbox{implicationbox}{colback=white,boxrule=0.4pt,colframe=black,fonttitle=\bfseries,top=1pt,bottom=1pt,left=1pt,right=1pt}

\newcommand{\revisioncolor}{black} 
\newcommand{\revision}[1]{{\color{\revisioncolor}#1}}

{\noindent\begin{minipage}[c]{\linewidth}%
\begin{bclogo}[couleur=gray!30,%
                arrondi=0.1,%
                logo=\bclampe,%
                ombre=true]{\normalsize Key Characteristic} {#1}}%
{\end{bclogo}\end{minipage}\vspace{2mm}}

\pgfplotsset{compat=1.11,
        /pgfplots/ybar legend/.style={
        /pgfplots/legend image code/.code={%
        \draw[##1,/tikz/.cd,bar width=3pt,yshift=-0.2em,bar shift=0pt]
                plot coordinates {(0cm,0.8em)};},
},
}
\AtBeginDocument{%
  \providecommand\BibTeX{{%
    \normalfont B\kern-0.5em{\scshape i\kern-0.25em b}\kern-0.8em\TeX}}}
    
\definecolor{mycolor}{rgb}{0.122, 0.435, 0.698}

\newcommand{\approach}{\texttt{Domland}}

\setcopyright{acmlicensed}
\acmJournal{TOSEM}

\begin{document}

\title{Revealing Domain-Spatiality Patterns for Configuration Tuning: Domain Knowledge Meets Fitness Landscapes}

\author{Yulong Ye}
\email{yxy382@student.bham.ac.uk}
\affiliation{
	\institution{IDEAS Lab, University of Birmingham}
	\country{United Kingdom}
}

\author{Hongyuan Liang}
\authornote{Hongyuan Liang is also supervised in the IDEAS Lab.}
\email{lianghy16@outlook.com}
\affiliation{
	\institution{University of Electronic Science and Technology of China}
	\country{China}
}

\author{Chao Jiang}
\email{cxj249@student.bham.ac.uk}
\affiliation{
	\institution{University of Birmingham}
	\country{United Kingdom}
}

\author{Miqing Li}
\affiliation{%
  \institution{University of Birmingham}
  \country{United Kingdom}
 }
\email{m.li.8@bham.ac.uk}

\author{Tao Chen}
\authornote{Corresponding author: Tao Chen, t.chen@bham.ac.uk.}
\email{t.chen@bham.ac.uk}

\affiliation{
	\institution{IDEAS Lab, University of Birmingham}
	\country{United Kingdom}
}



\begin{abstract}

Configuration tuning for better performance is crucial in quality assurance. Yet, there has long been a mystery on tuners' effectiveness, due to the black-box nature of configurable systems. Prior efforts predominantly adopt static domain analysis (e.g., static taint analysis), which often lacks generalizability, or dynamic data analysis (e.g., benchmarking performance analysis), limiting explainability. In this work, we embrace Fitness Landscape Analysis (FLA) as a bridge between domain knowledge and difficulty of the tuning. We propose \approach, a two-pronged methodology that synergizes the spatial information obtained from FLA and domain-driven analysis to systematically capture the hidden characteristics of configuration tuning cases, explaining how and why a tuner might succeed or fail. \revminor{This helps to better interpret and contextualize the behavior of tuners and inform tuner design.} To evaluate \approach, we conduct a case study of nine software systems and 93 workloads, from which we reveal several key findings: (1) configuration landscapes are inherently system-specific, with no single domain factor (e.g., system area, programming language, or resource intensity) consistently shaping their structure; (2) the core options (e.g., \texttt{pic-struct} of \textsc{x264}), which control the main functional flows, exert a stronger influence on landscape ruggedness (i.e. the difficulty of tuning) compared to resource options (e.g., \texttt{cpu-independent} of \textsc{x264}); (3) Workload effects on landscape structure are not uniformly tied to type or scale. Both contribute to landscape variations, but their impact is system-dependent.

\end{abstract}


\maketitle

\pgfplotsset{compat=1.18}
\usetikzlibrary{arrows.meta}
\usepgfplotslibrary{groupplots}

\newcommand{\RSDThreshold}{5} 

\definecolor{OverColor}{RGB}{52,101,164}   
\definecolor{UnderColor}{RGB}{206,92,0}    
\definecolor{DashMark}{RGB}{128,128,128}   

\newcommand{\RSDBarAxis}[1]{%
\begin{axis}[
  width=\textwidth,
  height=7cm,
  ymajorgrids,
  ymin=0,
  ylabel={RSD (\%)},
  xtick=data,
  xticklabel style={rotate=50, anchor=east, font=\scriptsize},
  enlarge x limits=0.01,
  bar width=5pt,
  legend style={at={(0.5,1.03)}, anchor=south, legend columns=-1, /tikz/every even column/.append style={column sep=6pt}},
  title={#1}
]
}

\newcommand{\FivePercentLine}[2]{
  \addplot+[black, dashed, very thick] coordinates { (#1,\RSDThreshold) (#2,\RSDThreshold) };
}

\section{Introduction}
\label{section:introduction}

Software systems are highly configurable, offering numerous tunable configuration options. The goal thereof is straightforward: by allowing flexible configuration, software systems can achieve greater applicability across a wide range of domains and cater to varying performance requirements, e.g., runtime and throughput~\cite{chen2015toward,chen2018survey,DBLP:conf/icse/WangChen26}. Yet, excessive configurability comes with its own costs: it has been shown that globally 59\% of the software performance issues---where performance requirements were severely violated---are attributed to poorly chosen configurations rather than code~\cite{han2016empirical}.

\begin{figure}[t]
    \centering
    \includegraphics[height=10 cm]{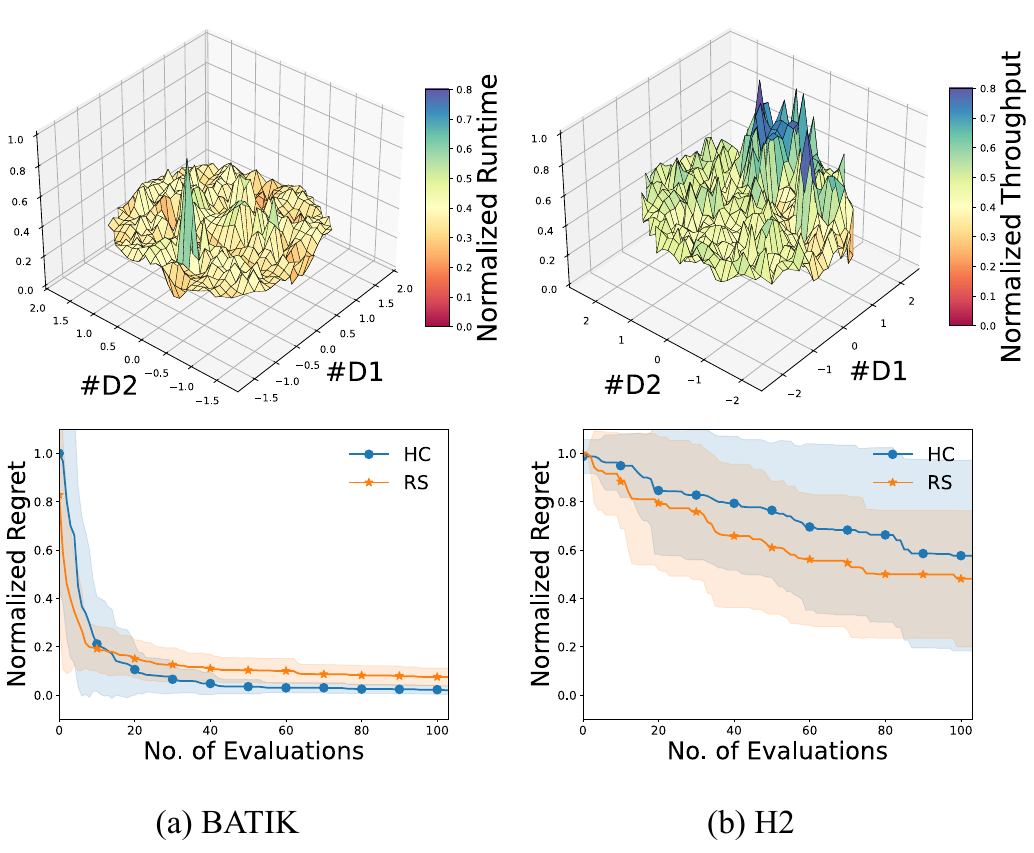}     
    \caption{The regret (i.e., gap to the optimal performance) using Random Search (RS) and Hill Climbing (HC) to tune \textsc{Batik} and \textsc{H2}. The top shows their landscapes processed by MDS~\cite{cox2000multidimensional}.}
    \label{fig:landscape_example}
    \vspace{-11pt}
\end{figure}

        
    

    

To mitigate these challenges, various automated configuration tuners, relying on algorithms such as Bayesian optimization (BO)~\cite{duan2009tuning,zhang2021restune,DBLP:conf/icse/ChenChen26,fekry2020tune} and genetic algorithms (GAs)~\cite{olaechea2014comparison,back1993overview,DBLP:conf/kbse/XiongC25}, have been developed to search for optimal configurations. However, it has been reported that the effectiveness of these tuners varies significantly across different contexts~\cite{DBLP:journals/pvldb/ZhangCLWTLC22}. For instance, Figure \ref{fig:landscape_example} depicts the tuning trajectories of two tuners, random search (RS)~\cite{oh2017finding} and hill climbing (HC)~\cite{li2014mronline,xi2004smart}, on \textsc{Batik} and \textsc{H2}---two systems that exhibit distinct configuration landscapes. Clearly, the convergence and tuning quality differ, with an advantage of HC over RS on \textsc{Batik}, as sharply opposed to on \textsc{H2}. 
The effectiveness of tuning algorithms heavily depends on the underlying characteristics of the system/workload, e.g., the landscape of \textsc{H2} is more rugged than that of \textsc{Batik} (see the top panel of Figure \ref{fig:landscape_example}), causing the ineffectiveness of HC but RS, as an random sampling, is immune to the ruggedness level. However, to the community, it remains unclear how the configuration landscape and domain-specific factors (e.g., system areas, structures, and workload characteristics) influence tuners' behaviors. Practitioners of configuration tuning need a systematic approach to understand the behaviors of tuners while linking/explaining such within the context of domain knowledge/characteristics of the system, which is a challenging desire.

Indeed, static domain analysis of configurable systems exist~\cite{sullivan2004using,he2022multi, DBLP:journals/pvldb/LaoWLWZCCTW24,chen2023diagconfig}, mainly leveraging \textit{domain-specific knowledge} or \textit{code-level analysis} without the need of running the system. The former refers to manually extracted knowledge from system documentation and expert-derived rules of thumb~\cite{sullivan2004using,he2022multi, DBLP:journals/pvldb/LaoWLWZCCTW24}; the latter uses code tracking, trace analysis, and software module reviews to understand how configuration options influence system behaviors~\cite{chen2023diagconfig}. {However, existing domain analysis are mostly static, heavily relying on expert assumptions, which may fail to provide accurate, in-depth insights with respect to the difficulty of tuning a system.}



To overcome the shortcomings of static domain analysis, an alternative way to study configurable systems is dynamic data analysis, which primarily relies on observing the runtime states of the system, understanding the data distributions, and their correlations, for the values of configuration options and performance~\cite{jamshidi2017transfer,muhlbauer2023analyzing}. However, this method only statistically considers the value of options and performance as isolated points, ignoring their spatial information in the configuration landscape, which is what a tuner's behaviors would be sensitive to~\cite{huang2024rethinking}. Recently, fitness landscape analysis (FLA)~\cite{malan2021survey,wright1932roles}, which explicitly incorporates spatial information, has been explored as a means to uncover key characteristics of configurable systems~\cite{huang2024rethinking}. \revminor{Specifically, it provides an analytical lens to understand the structural characteristics of configuration tuning cases, explaining how and why a tuner might succeed or fail.} Nevertheless, interpreting the system solely from a FLA perspective might not provide meaningful guidelines to the software practitioners, since the linkage between the outcomes of FLA and domain-specific factors is still missing. 



\revminor{In this paper, we present \approach, a holistic methodology for characterizing configuration tuning problems from both spatial and domain perspectives, and thus delivering insights for interpreting tuning behaviors/difficulty. Unlike the others, \approach~provides guidelines that synergize the spatial information of the configuration landscape, mined using FLA, with domain knowledge at the option, system, and workload levels. This provides comprehensive, intuitive, and actionable insights to practitioners of configuration tuning. In summary, our main contributions include:}


\begin{itemize}
    \item We provide a new dynamic data analysis guideline to spatially analyze the configuration landscape using the selected/designed categories of analysis from FLA, as well as the metrics, that are the most relevant to configurable systems.
    \item We document a procedure of system-specific domain analysis at three levels of abstraction, together with general categorization of the information.
    \item Drawing both the configuration spatiality and domain analysis above, we outline a method for domain-spatiality synergy, linking the information from two different sources of analysis, providing actionable insights to tuners and system designers.
\end{itemize}

To showcase how \approach~works and evaluates its usefulness, we conduct a case study by applying it to analyze nine widely-studied systems with a total of 93 workloads. From this, and the unique domain-spatiality synergy provided by \approach, we find several previously unknown/unconfirmed observations, such as:

\begin{itemize}
     \item the configuration landscapes across the systems are inherently system-specific, with no single domain factor (e.g., system area, programming language, or resource intensity) consistently shaping their structure;
    \item despite system-specific landscape variations, core options commonly exert a stronger influence on landscape ruggedness than resource options. This suggests that certain option properties universally impact optimization complexity across different systems;
    \item the workload-induced landscape shifts are system-dependent---some remain stable, while others exhibit significant changes. Workload type and scale exhibit no uniform influence, implying complex interactions with system attributes.
\end{itemize}

All code and data can be accessed at our repository: \textcolor{blue}{\texttt{\url{https://github.com/ideas-labo/domland}}}.

The remainder of this paper is as follows: Section \ref{section:preliminaries} introduces preliminaries. Section \ref{section:methodology} presents our methodology. Section \ref{section:empirical_setup} describes the empirical setup, followed by Section \ref{section:results_and_analysis} which reports the results and analysis. Section \ref{section:implication} discusses the implications of our findings. Threats to validity, related work, and conclusion are presented in Sections~\ref{section:threats_to_validity}, \ref{section:related_work}, and \ref{section:conclusion}, respectively.

\section{Preliminaries} 
\label{section:preliminaries}

\subsection{Software Configuration Tuning}
A configurable software system often comprises a set of configuration options, each taking categorical or numerical values. The objective of software configuration is to identify a configuration that optimizes the specific performance attribute, e.g., minimizing runtime or maximizing throughput, of the target system. Mathematically, this can be formulated as:

\begin{equation}
	\begin{array}{l}
		\arg\min f(\textbf{\textit{c}}) \text{ or } \arg\max f(\textbf{\textit{c}}), 
	\end{array}
\end{equation}where \textbf{\textit{c}} = ($c_1, c_2, \dots, c_n$) is a configuration with the values of $n$ options in search space $\mathcal{C}$. $f$ represents the performance attribute of the target system. Since the internal behavior of the system is often opaque and lacks of explicit analytical form of $f$, this is considered a black box optimization problem~\cite{weise2009global}, which has long been a fundamental challenge for software engineering~\cite{harman2012search}. 

\begin{figure}[t!]
    \centering
    \includegraphics[height=4.5 cm]{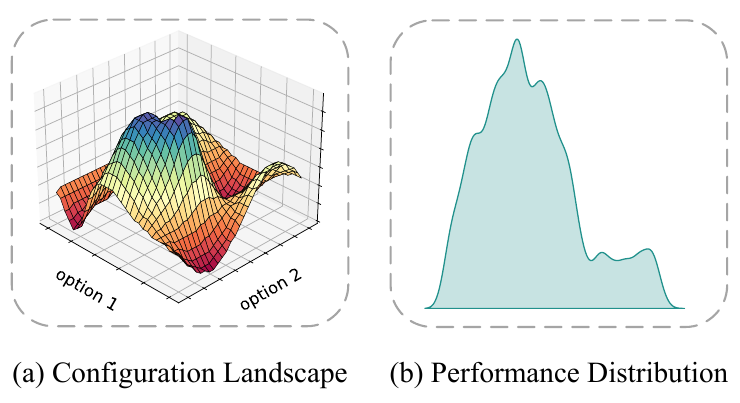}
    \caption{Illustration of configuration landscape and performance distribution.}
    \label{fig:spatial_vs_distribution}
\end{figure}

\subsection{Fitness Landscape Analysis}

A way to know the hardness of an optimization problem is to use fitness landscape analysis (FLA). By characterizing the configuration space, FLA can help understand the complexity of the problem through measuring key features of the configuration of a system. For example, by analyzing the configuration landscape in Figure~\ref{fig:spatial_vs_distribution}, FLA helps to answer several key questions regarding the configuration-performance landscape, each of which provides actionable insights for configuration tuning:

\begin{itemize}

    \item \textbf{Where is the optimum located?}  \revision{Understanding the position and accessibility of the optimum helps assess how easily a tuner can converge to the best-performing configuration(s). If the landscape exhibits strong global guidance (e.g., better configurations tend to lie closer to the optimum), exploitation-driven tuners can efficiently reach the optimum; otherwise, exploration-driven strategies may be required.} 
        \item \textbf{What does local structure look like?} \revision{Analyzing the local structure reveals the distribution and quality of local optima in the landscape, including how many exist and how good they are. This informs whether a tuner may benefit from escaping suboptimal regions through global exploration, or whether convergence to a high-quality local region is sufficient to achieve satisfactory performance.}
    
    \item \textbf{How rugged is the search space?} \revision{Ruggedness reflects how smoothly performance changes with small configuration perturbations. Smooth landscapes favor model-based tuners that rely on generalization, while rugged landscapes may demand exploration-driven strategies.}

\end{itemize}

\rev{By answering these questions, FLA provides not only a descriptive understanding of configuration–performance relationships but also actionable guidance for tuner design. 
For instance, even partial landscape knowledge derived from limited sampling or historical configuration–performance data can be used to approximate the structural properties of the configuration space. 
Such analysis helps practitioners anticipate the relative difficulty of optimization, explain/predict algorithmic behavior, and guide appropriate tuning algorithm selection/design: for example, model-based methods such as Bayesian optimization are suitable for smoother landscapes, whereas evolutionary algorithms are more effective for highly rugged ones.}



Formally, a fitness landscape can be represented as a triplet ($\mathcal{C,N},f$), where $\mathcal{C}$ defines the search space; $\mathcal{N}$ denotes a neighborhood structure that specifies the neighbors of each configuration; and $f$: $\mathcal{C}$ $\rightarrow$ $\mathbb{R}$ represents the fitness function~\cite{pitzer2012comprehensive}. As from Figure~\ref{fig:spatial_vs_distribution}, comparing with existing statistical distribution analysis~\cite{jamshidi2017transfer,muhlbauer2023analyzing}, FLA differs in the following aspects:

\begin{itemize}
    \item \textbf{Spatial awareness:} FLA can capture the spatial structure of search and performance spaces while statistical distribution analysis focuses solely on performance/option variation.
    \item \textbf{\revision{Topographical} insights:} FLA can additionally quantify critical spatial features in the landscape, e.g., local optima and ruggedness, to reveal landscape structure. 
    \item \textbf{Hardness quantification:} The landscape metrics can be used to assess the difficulty of a problem with respect to an optimization algorithm.
\end{itemize}


\subsubsection{Distance Measures and Neighborhood}

\label{subsection:distance_measures_and_neighborhood}
A key aspect of the fitness landscape is the concept of distance between two configurations $\textbf{\textit{c}}_i$ and $\textbf{\textit{c}}_j$ (i.e., $d(\textbf{\textit{c}}_i, \textbf{\textit{c}}_j)$), where  $\textbf{\textit{c}}_i$ and $\textbf{\textit{c}}_j$ $\in\mathcal{C}$. \revision{The most common distance metrics include Hamming distance, Manhattan distance, and Euclidean distance. Based on the distance measure(s), the neighborhood $\mathcal{N}(\boldsymbol{c})$ of a configuration $\boldsymbol{c}$ is commonly defined as the set of configurations differing from \textbf{\textit{c}} within a small distance $\mathcal{\epsilon}$, i.e., $\mathcal{N}(\boldsymbol{c})=\left\{\boldsymbol{c}^{\prime} \mid d\left(\boldsymbol{c}^{\prime}, \boldsymbol{c}\right)\leq \epsilon \right\}$~\cite{conway1994course}.}




\subsubsection{Local Optima}

Local optima capture the underlying local structural properties of a problem's landscape. Formally, a configuration \textbf{\textit{c}}$^{\ell}$ is considered as a local optimum if $f(\textbf{\textit{c}}^{\ell})$ is better than $f(\textbf{\textit{c}})$, $\forall \textbf{\textit{c}} \in \mathcal{N}\left(\textbf{\textit{c}}^{\ell}\right)$, where $\mathcal{N}\left(\textbf{\textit{c}}^{\ell}\right)$ is the neighborhood of \textbf{\textit{c}}$^{\ell}$.

\section{The \approach~Methodology}
\label{section:methodology}

As illustrated in Figure \ref{fig:methodology}, \approach~consists of three key phases: \textbf{Configuration Spatiality Analysis} (dynamic data analysis), \textbf{Configuration Domain Analysis} (static domain analysis), and \textbf{Domain-Spatiality Synergy}. The former two can run simultaneously, extracting the spatial information of the configuration landscape and domain-specific knowledge, respectively. Deriving from those, the last phase synergizes them to provide intuitive and comprehensive explanations to the configuration tuning for the relevant system(s).

The information analyzed in \textbf{Configuration Spatiality Analysis} is almost applicable to any scenarios and systems. In contrast, depending on the goal, the levels of abstraction in the \textbf{Configuration Domain Analysis} can be tailored to fit the requirements:

\begin{itemize}
    \item When there is needed to comparatively compare different systems and workloads, then this phase can be applied to the option, system, and workload levels.
    \item When only one system is concerned, one can emphasize on the option and workload levels.
    \item When there is only one system and one merely needs to understand the tuning under a specific workload, then \approach~can be adopted to the option level only.
\end{itemize}

In what follows, we delineate every phase in \approach.


    




\begin{figure}[t]
    \centering
    \includegraphics[width=0.9\linewidth]{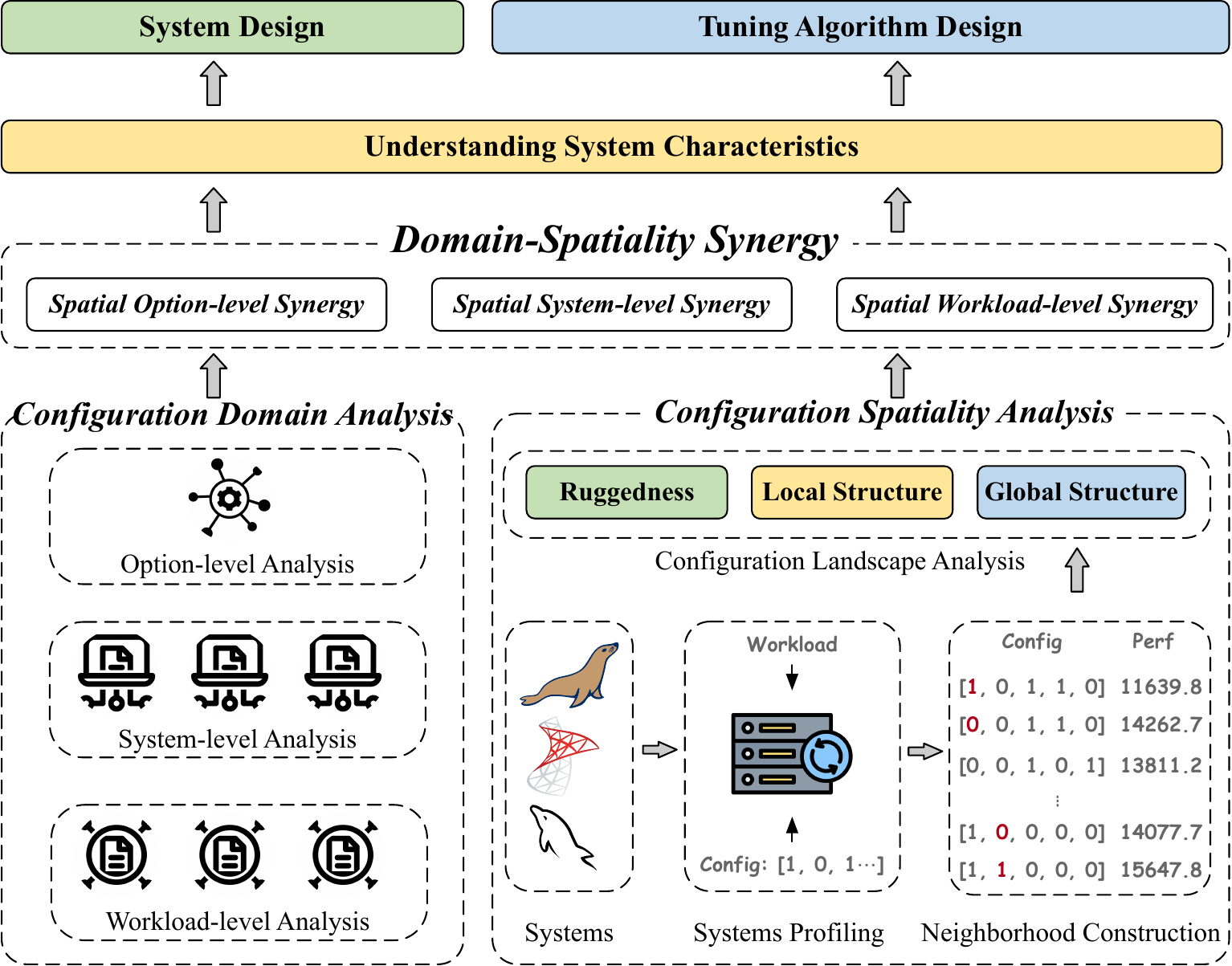}
    \caption{Overview of the \approach~methodology.}
    \label{fig:methodology}
    \vspace{-5pt}
\end{figure}

\subsection{Configuration Spatiality Analysis}
In data analysis, we leverage FLA to explore and quantify the topographical characteristics of configurable software systems. By treating the configuration-performance as a structured landscape, \approach~provides spatial insights into the relationships between configurations and their associated performances. 



\subsubsection{Systems Profiling}

An essential part of any dynamic data analysis is data collection, and for our problem, this means profiling the configurable system(s). Formally, given an arbitrary software system $\mathcal{S}$, the corresponding configuration space $\mathcal{C}$, and performance objective $f$, we need to obtain a representative set of data points $\{(\textbf{\textit{c}}, f(\textbf{\textit{c}}))\}$ from $\mathcal{C}$. 

To that end, several approaches can be employed. For example, for certain scenarios, it is possible to use grid search that comprehensively covers the configuration space. However, since measuring configurable systems is often profoundly expensive~\cite{chen2021multi,jamshidi2016uncertainty}, it makes sense to apply sampling-based methods~\cite{kaltenecker2019distance,siegmund2015performance} to select a subset of representative samples. In \approach, we support both categories of the sampling and it is up to the software engineers to choose depending on the cost and coverage trade-off~\cite{ henard2015combining}.

\subsubsection{Neighborhood Construction}
After collecting the configuration-performance data, constructing the configuration landscape requires defining a neighborhood structure $\mathcal{N}$, which determines how the configurations are connected based on a distance measure $d(\textbf{\textit{c}}_i, \textbf{\textit{c}}_j)$. The choice of this measure is crucial, as it directly influences the topographical properties of the landscape that will be measured. The most common distance metrics include Hamming distance, Manhattan distance, and Euclidean distance.



\subsubsection{Configuration Landscape Analysis}
\label{sec:cl-analysis}

Leveraging the definition of neighborhood, we can now analyze the spatial information of a configuration landscape in the following three aspects\footnote{The detailed specification of the chosen landscape metrics will be introduced in Section \ref{subsubsection:landscape_metrics}.}:




\textbf{\textit{Global structure}} captures a holistic view of configuration-performance spatial information, assessing how easily or hardly a tuner can navigate toward the global optimum. \approach~quantifies this aspect using two metrics:
\begin{itemize}
    \item \textbf{Fitness distance correlation (FDC, $\varrho$)}~\cite{jones1995fitness} measures the correlation between performance and proximity to the global optimum, capturing the overall searchability of the landscape. \revision{As suggested in~\cite{altenberg1997fitness,jones1995fitness, clergue2002fitness}, a correlation $\varrho$ $\geq$ 0.15 indicates a globally guided search, whereas  $\mathcal{-}0.15 \textless \varrho \textless 0.15$ and $\varrho \leq \mathcal{-} 0.15$ reflects irregular and deceptive structures, respectively.} 
    \item \textbf{Basin of attraction}~\cite{malan2021survey} refers to the set of points in the search space that eventually converge to a global/local optimum under a local search process. A larger attraction basin indicates that more configurations converge to the optimum, making it easier to reach. Here, we regard the basin size larger than $21\%$ of the whole search space as an ``easy'' problem, given the fact that randomly sampling $20$ configurations (same as many population-based search in software configuration tuning \cite{chen2024mmo,DBLP:journals/corr/abs-2501-00840}) for local search will lead to a probability of $99\%$ of locating at least one optimum in the search space (i.e., $1-(1-0.99)^{1/20}\approx0.21$).
\end{itemize}

\revision{Fitness distance correlation captures the directional guidance of the landscape, i.e., whether better configurations tend to be closer to the optimum, reflecting global searchability or deceptiveness. Basin of attraction reflects the likelihood that the global optimum can be reached from random initial configurations, quantifying the global optimum's attractiveness. Together, they provide an interpretable and practical assessment of the tuning difficulty from a global perspective~\cite{malan2013survey}.}
    
\textbf{\textit{Local structure}} examines the spatial distribution of suboptimal configurations in the landscape, assessing \textit{how frequently local traps appear; their quality against the global optimum; how strong their attractiveness is}. In \approach, we capture these properties through the following: 
\begin{itemize}
\item \textbf{The number of local optima} quantifies the modality of a landscape, indicating how many distinct suboptimal configurations exist.


\item \textbf{The quality of local optima} measures the performance of suboptimal configurations relative to the global optimum, indicating how competitive these configurations are within the landscape. {Here, we classify the performance of local optima into three tiers based on the relative quality metric $\ell_q$: high-quality ($\ell_q\geq0.67$); medium quality ($0.33 \textless \ell_q \textless 0.67 $); and low quality ($\ell_q \leq 0.33$).}


\item \textbf{Basins of attraction} of the local optima reflect the strength of their pull and how likely the tuning is to be trapped.
\end{itemize}

\revision{The number of local optima reflects how fragmented the landscape is, indicating how likely a tuner is to be caught in multiple competing regions~\cite{huang2024rethinking}. The quality of local optima indicates whether these regions are sufficiently competitive relative to the global optimum, which determines how acceptable a locally converged configuration may be. The basins of attraction quantify the attractiveness of local optima, indicating how likely a tuner is to reach a particular region~\cite{malan2021survey}. By jointly analyzing these dimensions, we can assess whether the landscape favors quick convergence to good enough configurations or demands sophisticated strategies to escape suboptimal traps.} \revision{For example, when local optima exhibit competitive performance and hold a large basin size, a tuner might easily locate satisfactory configurations. Conversely, an abundance of low-quality local optima with a large basin size can mislead the search and increase the risk of convergence to suboptimal regions, requiring stronger exploratory capabilities.}

    
\textbf{\textit{Ruggedness}} captures the degree of fluctuation in performance with respect to the neighboring configurations, indicating whether the landscape exhibits smooth trends or abrupt, irregular changes. {This reflects the difficulty of navigating the landscape.} \approach~measures this property using \textbf{autocorrelation} ($r$)~\cite{weinberger1990correlated}, which quantifies performance similarity across neighboring configurations. {A high autocorrelation ($r \geq 0.5$) suggests a smoother landscape; Moderate autocorrelation ($0.2 \textless r \textless 0.5$) indicates medium ruggedness; And low autocorrelation ($r \leq 0.2$) suggests a highly rugged landscape with sharp performance fluctuations~\cite{stadler1996landscapes,weinberger1990correlated,huang2024rethinking}. }


\revision{The landscape metrics used in \approach~and their implications for configuration tuning are summarised in Table~\ref{tb:metric_summary}. The detailed specification of these metrics will be introduced in Section \ref{subsubsection:landscape_metrics}. Notably, the metrics used in our analysis are chosen based on their well-established roles in the FLA literature~\cite {malan2013survey,zou2022survey,malan2013survey} and their demonstrated applicability in the software engineering community~\cite{huang2024rethinking, du2024contexts, chen2025accuracy,chen2022planning}. While we adopt a specific set of metrics in this study, our methodology is not limited to these choices. It remains extensible and can incorporate alternative metrics depending on the characteristics of the target systems (e.g., search-space type, dimensionality, sparsity) or the specific analysis goals (e.g., neutrality detection, evolvability assessment). We will further discuss this in Section \ref{section:threats_to_validity}.}

In summary, the above configuration spatiality analysis lays the groundwork for extracting spatial insights into system behaviors, performance trends, and landscape navigability, which, in turn, serve as a foundation for guiding higher-level decisions (e.g., tuner selection/design).

\begin{table*}[t]
    \centering
    \caption{\revision{Summary of landscape metrics and their implications for configuration tuning.}}
    \renewcommand{\arraystretch}{3.0}
    \scriptsize

\begin{tabular}{p{1.7 cm}| p{2.3 cm}| p{3 cm}| p{5.7 cm}}
\hline
\textbf{Landscape} & \textbf{Metric} & \textbf{Interpretation} & \textbf{Implications for tuner design/selection} \\
\hline

\multirow{5}{*}{\textbf{Global structure}} 
& \multirow{3}{*}{FDC ($\varrho$)~\cite{jones1995fitness}} 
& Guided: $\varrho \geq 0.15$ & 
\multirow{3}{=}{\vspace{0.1em}A high correlation indicates that better configurations lie closer to the optimum, revealing predictable structural characteristics that favour exploitation-driven strategies (e.g., hill climbing~\cite{li2014mronline,xi2004smart}). In contrast, weak or negative correlations illustrate that the landscape provides little or misleading structural signal, requiring exploration-driven methods (e.g., SWAY~\cite{chen2018sampling}, GA~\cite{olaechea2014comparison,back1993overview}).\vspace{0.1em}} \\
& & Irregular: $-0.15 \textless \varrho \textless 0.15$ & \\
& & Deceptive: $\varrho \leq -0.15$ & \\
\cline{2-4}

& \multirow{2}{*}{Basin of Attraction~\cite{malan2021survey}} 
& $> 0.21$: easy to reach optimum& 
\multirow{2}{=}{\vspace{0.1em}A large global basin implies that even simple local search can easily reach the optimum. A small basin, by contrast, calls for exploration-enhancing mechanisms such as restarts or diversity injection to avoid premature convergence (e.g., ParamILS~\cite{hutter2009paramils}).\vspace{0.1em}} \\
& & $\textless 0.01$: hard to reach optimum & \\
\hline

\multirow{7}{*}{\textbf{Local structure}} 
& \multirow{2}{*}{Number ($\ell_p$)} 
& Large: high multimodality & 
\multirow{2}{=}{\vspace{0.1em}A large number of local optima increases the risk of premature convergence, favoring diversity-preserving methods (e.g., multi-objectivization~\cite{chen2024mmo, chen2021multi}). Fewer optima allow effective use of deterministic local search (e.g., hill climbing~\cite{li2014mronline,xi2004smart}).} \\
& & Small: low multimodality & \\
\cline{2-4}

& \multirow{3}{*}{Quality ($\ell_q$)} 
& High: $\ell_q \geq 0.67$ & 
\multirow{3}{=}{\vspace{0.1em}High-quality local optima suggest that suboptimal configurations may still be acceptable, enabling smarter budget allocation and reducing unnecessary exploration. In contrast, low-quality local optima require escape strategies (e.g., multi-objectivization~\cite{chen2024mmo, chen2021multi}) to avoid stagnation in poor regions.\vspace{0.1em}} \\
& & Medium: $0.33 \textless \ell_q \textless 0.67$ & \\
& & Low: $\ell_q \leq 0.33$ & \\
\cline{2-4}

& \multirow{2}{*}{Basin of Attraction~\cite{malan2021survey}} 
& $\textgreater 0.21$: strong attractiveness & 
\multirow{2}{=}{\vspace{0.1em}The larger the basins of local optima, the more likely the search will be attracted and trapped, motivating the use of restart or perturbation mechanisms (e.g., ParamILS~\cite{hutter2009paramils}).\vspace{0.1em}} \\
& & $\textless 0.01$: weak attractiveness & \\
\hline

\multirow{3}{*}{\textbf{Ruggedness}} 
& \multirow{3}{*}{Autocorrelation ($r$)~\cite{weinberger1990correlated}} 
& Smooth: $r \geq 0.5$ & 
\multirow{3}{=}{\vspace{0.1em}Smooth landscapes imply gradual performance transitions between neighbors, enabling model-based methods to generalize effectively (e.g., FLASH~\cite{nair2018finding}, SMAC~\cite{lindauer2022smac3}, TPE~\cite{bergstra2011algorithms}). Highly rugged landscapes exhibit abrupt performance fluctuations between neighboring configurations, favoring exploration-driven strategies (e.g., GA~\cite{olaechea2014comparison,back1993overview}).\vspace{0.1em}} \\
& & Moderate: $0.2 \textless r \textless 0.5$ & \\
& & Rugged: $r \leq 0.2$ & \\
\hline
\end{tabular}

    \label{tb:metric_summary}
\end{table*}

\subsection{Configuration Domain Analysis}
\label{subsection:domain-level analysis}

Alongside configuration spatiality analysis, the other important phase in \approach~is the static domain analysis, mining characteristics specific to the system(s) being studied. In \approach~we provide generic ``domain features'' (or categories) extracted through our experiences and empirical observations. Those features, together with detailed guidelines, provide structural domain information to be synergized with the spatial information obtained previously.


To that end, three aspects are important as below.



\subsubsection{System Level}
\label{subsubsection:system-level}

For analyzing at the system level, we consider the following categories of features that are important for domain analysis and for cross-system comparisons:

\begin{itemize}
    \item \textbf{Area:} A nominal feature that covers different application areas of the system, e.g., \textsc{H2} is for database,  \textsc{X264} is for video encoding, \textsc{Kanzi} is for compression. 
    \item \textbf{Language:} A nominal feature describes the main languages that built a system, such as Java and C++.
    \item \textbf{Resource intensity:} According to the nominal feature of resource intensity, most systems can be classified into four features: CPU, I/O, memory, and storage. Note that a system might belong to more than one type.
    \item \textbf{Complexity:} This numeric feature reflects the system's complexity and configurability, e.g., features such as lines of code (LOC), number of options, and configuration space.

    
\end{itemize}

The process of representing a given system using the above features is straightforward, as this is the common domain knowledge that most software engineers would have. For example, it is easy to know that the system \textsc{H2} {is a database built in Java, characterized by high I/O, memory, and storage intensity, reflecting its resource demands for transactional operations}.


\subsubsection{Workload Level}
\label{subsubsection:workload-level}

Since a system often runs in diverse workloads, which can profoundly impact the performance~\cite{muhlbauer2023analyzing}, \approach~can analyze the workload level characteristic via two features:


\begin{itemize}
    \item \textbf{Workload Type:} A nominal feature that defines the nominal nature of workload inputs and processing tasks, for example, when tuning the system \textsc{Dconvert}, the workloads vary by image format such as \texttt{png} and \texttt{svg}; for system \textsc{H2}, they vary by transaction type (e.g., \texttt{tpcc} and \texttt{ycsb}).
    \item \textbf{Workload Scale:} An ordinal feature representing magnitude of the workload in terms of input size or operational intensity, e.g., for \textsc{Dconvert}, workloads are divided into small, medium, and large scales of the image; for system \textsc{x264}, workloads are classified into small, medium, and large scales of the resolution/file size.
\end{itemize}

Similar to the features at the system level, representing a workload with respect to the above features can often be easily achieved with domain understanding of the system.


\subsubsection{Option Level}
\label{subsubsection:option-level}

\approach~encourages domain analysis at the option level, relying on understanding the code logic related to the configuration options. As such, the results would provide fine-grained details on the system's internal working structure. Particularly, we categorize the options using a single nominal feature, representing two groups of types, i.e., \textbf{\textit{functional feature}} and \textbf{\textit{resource feature}}, as shown in Table~\ref{tb:option-taxonomy}. \rev{Specifically, functional options capture logic- or behavior-related mechanisms that determine what a system does, whereas resource options capture control parameters that influence how efficiently the system executes (e.g., CPU, memory, or storage).}

\begin{table*}[t]
    \centering
    \caption{Domain features of option for configurable systems.}
    \begin{adjustbox}{width=1\textwidth,center}
    
\footnotesize
\begin{tabular}{llp{7cm}l}
    \toprule
    \multicolumn{2}{c}{\textbf{Type}} & \textbf{Descriptions} & \textbf{Exampled Option}  \\ 
    \midrule
    \multirow{2}{*}{\rotatebox{0}{Functional}} 
    &  {Core ($F_1$)} & Governs the system’s core logic, directly impacting execution behavior.  
            & \texttt{MVSTORE} in \textsc{H2}  \\ 
    & {Utility ($F_2$)} & Provides auxiliary functionalities applicable across multiple modules.  
              & \texttt{anyScriptOrigin} in \textsc{Batik} \\ 
    \midrule
    \multirow{4}{*}{\rotatebox{0}{Resource}} 
    & {CPU ($R_1$)} & Controls CPU usage, parallel processing, or multi-threading.  
                    & \texttt{threads} in \textsc{Dconvert} \\ 
    & {Storage ($R_2$)} & Manages persistent data storage.  
              & \texttt{VBRRange} in \textsc{Jump3r}  \\ 
    & {Memory ($R_3$)} & Handles cache, memory pools, and buffer.  
             & \texttt{blocksize} in \textsc{Kanzi}  \\ 
    & {Queue ($R_4$)} & Regulates scheduling, timeouts, and delays.  
            & \texttt{FlushTimeout} in \textsc{Xz} \\ 
    \bottomrule
\end{tabular}

    \end{adjustbox}
    \label{tb:option-taxonomy}
\end{table*}

When executing the option level analysis, it might require some efforts to categorize a system with respect to the above. Therefore, \approach~also provides guidelines to that end using both \textbf{\textit{manual comprehension}} and \textbf{\textit{code analysis}}~\cite{DBLP:conf/icse/LiangChen25}. Manuals (or API) are software documentation that contains rich information about the configuration options and their implications on the system~\cite{DBLP:conf/icse/LiangChen25}. In manual comprehension, one can follow the steps below:

\begin{enumerate}
    \item \textbf{Screening:} Given a list of keywords related to the features above\footnote{Details can be accessed at: \textcolor{blue}{\texttt{\url{https://tinyurl.com/7dt2zrd4}}}.}, one can filter which function descriptions are relevant to a feature. For example, ``compress'' and ``storing'' are common keywords in the manual of \textsc{H2} for storage options.
    
    \item \textbf{Option backtracking:} Once the functions are categorized into the above features, we can then examine whether each function's description mentions the name of the options (which is often available on \texttt{yaml} or other files). For example, \texttt{FlushTimeout} is an option for \textsc{Xz} named in its API.
    \item \textbf{Feature assignment:} Finally, we say an option belonging to a feature that contains the most relevant functions to that option. For example, in \textsc{Xz}, \texttt{FlushTimeout} has most functions belongs to \textit{Queue}, hence it is also of this type.
    
\end{enumerate}

While manual is important, it can still be misleading. To ensure precision, \approach~also suggests a semi-automatic code analysis:

\begin{enumerate}
    \item \textbf{Variable identification:} Knowing the names of the options does not mean that we know exactly where they are referred to as variables in the code. To that end, we use three methods to create the option-variable mappings: (1) finding in a centralized configuration file; (2) discovering the variables in the \texttt{main} file; and (3) searching the keywords on \texttt{setter} and \texttt{getter}, then track back to the variables. 
    \item \textbf{Taint analysis:} When the variables are localized, we then use taint analysis tools~\cite{javaparser,llvm} to mine all functions that can be affected by a variable, linking functions to the options. 
    
        
   \item \textbf{Feature assignment:} Same as the manual comprehension.
\end{enumerate}

\rev{We unionize the features assigned to the same options from both manual comprehension and code analysis, e.g., when an option $O$ is linked to $F_1$ from manual and to $R_1$ from code, then eventually we say $O$ is associated with both $F_1$ and $R_1$. As such, an option might be linked to more than one feature above. Since manual comprehension and code analysis involve human judgment rather than full automation, we adopt a reliability-controlled labeling process to ensure rigor. Specifically, multiple annotators independently performed the classification and resolved disagreements through discussion or, when necessary, external consultation. We quantified the inter-rater agreement using Cohen's Kappa $\kappa$~\cite{emam1999benchmarking} and required $\kappa \geq 0.7$ before finalizing the labels. This multi-annotator, reliability-validated process helps mitigate subjectivity of the option-level categorization.}




\rev{Considering that manual option classification can be labor-intensive and difficult to scale across large systems, especially when multiple annotators are required to ensure reliability, \approach~optionally incorporates large language models (LLMs) as additional annotators to reduce human workload and improve scalability. In this setting, the LLM acts as an independent labeling agent, analogous to a human annotator in the multi-annotator workflow. The LLM’s outputs are then compared with the human-derived labels, where any conflicts will trigger human review.  This governance structure preserves consistency, traceability, and reliability while reducing manual burden.}




\rev{Specifically, the LLM-assisted workflow consists of the following steps:}

\begin{itemize}
  \item [(1)] \textbf{Label/rule definition:} Before applying LLM-based classification, we should first explicitly define the labeling criteria using the domain taxonomy in Table~\ref{tb:option-taxonomy}. This includes: (i) the top-level distinction between functional options ($F_1:$ Core, $F_2:$ Utility) and resource-related options ($R_1-R_4:$ CPU, Storage, Memory, Queue), and (ii) the decision rules for determining which category an option belongs to. These definitions are embedded directly into the prompt template (see Figure~\ref{fig:Identifier-prompt}), guiding the LLM through a transparent reasoning process.
  \item [(2)] \textbf{Input preparation:} For each configuration option, we then construct a structured input that aggregates all available evidence: the option name, a semantic description extracted from manuals/APIs, representative code snippets, and brief system background. These elements are organized in a fixed order (system context $\rightarrow$  option name $\rightarrow$ documentation description $\rightarrow$ code snippets), with the description serving as the primary decision basis and the other information, if any, as auxiliary context.  
  \item [(3)] \textbf{LLM inference:} The prepared evidence is then fed into the prompt template (see Figure~\ref{fig:Identifier-prompt}), which already includes the labeling rules defined before and a few short illustrative examples. Given this structured input and guidance, the LLM will produce one label from the closed set \{$F_1, F_2, R_1, R_2, R_3, R_4$\}, encoded as a single ASCII label. 
  \item [(4)] \textbf{Label adjudication:} After LLM inference, the LLM-generated labels are compared with the human-derived labels. Any disagreement between the two is explicitly flagged and routed for adjudication. Notably, when clear documentary/code evidence supports one label over the other, the label supported by that evidence should be adopted. When no such decisive evidence exists, the disagreement is simply recorded and both labels are retained for subsequent agreement analysis. This ensures that corrections occur only when justified, preserving both the reliability of the labels while keeping manual intervention limited.
\end{itemize}

\begin{figure}[t] 
\input{figures/prompt}
\caption{Prompt template. \mytag{\phantom{ab}} denotes the content that will be substituted by actual data.}
\label{fig:Identifier-prompt}
\end{figure}

\rev{Similarly, to maintain consistency with the reliability-controlled process described above, we treat the labels generated across all annotators, including human annotators and the LLMs, as independent labeling sources and assess their agreement using Cohen's Kappa $\kappa$. Specifically, we first compute $\kappa$ among the human annotators to quantify baseline human-human agreement. This yields an initial reliability measure based solely on human judgments. Then, the LLM annotators are incorporated sequentially. Let the consensus label set produced by the human annotators be denoted as $\mathbfcal{H}$. For each LLM annotator, we compute a separate agreement score against this human consensus. For example, with three LLMs, we can obtain $\kappa(\mathbfcal{H},LLM_1)$, $\kappa(\mathbfcal{H},LLM_2)$, and $\kappa(\mathbfcal{H},LLM_3)$. The overall inter-rater consistency is then defined as the average of the human-human $\kappa$ and these human-LLM $\kappa$ values, providing a measure of agreement across all annotators while preventing the disagreements between LLMs from dominating the results. If the overall agreement falls below the commonly accepted threshold ($\kappa \textless 0.7$), we follow a standard reconciliation procedure used in prior studies~\cite{wang2022machine,cheriyan2021towards}:
\begin{itemize}
    \item [(1)] \textbf{Identify disagreements:} We locate the specific options where annotators assigned different categories.
    \item [(2)] \textbf{Evidence-guided review:} For each conflicting option, the annotators jointly review the corresponding manuals and code snippets, discussing the conflicting interpretations and striving to reach a consensus grounded in the documentary/code evidence. 
    \begin{itemize}
        \item When the evidence clearly supports one category, the label is corrected accordingly.
        \item When the evidence remains ambiguous, no forced correction is made at this stage.
    \end{itemize}
    \item [(3)] \textbf{Reassessment:} $\kappa$ is recalculated over all options to assess whether the disagreements have been sufficiently reduced.
\end{itemize}
}

\rev{This reconciliation may be repeated when necessary, but only evidence-supported corrections, i.e., those justified by documentation/code, are introduced in each round. Once the overall agreement reaches $\kappa \geq 0.7$, the labeling process is deemed reliable. For any remaining options without decisive evidence, the majority (or most consistently assigned) label across annotators is used. This procedure ensures that every option receives an evidence-grounded and reliable final label.
}

\subsection{Domain-Spatiality Synergy}

To synergize domain-spatiality information, we use the three levels of domain knowledge as the root (option, system, and workload), associating each of them with the one (or three) aspect of spatial information from landscape, i.e., global structure, local structure, and ruggedness. \rev{This domain-spatiality synergy provides an analytical integration between domain contextual and spatial landscape characteristics. Specifically, domain knowledge (e.g., option category, system property, or workload feature) serves as contextual grounding for interpreting spatial landscape metrics, allowing us to observe and characterize how domain-related factors correspond to variations in landscape properties (e.g., certain option types or systems being associated with higher ruggedness or distinct local structures).} The results can be concluded to form an explanation, linking the domain knowledge and the design/selection of tuner/system via properties of the configuration landscape.

\subsubsection{Spatial Option-level Synergy}
\label{subsubsection:spatial_option_level_synergy}
Here, \approach~suggests to only measure the ruggedness of landscape (via autocorrelation). This is because ruggedness inherently reflects how an option influences performance variability, with certain settings simplifying dramatic neighboring performance fluctuations. \rev{While traditional sensitivity and feature-importance analyses (e.g., ANOVA~\cite{st1989analysis}, SHAP~\cite{lundberg2017unified}, and LIME~\cite{ribeiro2016should}) focus on how much a configuration option contributes to overall performance variation, often through mean differences (ANOVA) or model-based attributions (SHAP/LIME), our notion of ruggedness-sensitivity instead captures how an option reshapes the structure of the configuration landscape. Specifically, it measures how the option alters fluctuation intensity across neighboring configurations, as reflected by autocorrelation under the neighborhood structure. From a landscape perspective, this option-level analysis goes beyond identifying performance-critical options; it reveals how certain options affect the search dynamics of different tuning strategies. For example, exploitation-driven tuners (e.g., Hill Climbing~\cite{li2014mronline,xi2004smart}) are more sensitive to rugged landscapes, where prioritizing ruggedness-sensitive options might improve convergence efficiency. Tuners with strong exploration (e.g., Genetic Algorithm~\cite{olaechea2014comparison,back1993overview}) are inherently more resilient to ruggedness and may show limited benefit from such prioritization. In contrast, with the classic ANOVA, it is likely to recommend that any tuner that takes this advantage (important options) would be benefited, e.g., both HC and GA, i.e., ANOVA cannot discriminate/link the behaviors of different tuner designs. Hence, our spatial option-level analysis complements feature-importance approaches by uncovering when and why different search strategies succeed or fail under different landscape structures. In this work, we focus on option-specific ruggedness, analyzing how each individual configuration option affects the smoothness of the landscape. This choice allows for interpretable insights into option-level structural influences while keeping the analysis consistent with the option-level perspective in Section~\ref{subsubsection:option-level} where each configuration option is examined independently.} To evaluate how a single option's change influences the ruggedness of the configuration landscape, we perform the following on the profiled configuration data for each option:

\begin{enumerate}
    \item \textbf{Data partitioning:} Given a configuration option taking different values, we partition the dataset into symmetric subsets based on the option's values, ensuring all other factors remain identical except for the investigated option. 
    \item \textbf{Autocorrelation computation:} Once the subsets are obtained, we can then calculate the autocorrelation metric for each subset, quantifying the ruggedness of the corresponding configuration landscape.
    \item \textbf{Impact assessment:} Finally, we evaluate the option's effect by analyzing the difference in autocorrelation values between subsets and computing the relative standard deviation (RSD) to measure the variability in ruggedness.
\end{enumerate}


To link the outcome of ruggedness assessment to the types of options, one can summarize the common patterns on the interpretation of the ruggedness and the option's type in several ways:

\begin{itemize}
    \item For a single system-workload pair. 
    \item For all workloads of a system.
    \item For all workloads across all systems.
\end{itemize}

There is no requirement to complete all of the above actions, but a subset of them can be selected depending on the needs. Each of the above could lead to a finding. Even if no clear pattern can be observed, a conclusion related to all systems in general can also be made. For example, {if it is observed that for a system $\mathcal{S}$ changing the core options ($F_1$), even across workloads, can lead to rugged configuration landscape ($r \textless 0.2$), we can find that: ``\textit{Core options exert a stronger impact on shaping the ruggedness of system $\mathcal{S}$''.}}

\subsubsection{Spatial System-level Synergy}

Here, we recommend using all landscape metrics included in \approach. Since each workload can significantly impact the configuration landscapes, we have a total of $T$ landscapes to analyze where $T$ is the total number of system-workload pairs.

To link the outcomes of landscape metrics to the features of systems, one can summarize the common patterns concerning each landscape metric with respect to system-level features for all systems. For example, we might find that ``\textit{the configuration landscape is generally highly rugged regardless of the systems and their features}''; yet, in another example, {if it is found that most systems within an area (e.g., database) have a larger proportion of local optima than others (e.g., compression), the finding could be: ``\textit{database systems are prone to getting trapped into local optima compared to compression systems, requiring tuners with stronger global exploration capabilities to avoid early stagnation''.}}


Additionally, multiple landscape metrics can be combined to create a new connection to the domain knowledge. {For example, if it is observed that in most systems with different workloads, the performance of local optima demonstrate a strong Spearman correlation ($\rho \textgreater 0.69$) with their basins of attraction, our conclusion is: ``\textit{in most software systems, high-quality local optima tend to exhibit larger basins''.}}

Since the system and option features can be correlated, the above outcomes can also be linked to the commonality of option types, but we are less concerned about the options' changes. For example, {if it is found that those systems with a larger proportion of core options consistently show more rugged landscapes, the finding could be: \textit{``tailoring tuners equipped with global exploration might be promising for tuning those systems with a higher proportion of core options''.}}



\subsubsection{Spatial Workload-level Synergy}

Again, for workload types and scales, all landscape metrics in \approach~can be used. To correlate the outcomes of landscape metrics to those domain features, one can summarize the common patterns concerning each landscape metric with respect to workload level features in two ways:

\begin{itemize}
    \item For all workloads of a system.
    \item For all workloads across all systems.
\end{itemize}

Each of the above actions can be chosen depending on the needs. For example, if no clear patterns can be observed between workload features and the landscape metrics, then the finding can be: {``\textit{workload effects on ruggedness are not uniformly tied to type or scale. Both contribute to variations, but their impact is system-dependent''}}. It is also possible to link the conclusions to the commonality of system/option level features. For example, {if a system $\mathcal{S}$ exhibits a greater ruggedness fluctuation only in particular contexts (e.g., under a workload feature and an option type), the finding can be: “\textit{there might be interactions between a particular workload feature and option type, making tuning under such extremely challenging~\cite{DBLP:journals/corr/abs-2501-00840}.”}}



Again, the metrics can be combined as, e.g., correlation, to further link to the domain knowledge.

\section{How to Use \approach? A Case Study}
\label{section:empirical_setup}

To evaluate how \approach~can be used, we conduct a case study on real-world configurable systems, aiming to answer the following research questions (RQs):

\begin{itemize}
    \item \textbf{RQ1 \rev{(Option-level)}:} How do options shape the spatiality of configuration landscape?

    \item \textbf{RQ2 \rev{(System-level)}:} What spatiality can be extracted across the systems?

    \item \textbf{RQ3 \rev{(Workload-level)}:} How do the workloads impact configuration landscapes?
    
    

\end{itemize}

\revminor{Overall, these research questions focus on characterizing configuration tuning problems by jointly examining configuration landscape spatiality and domain knowledge at the option, system, and workload levels, with the goal of deriving insights for interpreting tuning difficulty and informing tuning algorithm design.}

\rev{Specifically, \textbf{RQ1} seeks to explore how configuration options and their domain characteristics (\textit{functional} and \textit{resource} options) shape the spatiality of the configuration landscape. \textbf{RQ2} investigates spatial characteristics across different configurable systems. By comparing multiple systems, this question examines whether domain attributes at the system level (e.g., system types, programming language, or resource intensity) lead to specific spatial patterns in their configuration landscapes. \textbf{RQ3} analyzes how workload features influence configuration landscapes within the same system. By comparing landscape metrics across workloads of different types and scales, this question aims to uncover whether specific workload features consistently shape the landscape structure.} \revminor{Through answering these research questions, we reveal domain-spatiality patterns in configuration tuning and demonstrate how these patterns can be translated into actionable insights (see Sections~\ref{subsection:insights} and \ref{subsection:validation}) and practical implications (see Section~\ref{section:implication}).}

\subsection{Systems Profiling}
\label{subsection:subject system}

\rev{Our methodology is designed to be generally applicable to any configurable system. In practice, however, conducting large-scale configuration evaluations from scratch would be prohibitively time-consuming and resource-intensive. For example, measuring a single configuration of \textsc{MySQL} often takes several minutes even under controlled conditions~\cite{gong2024deep}. To ensure both representativeness and practical feasibility, our case study uses data collected by Muhlbauer et al.~\cite{muhlbauer2023analyzing} using selective sampling~\cite{siegmund2015performance,kaltenecker2019distance}, which provides extensive performance measurements across a wide range of configurable systems and diverse workload settings. In total, this dataset covers nine systems and 93 workloads, including roughly {25,258} configurations. We chose this dataset for two key reasons:}

\begin{itemize}
    \item \rev{\textbf{Comprehensive coverage:} Compared with other publicly available datasets, which often evaluate configurations under a single workload~\cite{nair2018finding,weber2023twins} or focus on less diverse systems~\cite{jamshidi2018learning,jamshidi2017transfer,huang2024rethinking}, the dataset collected by Muhlbauer et al.~\cite{muhlbauer2023analyzing} spans multiple domains and provides 6-13 diverse workloads per system. Such diversity enables our multi-level analysis across options, systems, and workloads.}
    \item \rev{\textbf{Transparency:} The dataset provides detailed metadata for each system, including system versions, configuration options, workload sources, and corresponding system source code. Such transparency allows us to precisely trace how each configuration and workload is defined and executed, enabling domain-level analyses such as taint analysis in our option-level study.}
\end{itemize}

{Details can be found at the repository of Muhlbauer et al.: \textcolor{blue}{\texttt{\url{https://zenodo.org/records/7504284}}}.
}

\begin{table*}[t!]
	\centering
	\footnotesize
	\tabcolsep=0.1 cm
	\caption{Subject system characteristics. The workload details can be accessed via \textcolor{blue}{\texttt{\url{https://tinyurl.com/546svkb2}}}.}
	\label{tb:subject_sas}
	\begin{threeparttable}
		
\begin{tabular}{llllllcccc}
\toprule
\textbf{System} & \textbf{Language} & \textbf{Area} & \textbf{Performance} & \textbf{Version}             & \textbf{Resource Intensity}  & \textbf{LOC} & \textbf{\#O} & \textbf{\#C} & \textbf{\#W} \\ \midrule
\textsc{Jump3r}~\cite{subjectsys_jump3r}          & Java              & Audio Encoder   & Runtime              & 1.0.4                        & CPU           & 29685        & 13           & 4196         & 6            \\
\textsc{Kanzi}~\cite{subjectsys_kanzi}           & Java              & File Compressor & Runtime              & 1.9                          & I/O; Memory         & 28614        & 18           & 4112         & 9            \\
\textsc{Dconvert}~\cite{subjectsys_dconvert}        & Java              & Image Scaling   & Runtime              & 1.0.0-$\alpha7$ & I/O; Memory         & 6888         & 10           & 6764         & 12           \\
\textsc{H2}~\cite{subjectsys_h2}              & Java              & Database        & Throughput           & 1.4.200                      & I/O; Memory; Storage & 333539       & 16           & 1954         & 8            \\
\textsc{Batik}~\cite{subjectsys_batlik}          & Java              & SVG Rasterizer  & Runtime              & 1.14                         & CPU           & 360924       & 8           & 1919         & 11           \\ \midrule
\textsc{Xz}~\cite{subjectsys_xz}              & C/C++             & File Compressor & Runtime              & 5.2.0                        & CPU           & 43130        & 12           & 1999         & 13           \\
\textsc{Lrzip}~\cite{subjectsys_lrzip}           & C/C++             & File Compressor & Runtime              & 0.651                        & CPU           & 20797        & 7           & 190          & 13           \\
\textsc{X264}~\cite{subjectsys_x264}            & C/C++             & Video Encoder   & Runtime              & baee400…                     & CPU; I/O      & 86740        & 25           & 3113         & 9            \\
\textsc{Z3}~\cite{subjectsys_z3}              & C/C++             & SMT Solver      & Runtime              & 4.8.14                       & CPU           & 636268       & 12           & 1011         & 12           \\ \bottomrule
\end{tabular}

		\footnotesize
		\#\textbf{O:} No. of options;
		\#\textbf{C:} No. of configurations;
		\#\textbf{W:} No. of workloads tested.
	\end{threeparttable}
    \vspace{-8pt}
\end{table*}

\subsubsection{Systems}

The datasets cover well-known configurable systems, which have been widely studied in the literature~\cite{muhlbauer2023analyzing,alves2020sampling,velez2020configcrusher,weber2021white}. The chosen systems span diverse application areas, performance objectives, and languages, providing a robust foundation for evaluating system characteristics, as shown in Table \ref{tb:subject_sas}. The system features can be easily summarized based on domain expertise. More details on how to use these systems for experiments can be found in \cite{muhlbauer2023analyzing,workload_similarity_dataset}.

\subsubsection{Workloads}

We use data measured on various workloads as collected by Muhlbauer et al.~\cite{muhlbauer2023analyzing} (see Table \ref{tb:subject_sas}). These workloads again have diverse types and scales, which can be directly summarized by checking the specifications of those workloads. For example, the system \textsc{X264} has various workloads that consist of different raw video frames, diverse resolutions, and file sizes.

\rev{Specifically, the workloads used for each system are summarized as follows:
\begin{itemize}
    \item \textbf{\textsc{Jump3r}} encodes raw \texttt{WAVE} audio signals to MP3, with variations in file size, signal length, sampling rate, and number of channels.
    \item \textbf{\textsc{X264}} encodes raw video frames (y4m format) across different resolutions (low/DS up to 4K) and file sizes.
    \item \textbf{\textsc{Kanzi}, \textsc{Xz}, and \textsc{Lrzip}} compress mixed-type files (text, binary, structured data, etc) or single-type files from community benchmarks.  Additionally, custom datasets, including the Hubble Deepfield image and a Linux kernel binary, are used to augment the workloads. For \textsc{Xz} and \textsc{Lrzip}, different parameterizations of UIQ2 benchmark\footnote{\url{http://mattmahoney.net/dc/uiq/}} are added to study the effect of varying file size. 
    \item \textbf{\textsc{Z3}} evaluates the satisfiability of logical problems in SMT2 format. The workload consists of the six longest-running instances from Z3’s performance test suite, augmented with additional instances from the SMT2-Lib\footnote{\url{https://smt-comp.github.io/2017/benchmarks.html}} repository to increase diversity in logic types.
    \item \textbf{\textsc{Batik}} transforms \texttt{SVG} vector graphic into a bitmap. The workload consists of resources from the Wikimedia Commons collection\footnote{\url{https://commons.wikimedia.org/wiki/Category:Images}}, varying primarily in file size.
    \item \textbf{\textsc{H2}} executs database transactions using different \textsc{OLTPBench}~\cite{difallah2013oltp} benchmarks (\texttt{Voter}, \texttt{SmallBank}, \texttt{TPC-H}, \texttt{YCSB}). The scale factor is varied to adjust transaction complexity.
    \item \textbf{\textsc{Dconvert}} transforms resources (e.g., image files) at different scales. The workload includes various input formats (\texttt{JPEG}, \texttt{PNG}, \texttt{PSD}, \texttt{SVG}) with variations in file size.
\end{itemize}
}
\rev{In summary, compared with other publicly available datasets that usually evaluate configuration under a single workload or a narrow set of input scenarios~\cite{nair2018finding,weber2023twins,jamshidi2018learning,jamshidi2017transfer}, the dataset collected by Muhlbauer et al.~\cite{muhlbauer2023analyzing} provides a substantially more comprehensive workload coverage. Each of the nine systems is tested under 6-13 distinct workloads, spanning different workload types, scales, and complexities.}

\subsubsection{Options}

We use the same options as Muhlbauer et al.~\cite{muhlbauer2023analyzing}. Each system has varying option types (e.g., integer, boolean, and enumerated) and dimensions. To categorize the options, we follow the procedure of \approach~as discussed in Section~3.2.3. \rev{Specifically, the options were independently labeled by the authors and by three LLM-based annotators (DeepSeek\footnote{https://www.deepseek.com/en}, Qwen\footnote{https://qwen.ai/home}, and GPT-5\footnote{https://chatgpt.com/}). The LLMs received the domain taxonomy and prompt template introduced in Section~3.2.3, while human annotators applied the same taxonomy and decision rules when reading manuals and source code. We then computed the inter-rater agreement over these labels using Cohen’s Kappa $\kappa$, following the procedures described in Section 3.2.3. First, we assessed the agreement between the two human annotators across all 121 options. In this initial round, the annotators agreed on 93 options and disagreed on 28 options, yielding an initial $\kappa$ of 0.6700. To resolve these disagreements, we followed the adjudication procedure described in Section 3.2.3: examining disagreements, checking the supporting documentation and code snippets, and updating labels only when the evidence clearly favored a particular category. After this reconciliation step, the two annotators reached a $\kappa$ of 0.7059, forming the evidence-grounded human consensus label set. Next, we compared this human consensus label set with the labels produced by each LLM-based annotator (i.e., DeepSeek, Qwen, and GPT-5), obtaining individual Cohen's Kappa $\kappa$ values of 0.6870, 0.7513, and 0.7357, respectively. Finally, we summarize the overall agreement by averaging the consistency scores across all annotators, including humans and LLMs, and obtain an overall $\kappa = 0.718$, demonstrating substantial agreement among annotators. The full labeled option set is available in our repository: \textcolor{blue}{\texttt{\url{https://tinyurl.com/4stvjurm}}}.}


\subsection{Customizing Fitness Landscape Analysis}


\subsubsection{Neighborhood Construction}
\label{subsubsection:neighborhood_construction}

In this case study, we use Hamming distance to define a configuration's neighborhood. \revision{Generally, when using Hamming distance as the distance metric, the neighborhood radius is set to $\epsilon$ = 1, i.e., $\mathcal{N}(\boldsymbol{c})=\left\{\boldsymbol{c}^{\prime} \mid d\left(\boldsymbol{c}^{\prime}, \boldsymbol{c}\right) \leq 1\right\}$ (such that the number of neighbors is $n$, where $n$ is the configuration dimensionality)~\cite{malan2013survey}.} However, when applying configuration spatiality analysis of \approach~to sampling-based datasets~\cite{muhlbauer2023analyzing}, the strict definition of a neighborhood as $d(\boldsymbol{c}^{\prime}, \boldsymbol{c}) \leq 1$ can leave many configurations without neighbors, particularly in sparse regions. Thus, in this case study, the neighborhood radius is expanded to ensure that the expected average number of neighbors for configurations is $n$, where $n$ is the configuration dimensionality.  This adjustment mitigates sparsity while ensuring meaningful connectivity for analysis and aligns with the neighborhood size in a complete binary configuration space, where each configuration has exactly $n$ neighbors at Hamming distance 1. Notably, in our case study, all metrics are computed following this neighborhood structure, ensuring consistency in landscape analysis. 

\rev{To evaluate the validity of adaptive neighborhood construction and examine whether it can effectively mitigate sparsity-induced distortion in landscape characterization, we conducted an empirical analysis comparing it with the conventional fixed radius approach. In this validation, we selected two systems, \textsc{7z} and \textsc{LLVM}, from an open-source dataset~\cite{weber2023twins}, as they possess high-dimensional configuration space representative of real-world tuning scenarios and provide complete configuration-performance data for compute ground-truth landscape metrics.}

\begin{table*}[t!]
	\centering
	\footnotesize
	\tabcolsep=0.093 cm
	\caption{\rev{Comparison of fitness landscape metrics computed under different neighborhood definitions and sampling granularities, where results from the full-scale dataset serve as the ground truth. $\ell_p$ denotes the proportion of local optima; $\ell_q$ represents the quality relative to the global optimum; $d$ is neighborhood radius. ``Fixed'' uses a constant neighborhood radius, i.e., $d = 1$, whereas ``Adaptive'' applies the proposed sparsity-aware radius.}}
	\label{tb:adaptive_validation}
    
\begin{tabular}{cc|cccc|cccc}
\hline
\multicolumn{2}{c|}{\textbf{System}} & \multicolumn{4}{c|}{\textbf{7z}} & \multicolumn{4}{c}{\textbf{LLVM}} \\ \hline
\multicolumn{1}{c|}{\textbf{Method}} & \textbf{Data Scale} & \multicolumn{1}{c|}{\textbf{$\ell_p$}} & \multicolumn{1}{c|}{\textbf{$\ell_q$}} & \multicolumn{1}{c|}{\textbf{Basin of Attraction}} & \textbf{$d$} & \multicolumn{1}{c|}{\textbf{$\ell_p$}} & \multicolumn{1}{c|}{\textbf{$\ell_q$}} & \multicolumn{1}{c|}{\textbf{Basin of Attraction}} & \textbf{$d$} \\ \hline
\multicolumn{1}{c|}{} & Large (50\%) & \multicolumn{1}{c|}{0.00758} & \multicolumn{1}{c|}{0.85737} & \multicolumn{1}{c|}{0.02092} & 1 & \multicolumn{1}{c|}{0.01230} & \multicolumn{1}{c|}{0.32979} & \multicolumn{1}{c|}{0.08344} & 1 \\
\multicolumn{1}{c|}{} & Medium (25\%) & \multicolumn{1}{c|}{0.03963} & \multicolumn{1}{c|}{0.54504} & \multicolumn{1}{c|}{0.00804} & 1 & \multicolumn{1}{c|}{0.12177} & \multicolumn{1}{c|}{0.15062} & \multicolumn{1}{c|}{0.02307} & 1 \\
\multicolumn{1}{c|}{\multirow{-3}{*}{Fixed}} & Small (5\%) & \multicolumn{1}{c|}{0.43036} & \multicolumn{1}{c|}{0.11389} & \multicolumn{1}{c|}{0.00000} & 1 & \multicolumn{1}{c|}{0.66575} & \multicolumn{1}{c|}{0.02754} & \multicolumn{1}{c|}{0.00000} & 1 \\ \hline
\multicolumn{1}{c|}{} & Large (50\%) & \multicolumn{1}{c|}{0.00758} & \multicolumn{1}{c|}{0.85737} & \multicolumn{1}{c|}{0.02092} & 1 & \multicolumn{1}{c|}{0.00009} & \multicolumn{1}{c|}{0.92532} & \multicolumn{1}{c|}{0.68311} & 2 \\
\multicolumn{1}{c|}{} & Medium (25\%) & \multicolumn{1}{c|}{0.03963} & \multicolumn{1}{c|}{0.54504} & \multicolumn{1}{c|}{0.00804} & 1 & \multicolumn{1}{c|}{0.00018} & \multicolumn{1}{c|}{0.90927} & \multicolumn{1}{c|}{0.82104} & 2 \\
\multicolumn{1}{c|}{\multirow{-3}{*}{Adaptive}} & Small (5\%) & \multicolumn{1}{c|}{0.00932} & \multicolumn{1}{c|}{0.91993} & \multicolumn{1}{c|}{0.13520} & 2 & \multicolumn{1}{c|}{0.00061} & \multicolumn{1}{c|}{0.93064} & \multicolumn{1}{c|}{0.92247} & 3 \\ \hline
\rowcolor[HTML]{EFEFEF} 
\multicolumn{1}{c|}{\cellcolor[HTML]{EFEFEF}Ground Truth} & Full & \multicolumn{1}{c|}{\cellcolor[HTML]{EFEFEF}0.00178} & \multicolumn{1}{c|}{\cellcolor[HTML]{EFEFEF}0.91104} & \multicolumn{1}{c|}{\cellcolor[HTML]{EFEFEF}0.15392} & 1 & \multicolumn{1}{c|}{\cellcolor[HTML]{EFEFEF}0.00003} & \multicolumn{1}{c|}{\cellcolor[HTML]{EFEFEF}0.91490} & \multicolumn{1}{c|}{\cellcolor[HTML]{EFEFEF}0.97508} & 1 \\ \hline
\end{tabular}

\end{table*}

\rev{Specifically, the validation process consists of four main steps:
\begin{itemize}
    \item [1)] We first computed a set of representative fitness landscape metrics on the complete datasets, including the proportion of local optima $\ell_p$, the quality of local optima $\ell_q$, and the basin size of the global optimum (against the whole space). These results, obtained using the standard neighborhood radius of Hamming distance $d=1$, serve as the ground truth.
    \item [2)] From each full dataset, we then generated three subsets using Latin Hypercube Sampling (LHS)~\cite{mckay1992latin} with sampling rates of 50\%, 25\%, and 5\%, corresponding to large, medium, and small data scales, respectively.
    \item [3)] On each subset, we applied two neighborhood construction strategies: the conventional fixed radius ($d$ = 1) and the adaptive radius, where the neighborhood radius is adjusted according to data sparsity to preserve meaningful connectivity.
    \item [4)] We computed the same landscape metrics for all settings and compared them with the ground truth; 
\end{itemize}}

\rev{As shown in Table~\ref{tb:adaptive_validation}, applying a fixed neighborhood radius under sparse sampling leads to substantial distortion in the computed landscape metrics, and this distortion becomes more pronounced as the sampling density decreases. At the small data scale, for example, the number of detected local optima increases sharply, hundreds (0.43036/0.00178) or even thousands (0.66575/0.00003) of times greater than the ground truth, because many isolated configurations are incorrectly identified as local optima, which in turn degrades the overall estimation of local optima quality. Moreover, the measured basin of attraction of the global optimum collapses to zero for both \textsc{7z} and \textsc{LLVM}, indicating that sparse sampling disrupts the continuity of configuration transitions toward the global optimum. In contrast, the adaptive neighborhood radius effectively mitigates this sparsity-induced bias, yielding landscape metrics that closely approximate those obtained from the full configuration space. Meanwhile, we also observed similar results on other systems from the same open-source dataset that provides complete configuration-performance data. These findings confirm that the adaptive neighborhood radius maintains meaningful connectivity and preserves the structural characteristics of the configuration landscape even under incomplete sampling. Notably, when the dataset itself already exhibits sufficient connectivity, as in the case of full or dense sampling, the adaptive neighborhood radius naturally switches to the standard definition with Hamming distance $d = 1$, thereby avoiding the introduction of any additional bias.}


\subsubsection{Landscape Metrics}
\label{subsubsection:landscape_metrics}
\begin{figure}[t]
  \begin{subfigure}[t]{0.455\textwidth}
    \centering \begin{tikzpicture}
\begin{axis}[
  width=6cm,
  height=5cm,
  scale only axis,
  ybar,
  axis lines=left,
  ymin=0.00,
  ymax=0.19,
  ymajorgrids,
  xmajorgrids=false,
  grid style={densely dashed, gray!60},
  bar width=20pt,
  enlarge x limits=0.2,
  xtick=data,
  xticklabel style={font=\small, rotate=45, anchor=east},
  yticklabel style={font=\small},
  ylabel={FDC ($\varrho$)},
  xlabel={Data Scale},
  symbolic x coords={
    {Large (50\%)},
    {Medium (25\%)},
    {Small (5\%)}
  },
  nodes near coords,
  point meta=y,
  every node near coord/.append style={
    rotate=90, anchor=west, font=\small, text=black, yshift=1.5pt,
    /pgf/number format/fixed, /pgf/number format/precision=4
  },
  extra y ticks={0.1424},
  extra y tick labels={},
  extra y tick style={grid=major, grid style={red, dashed, thick}, tick style={draw=none}},
]

\addplot+[draw=black, fill=blue!20]
coordinates {
  (Large (50\%),0.1348)
  (Medium (25\%),0.1486)
  (Small (5\%),0.1152)
};

\end{axis}
\end{tikzpicture}      
    \caption{7z}
  \end{subfigure}\hfill
  \begin{subfigure}[t]{0.435\textwidth}
    \centering \begin{tikzpicture}
\begin{axis}[
  width=6cm,
  height=5cm,
  scale only axis,
  ybar,
  axis lines=left,
  ymin=0.00,
  ymax=0.90,
  ymajorgrids,
  xmajorgrids=false,
  grid style={densely dashed, gray!60},
  bar width=20pt,
  enlarge x limits=0.2,
  xtick=data,
  xticklabel style={font=\small, rotate=45, anchor=east},
  yticklabel style={font=\small},
  ylabel={FDC ($\varrho$)},
  xlabel={Data Scale},
  symbolic x coords={
    {Large (50\%)},
    {Medium (25\%)},
    {Small (5\%)}
  },
  nodes near coords,
  point meta=y,
  every node near coord/.append style={
    rotate=90, anchor=west, font=\small, text=black, yshift=1.5pt,
    /pgf/number format/fixed, /pgf/number format/precision=4
  },
  extra y ticks={0.6594},
  extra y tick labels={},
  extra y tick style={grid=major, grid style={red, dashed, thick}, tick style={draw=none}},
]

\addplot+[draw=black, fill=blue!20]
coordinates {
  (Large (50\%),0.6412)
  (Medium (25\%),0.6454)
  (Small (5\%),0.6573)
};

\end{axis}
\end{tikzpicture}       
    \caption{LLVM}
  \end{subfigure}
  \caption{\rev{Comparison of FDC ($\varrho$) under different sampling densities. 
  The red dashed line indicates the FDC value computed from the full-scale dataset (ground truth).}}
  \label{fig:fdc_comparison}
\end{figure}
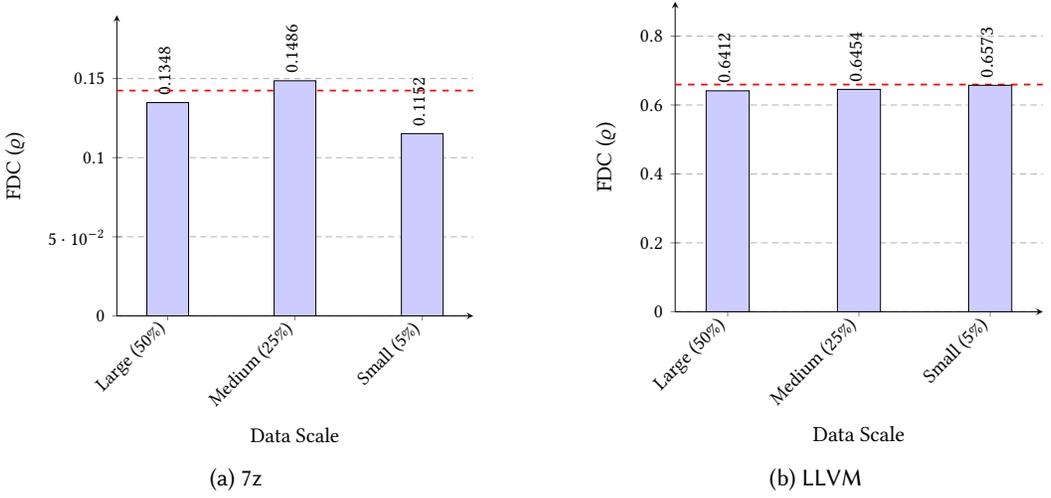

The selection of landscape metrics can follow what has been listed in Section~\ref{sec:cl-analysis} as part of \approach. The detailed definitions and usage are as follows:



\paragraph{\textbf{(1) Fitness Distance Correlation (FDC)}}

FDC is a widely used metric for evaluating the relationship between the fitness (e.g., runtime and throughput) of solutions (i.e., configurations) and their distance to the global optimum in a given landscape~\cite{jones1995fitness}. It helps to characterize the global structure of the fitness landscape and evaluate the ease or difficulty of navigating toward the global optimum. Formally, FDC is calculated as:

\begin{equation}
\varrho(f, d)=\frac{1}{\sigma_f \sigma_d} \frac{1}{s}\sum_{i=1}^s\left(f_i-\bar{f}\right)\left(d_i-\bar{d}\right),
\end{equation}
where $s$ is the number of solutions considered in FDC. Within such a solution set, $d_i$ denotes the distance of $i$th solution to the nearest global optimum. $\bar{f}$ ($\bar{d}$) and $\sigma_f$ ($\sigma_d$) are the mean and standard deviation of fitness (and distance). \rev{In practice, since the true global optimum is typically unknown in high-dimensional configuration spaces, we approximate it using the best-performing configuration observed within the sampled configurations, following the setting adopted in the literature~\cite{huang2018self,tanabe2020analyzing}.} \rev{To verify the reliability of using the observed best-performing configuration as the proxy of global optimum, we conducted an additional validation using the same systems and sampled subsets as in Section~\ref{subsubsection:neighborhood_construction}. We first computed the FDC on the full dataset as the ground truth. Then, for the three sampled subsets (with 50\%, 25\%, and 5\% sampling rates), we calculated FDC by using the best-performing configuration within each subset as the proxy for the global optimum. As shown in Figure~\ref{fig:fdc_comparison}, the red dashed line denotes the FDC value computed from the full-scale dataset, which serves as the ground truth reference. The FDC values obtained across the three sampled subsets (50\%, 25\%, and 5\%) remain close to this ground truth, confirming that using the best-performing configuration observed within a subset provides a reliable approximation even under a sparse sampling subset. Similar patterns were also observed in other systems with complete configuration–performance data~\cite{weber2023twins}.}


\paragraph{\textbf{(2) Local optima}} Local optima capture the underlying local structural properties of a problem's landscape. Formally, a configuration \textbf{\textit{c}}$^{\ell}$ is considered as a local optima if $f(\textbf{\textit{c}}^{\ell})$ is better than $f(\textbf{\textit{c}})$, $\forall \textbf{\textit{c}} \in \mathcal{N}\left(\textbf{\textit{c}}^{\ell}\right)$, where $\mathcal{N}\left(\textbf{\textit{c}}^{\ell}\right)$ is the neighborhood of \textbf{\textit{c}}$^{\ell}$. Two key aspects are considered when analyzing local optima: (1) the number of local optima; and (2) the quality of local optima.

\begin{itemize}
    \item \textbf{The number of local optima.} The number of local optima reflects landscape multimodality, indicating the presence of multiple high-performing regions and complexity of the problem. To quantify this, we define $\ell_p$ as the proportion of local optima among all sampled configurations:
    \begin{equation}
        \ell_p = \frac{|\mathcal{L}|}{\mathcal{|S|}},
    \end{equation}
    where $\mathcal{L}$ and $\mathcal{S}$ denote the sets of local optima and all sampled configurations, respectively.
    
    \item \textbf{The quality of local optima.} The quality of local optima, particularly their relative performance compared to the global optima, indicates the prominence of local peaks and the susceptibility of the landscape to trapping suboptimal solutions. For minimization problems, we define the relative quality of local optima as:
    \begin{equation}
    \ell_q = \frac{\frac{1}{|\mathcal{S}|} \sum\limits_{\textbf{\textit{c}} \in S} f(\textbf{\textit{c}}) - \frac{1}{|\mathcal{L}|} \sum\limits_{\textbf{\textit{c}}^\ell \in \mathcal{L}} f(\textbf{\textit{c}}^\ell)}
    {\frac{1}{|\mathcal{S}|} \sum\limits_{\textbf{\textit{c}} \in \mathcal{S}} f(\textbf{\textit{c}}) - f(\textbf{\textit{c}}^g)},
    \end{equation}
    where $\textbf{\textit{c}}^g$ represents the global optimum, and $\textbf{\textit{c}}^\ell$ denotes a local optima from $\mathcal{L}$. A higher $\ell_q$ indicates that local optima are of higher quality, being closer to the global optimum, while a lower $\ell_q$ suggests an abundance of low-quality local optima.
\end{itemize}

\paragraph{\textbf{(3) Basin of attraction}} The basin of attraction~\cite{malan2021survey} of a local optimum \( \textbf{\textit{c}}^\ell \), denoted as \( \mathcal{B}(\textbf{\textit{c}}^\ell) \), is the set of all configurations that converge to \( \textbf{\textit{c}}^\ell \) under a local search process. Formally, it can be defined as:

\begin{equation}
\mathcal{B}(\textbf{\textit{c}}^\ell) = \{ \textbf{\textit{c}} \in \mathcal{C} \mid \text{LocalSearch}(\textbf{\textit{c}}) \to \textbf{\textit{c}}^\ell \}.
\end{equation}

\revision{Here, we adopt hill climbing with the first-improvement strategy as the local search algorithm. The step size follows the neighborhood structure defined in Section \ref{subsubsection:neighborhood_construction}. In other words, if the neighborhood radius is set to $d$, the local search step size is also defined as $d$ to ensure consistency in basin identification. As a result, the basin of attraction of a local optimum is identified as the set of all configurations that lead to it under this search process.}

\paragraph{\textbf{(4) Autocorrelation}} Autocorrelation~\cite{weinberger1990correlated} quantifies the ruggedness of a fitness landscape by measuring how fitness values change across neighboring configurations. Formally, it is defined as:

\begin{equation}
\label{eq:autocorrelation}
r(d) =\frac{\sum_{i=1}^{N-d}\left(f\left(\textbf{\textit{c}}_i\right)-\bar{f}\right)\left(f\left(\textbf{\textit{c}}_{i+d}\right)-\bar{f}\right)}{\sum_{i=1}^{N-d}\left(f\left(\textbf{\textit{c}}_i\right)-\bar{f}\right)^2},
\end{equation}
where $f(\textbf{\textit{c}}_i)$ is the fitness of $i$th visited configuration $\textbf{\textit{c}}_i$ in the random walk~\cite{tavares2008multidimensional}, $\textbf{\textit{c}}_{i+d}$ is the next visited neighbor at step size $d$, and $N$ is the walk length. A high $r(d)$ indicates a smooth landscape, where similar configurations yield comparable fitness values, while a low autocorrelation suggests a rugged landscape with abrupt fitness variations. Notably, to accommodate analysis on sampled datasets, our random walk starts with a step size of $d$ = 1. If no neighbor is found, $d$ is incrementally increased until a valid neighbor is located, ensuring continuity. The walk proceeds until all configurations in the dataset are visited, recording the total walk length $N$. Finally, autocorrelation is computed using Equation (\ref{eq:autocorrelation}).

\begin{table*}[t]
    \centering
    
    \caption{The relative standard deviation (RSD) of autocorrelation across configuration options in nine software systems. For each case, \colorbox{steel!10}{blue cells} indicate options with significant influence on landscape ruggedness (RSD $\geq$ 5\%); or \colorbox{red-}{red cells} otherwise. The “Type” column denotes the functional-related or resource-related roles of options as categorized in Table~\ref{tb:option-taxonomy}. The \textcolor{red}{\ding{55}} markers indicate that there are insufficient data samples for analysis.}
    \tiny
    \setlength{\tabcolsep}{0.5 mm}
    
\begin{tabular}{c|cc|cc|cc|cc|cc|cc|cc|cc|cc}
\toprule
                                  & \multicolumn{2}{c|}{\textsc{\textbf{Lrzip}}}                                   & \multicolumn{2}{c|}{\textsc{\textbf{Xz}}}                                     & \multicolumn{2}{c|}{\textsc{\textbf{Z3}}}                                    & \multicolumn{2}{c|}{\textsc{\textbf{Dconvert}}}                               & \multicolumn{2}{c|}{\textsc{\textbf{Batik}}}                                & \multicolumn{2}{c|}{\textsc{\textbf{Kanzi}}}                                   & \multicolumn{2}{c|}{\textsc{\textbf{X264}}}                                   & \multicolumn{2}{c|}{\textsc{\textbf{H2}}}                                     & \multicolumn{2}{c}{\textsc{\textbf{Jump3r}}}                                  \\ \cline{2-19} 
\multirow{-2}{*}{\textbf{Option}} & \textbf{Type}               &  \textbf{RSD (\%)}               & \textbf{Type}               & \textbf{RSD (\%)}              & \textbf{Type}               & \textbf{RSD (\%)}             & \textbf{Type}               & \textbf{RSD (\%)}              & \textbf{Type}               & \textbf{RSD (\%)}             & \textbf{Type}               & \textbf{RSD (\%)}               & \textbf{Type}               & \textbf{RSD (\%)}              & \textbf{Type}               & \textbf{RSD (\%)}              & \textbf{Type}               & \textbf{RSD (\%)}              \\ \midrule
O1                                & \cellcolor{steel!10}$F_1$ & \cellcolor{steel!10}107.254 & \cellcolor[HTML]{FCE4D6}$R_3$ & \cellcolor[HTML]{FCE4D6}2.118  & \cellcolor[HTML]{FCE4D6}$F_2$ & \cellcolor[HTML]{FCE4D6}1.333 & \cellcolor{steel!10}$F_1$ & \cellcolor{steel!10}7.422  & \cellcolor[HTML]{FCE4D6}$F_2$ & \cellcolor[HTML]{FCE4D6}0.14  & \cellcolor[HTML]{FCE4D6}$F_1$ & \cellcolor[HTML]{FCE4D6}1.136   & \cellcolor[HTML]{FCE4D6}$F_1$ & \cellcolor[HTML]{FCE4D6}1.162  & \cellcolor{steel!10}$R_1$ & \cellcolor{steel!10}6.703  & \cellcolor[HTML]{FCE4D6}$F_1$ & \cellcolor[HTML]{FCE4D6}1.695  \\
O2                                & \cellcolor{steel!10}$F_2$ & \cellcolor{steel!10}210.332 & \cellcolor{steel!10}$F_1$ & \cellcolor{steel!10}7.29   & \cellcolor[HTML]{FCE4D6}$F_2$ & \cellcolor[HTML]{FCE4D6}3.881 & \cellcolor[HTML]{FCE4D6}$F_1$ & \cellcolor[HTML]{FCE4D6}1.105  & \cellcolor[HTML]{FCE4D6}$F_2$ & \cellcolor[HTML]{FCE4D6}0.376 & \cellcolor{steel!10}$F_1$ & \cellcolor{steel!10}121.468 & \cellcolor{steel!10}$F_1$ & \cellcolor{steel!10}6.467  & \cellcolor{steel!10}$R_1$ & \cellcolor{steel!10}5.301  & $F_1$                          & \textcolor{red}{\ding{55}}                               \\
O3                                & \cellcolor{steel!10}$F_2$ & \cellcolor{steel!10}83.073  & \cellcolor{steel!10}$F_1$ & \cellcolor{steel!10}19.927 & \cellcolor[HTML]{FCE4D6}$F_1$ & \cellcolor[HTML]{FCE4D6}0.276 & \cellcolor[HTML]{FCE4D6}$F_2$ & \cellcolor[HTML]{FCE4D6}0.2    & \cellcolor[HTML]{FCE4D6}$F_1$ & \cellcolor[HTML]{FCE4D6}0.705 & \cellcolor[HTML]{FCE4D6}$F_1$ & \cellcolor[HTML]{FCE4D6}0.081   & \cellcolor{steel!10}$F_1$ & \cellcolor{steel!10}11.448 & \cellcolor{steel!10}$R_1$ & \cellcolor{steel!10}5.947  & \cellcolor[HTML]{FCE4D6}$F_1$ & \cellcolor[HTML]{FCE4D6}1.412  \\
O4                                & \cellcolor{steel!10}$R_1$ & \cellcolor{steel!10}499.477 & \cellcolor{steel!10}$R_3$ & \cellcolor{steel!10}20.944 & \cellcolor[HTML]{FCE4D6}$F_1$ & \cellcolor[HTML]{FCE4D6}1.182 & \cellcolor[HTML]{FCE4D6}$F_1$ & \cellcolor[HTML]{FCE4D6}0.099  & \cellcolor[HTML]{FCE4D6}$F_2$ & \cellcolor[HTML]{FCE4D6}1.224 & \cellcolor{steel!10}$F_1$ & \cellcolor{steel!10}164.22  & \cellcolor{steel!10}$F_1$ & \cellcolor{steel!10}8.785  & \cellcolor[HTML]{FCE4D6}$R_1$ & \cellcolor[HTML]{FCE4D6}0.133  & $F_1$                         &   \textcolor{red}{\ding{55}}                             \\
O5                                & \cellcolor{steel!10}$F_1$ & \cellcolor{steel!10}306.08  & \cellcolor[HTML]{FCE4D6}$F_1$ & \cellcolor[HTML]{FCE4D6}2.042  & \cellcolor[HTML]{FCE4D6}$F_1$ & \cellcolor[HTML]{FCE4D6}2.047 & \cellcolor[HTML]{FCE4D6}$F_1$ & \cellcolor[HTML]{FCE4D6}0.615  & \cellcolor[HTML]{FCE4D6}$F_2$ & \cellcolor[HTML]{FCE4D6}0.476 & \cellcolor{steel!10}$F_1$ & \cellcolor{steel!10}32.901  & \cellcolor{steel!10}$F_1$ & \cellcolor{steel!10}13.159 & \cellcolor[HTML]{FCE4D6}$R_1$ & \cellcolor[HTML]{FCE4D6}0.152  & \cellcolor{steel!10}$F_1$ & \cellcolor{steel!10}9.008  \\
O6                                & \cellcolor{steel!10}$F_1$ & \cellcolor{steel!10}67.485  & \cellcolor[HTML]{FCE4D6}$R_1$ & \cellcolor[HTML]{FCE4D6}0.288  & \cellcolor[HTML]{FCE4D6}$F_2$ & \cellcolor[HTML]{FCE4D6}3.192 & \cellcolor[HTML]{FCE4D6}$F_1$ & \cellcolor[HTML]{FCE4D6}0.679  & \cellcolor[HTML]{FCE4D6}$F_2$ & \cellcolor[HTML]{FCE4D6}1.354 & \cellcolor[HTML]{FCE4D6}$F_1$ & \cellcolor[HTML]{FCE4D6}3.946   & \cellcolor{steel!10}$F_1$ & \cellcolor{steel!10}10.602 & \cellcolor{steel!10}$R_3$ & \cellcolor{steel!10}23.42  & \cellcolor[HTML]{FCE4D6}$F_1$ & \cellcolor[HTML]{FCE4D6}1.418  \\
O7                                & \cellcolor{steel!10}$F_2$ & \cellcolor{steel!10}57.613  & \cellcolor{steel!10}$R_3$ & \cellcolor{steel!10}23.014 & \cellcolor[HTML]{FCE4D6}$F_2$ & \cellcolor[HTML]{FCE4D6}0.875 & \cellcolor[HTML]{FCE4D6}$R_1$ & \cellcolor[HTML]{FCE4D6}0.086  & \cellcolor[HTML]{FCE4D6}$F_2$ & \cellcolor[HTML]{FCE4D6}0.556 & \cellcolor{steel!10}$F_1$ & \cellcolor{steel!10}18.394  & \cellcolor{steel!10}$F_1$ & \cellcolor{steel!10}7.546  & \cellcolor{steel!10}$F_1$ & \cellcolor{steel!10}18.454 & \cellcolor{steel!10}$F_2$ & \cellcolor{steel!10}6.916  \\
O8                                &                          &                               & \cellcolor{steel!10}$R_4$ & \cellcolor{steel!10}5.157  & \cellcolor[HTML]{FCE4D6}$F_2$ & \cellcolor[HTML]{FCE4D6}1.07  & \cellcolor[HTML]{FCE4D6}$F_2$ & \cellcolor[HTML]{FCE4D6}2.224  & \cellcolor[HTML]{FCE4D6}$F_2$ & \cellcolor[HTML]{FCE4D6}1.38  & \cellcolor{steel!10}$F_1$ & \cellcolor{steel!10}11.745  & \cellcolor{steel!10}$F_1$ & \cellcolor{steel!10}9.864  & \cellcolor{steel!10}$F_1$ & \cellcolor{steel!10}7.833  & \cellcolor{steel!10}$F_2$ & \cellcolor{steel!10}10.665 \\
O9                                &                          &                               & \cellcolor[HTML]{FCE4D6}$F_1$ & \cellcolor[HTML]{FCE4D6}2.822  & \cellcolor[HTML]{FCE4D6}$F_2$ & \cellcolor[HTML]{FCE4D6}1.103 & \cellcolor[HTML]{FCE4D6}$F_2$ & \cellcolor[HTML]{FCE4D6}3.458  &                          &                             & \cellcolor{steel!10}$F_1$ & \cellcolor{steel!10}20.259  & \cellcolor[HTML]{FCE4D6}$F_1$ & \cellcolor[HTML]{FCE4D6}1.393  & \cellcolor{steel!10}$F_2$ & \cellcolor{steel!10}5.601  & $R_1$                         &  \textcolor{red}{\ding{55}}                              \\
O10                               &                          &                               & \cellcolor{steel!10}$F_2$ & \cellcolor{steel!10}84.806 & \cellcolor[HTML]{FCE4D6}$F_2$ & \cellcolor[HTML]{FCE4D6}0.302 & \cellcolor{steel!10}$F_2$ & \cellcolor{steel!10}35.124 &                          &                             & \cellcolor{steel!10}$F_1$ & \cellcolor{steel!10}58.027  & \cellcolor{steel!10}$F_1$ & \cellcolor{steel!10}15.91  & \cellcolor[HTML]{FCE4D6}$R_1$ & \cellcolor[HTML]{FCE4D6}0.809  & \cellcolor{steel!10}$R_1$ & \cellcolor{steel!10}24.122 \\
O11                               &                          &                               & \cellcolor{steel!10}$F_1$ & \cellcolor{steel!10}23.679 & \cellcolor[HTML]{FCE4D6}$F_1$ & \cellcolor[HTML]{FCE4D6}1.343 &                          &                              &                          &                             & \cellcolor[HTML]{FCE4D6}$F_2$ & \cellcolor[HTML]{FCE4D6}2.179   & \cellcolor[HTML]{FCE4D6}$F_1$ & \cellcolor[HTML]{FCE4D6}1.717  & \cellcolor[HTML]{FCE4D6}$F_1$ & \cellcolor[HTML]{FCE4D6}4.294  & \cellcolor[HTML]{FCE4D6}$F_2$ & \cellcolor[HTML]{FCE4D6}1.796  \\
O12                               &                          &                               & $F_1$                         & \textcolor{red}{\ding{55}}                              & \cellcolor[HTML]{FCE4D6}$F_2$ & \cellcolor[HTML]{FCE4D6}0.312 &                          &                              &                          &                             & \cellcolor{steel!10}$F_1$ & \cellcolor{steel!10}8.859   & \cellcolor[HTML]{FCE4D6}$F_2$ & \cellcolor[HTML]{FCE4D6}0.168  & \cellcolor[HTML]{FCE4D6}$R_2$ & \cellcolor[HTML]{FCE4D6}0.359  & \cellcolor{steel!10}$R_2$ & \cellcolor{steel!10}18.146 \\
O13                               &                          &                               &                          &                              &                          &                             &                          &                              &                          &                             & \cellcolor{steel!10}$F_1$ & \cellcolor{steel!10}25.665  & \cellcolor[HTML]{FCE4D6}$F_1$ & \cellcolor[HTML]{FCE4D6}3.63   & \cellcolor[HTML]{FCE4D6}$F_2$ & \cellcolor[HTML]{FCE4D6}0.286  & \cellcolor{steel!10}$F_1$ & \cellcolor{steel!10}9.132  \\
O14                               &                          &                               &                          &                              &                          &                             &                          &                              &                          &                             & \cellcolor{steel!10}$F_1$ & \cellcolor{steel!10}18.236  & \cellcolor[HTML]{FCE4D6}$F_1$ & \cellcolor[HTML]{FCE4D6}0.196  & \cellcolor{steel!10}$R_2$ & \cellcolor{steel!10}6.037  &                          &                              \\
O15                               &                          &                               &                          &                              &                          &                             &                          &                              &                          &                             & \cellcolor{steel!10}$F_1$ & \cellcolor{steel!10}216.355 & \cellcolor{steel!10}$F_1$ & \cellcolor{steel!10}5.904  & \cellcolor{steel!10}$F_2$ & \cellcolor{steel!10}16.952 &                          &                              \\
O16                               &                          &                               &                          &                              &                          &                             &                          &                              &                          &                             & \cellcolor{steel!10}$R_1$ & \cellcolor{steel!10}142.712 & \cellcolor{steel!10}$F_2$ & \cellcolor{steel!10}12.122 & \cellcolor{steel!10}$F_1$ & \cellcolor{steel!10}28.767 &                          &                              \\
O17                               &                          &                               &                          &                              &                          &                             &                          &                              &                          &                             & \cellcolor{steel!10}$R_3$ & \cellcolor{steel!10}46.585  & \cellcolor{steel!10}$F_1$ & \cellcolor{steel!10}10.031 &                          &                              &                          &                              \\
O18                               &                          &                               &                          &                              &                          &                             &                          &                              &                          &                             & \cellcolor{steel!10}$F_1$ & \cellcolor{steel!10}42.856  & \cellcolor{steel!10}$F_1$ & \cellcolor{steel!10}11.83  &                          &                              &                          &                              \\
O19                               &                          &                               &                          &                              &                          &                             &                          &                              &                          &                             &                          &                               & \cellcolor[HTML]{FCE4D6}$R_1$ & \cellcolor[HTML]{FCE4D6}1.692  &                          &                              &                          &                              \\
O20                               &                          &                               &                          &                              &                          &                             &                          &                              &                          &                             &                          &                               & \cellcolor[HTML]{FCE4D6}$R_1$ & \cellcolor[HTML]{FCE4D6}4.996  &                          &                              &                          &                              \\
O21                               &                          &                               &                          &                              &                          &                             &                          &                              &                          &                             &                          &                               & \cellcolor{steel!10}$R_1$ & \cellcolor{steel!10}19.874 &                          &                              &                          &                              \\
O22                               &                          &                               &                          &                              &                          &                             &                          &                              &                          &                             &                          &                               & \cellcolor{steel!10}$F_1$ & \cellcolor{steel!10}15.94  &                          &                              &                          &                              \\
O23                               &                          &                               &                          &                              &                          &                             &                          &                              &                          &                             &                          &                               & \cellcolor{steel!10}$F_1$ & \cellcolor{steel!10}6.854  &                          &                              &                          &                              \\
O24                               &                          &                               &                          &                              &                          &                             &                          &                              &                          &                             &                          &                               & \cellcolor[HTML]{FCE4D6}$R_1$ & \cellcolor[HTML]{FCE4D6}0.671  &                          &                              &                          &                              \\
O25                               &                          &                               &                          &                              &                          &                             &                          &                              &                          &                             &                          &                               & \cellcolor{steel!10}$R_1$ & \cellcolor{steel!10}5.52   &                          &                              &                          &                              \\ \midrule
\multicolumn{1}{l|}{\textbf{RSD $\geq$ 5\%}} & \multicolumn{2}{c|}{\textbf{7/7}}                            & \multicolumn{2}{c|}{\textbf{7/11}}                          & \multicolumn{2}{c|}{\textbf{0/12}}                         & \multicolumn{2}{c|}{\textbf{2/10}}                          & \multicolumn{2}{c|}{\textbf{0/8}}                          & \multicolumn{2}{c|}{\textbf{14/18}}                          & \multicolumn{2}{c|}{\textbf{16/25}}                         & \multicolumn{2}{c|}{\textbf{10/16}}                         & \multicolumn{2}{c}{\textbf{6/10}}                           \\ \bottomrule
\end{tabular}

    \label{tb:option-sensitivity}
\end{table*}

\section{Results and Analysis}
\label{section:results_and_analysis}

\subsection{RQ1: Spatial Option-level Synergy}

Here, we follow the procedures in Section~\ref{subsubsection:spatial_option_level_synergy} and summarize the common patterns for all workloads across all systems. To simplify exposition, we aggregate results across workloads by taking the median autocorrelation. As such, we focus on findings for \textit{all workloads of a system} and \textit{all workloads across all systems} in \approach.

The results are summarized in Table \ref{tb:option-sensitivity}. \rev{Here, we adopt a 5\% variation threshold to characterize whether a configuration option has a significant impact on ruggedness, aligning with the setting of the significant variation factor in the field~\cite{DBLP:conf/icse/LiangChen25,chen2023diagconfig}. This threshold is widely accepted in configurable system analysis because performance fluctuations below 5\% are often attributed to measurement noise or environmental variability rather than genuine configuration effects. Therefore, adopting this level helps distinguish substantive configuration/option-induced changes from normal experimental variance, ensuring both consistency with prior work and practical interpretability of ruggedness sensitivity.} As can be seen, in \textsc{Lrzip}, all investigated options exhibit a significant effect (RSD $\geq 5\%$) on ruggedness. Systems such as \textsc{Xz}, \textsc{Kanzi}, \textsc{X264}, \textsc{H2}, and \textsc{Jump3r} show a relatively high proportion of impactful options, taking more than half of the options. In contrast, \textsc{Dconvert} displays a lower proportion of options significantly affecting ruggedness. Notably, the ruggedness of \textsc{Z3} and \textsc{Batik} demonstrate a striking insensitivity to option variations, with nearly all options having negligible impact on landscape ruggedness. On the other hand, the degree of impact also differs considerably---\textsc{Lrzip} exhibits extreme sensitivity with RSD values reaching up to 499\%, whereas in \textsc{Batik}, the most influential option only induces an RSD of 1.38\%.


\begin{quotebox}
   \noindent
   \textit{\textbf{Finding 1:} The proportion and magnitude of ruggedness-sensitive options vary greatly within and across systems. 
   } 
\end{quotebox}




A further observation is that core options ($F_1$) account for the majority (37 out of 62) of impactful options where RSD $\geq$ 5\%, far exceeding other types \rev{($F_2$: 10, $R_1$: 8, $R_2$: 2, $R_3$: 4, $R_4$: 1).} This implies a strong connection between core options and landscape ruggedness.

\begin{quotebox}
   \noindent
   \textit{\textbf{Finding 2:} Core options ($F_1$) exhibit a stronger impact on landscape ruggedness than the other types. 
   }
\end{quotebox}

\subsection{RQ2: Spatial System-level Synergy}


\subsubsection{Fitness Distance Correlation ($\varrho$)}
The FDC results across different software systems and workloads are summarized in Table \ref{tb:FDC}. As observed, the majority of systems (7 out of the 9) exhibit $\varrho \geq 0.15$, indicating a positive correlation between performance and proximity to the global optimum. However, the FDC values of \textsc{Lrzip} and \textsc{X264} tend to fall into the range of $-0.15 \leq \varrho \leq 0.15$, suggesting irregular landscapes where proximity offers little guidance.


\vspace{-2pt}
\begin{quotebox}
   \noindent
   \textit{\textbf{Finding 3:} Apart from a few edge cases, configuration landscapes can generally provide rich guidance to a tuner.
   }
\end{quotebox}
\vspace{-2pt}



\begin{table}[t]
    \centering
    \caption{The FDC ($\varrho$) across nine systems with different workloads (\colorbox{steel!10}{blue cells}: $\varrho \geq$ 0.15; \colorbox{red-}{red cells}: $\mathcal{-}0.15$ \textless $\varrho$ \textless 0.15; \colorbox{gold}{gold cells}: $\varrho \leq \mathcal{-}$ 0.15). }
    \scriptsize
    \setlength{\tabcolsep}{0.85mm}
    
    \setlength{\tabcolsep}{9pt}

\begin{tabular}{c|c|c|c|c|c|c|c|c|c}
\toprule
\textbf{Workload} & \textbf{\textsc{Lrzip}}                 & \textbf{\textsc{Xz}}                    & \textbf{\textsc{Z3}}                   & \textbf{\textsc{Dconvert}}              & \textbf{\textsc{Batik}}                & \textbf{\textsc{Kanzi}}                & \textbf{\textsc{X264}}                  & \textbf{\textsc{H2}}                   & \textbf{\textsc{Jump3r}}               \\ \midrule
W1                & \cellcolor[HTML]{FCE4D6}0.025  & \cellcolor{steel!10}0.381  & \cellcolor{steel!10}0.488 & \cellcolor{steel!10}0.461  & \cellcolor{steel!10}0.319 & \cellcolor{steel!10}0.207 & \cellcolor[HTML]{FCE4D6}-0.083 & \cellcolor{steel!10}0.637 & \cellcolor{steel!10}0.439 \\
W2                & \cellcolor{steel!10}0.186  & \cellcolor{steel!10}0.460  & \cellcolor{steel!10}0.535 & \cellcolor{steel!10}0.524  & \cellcolor{steel!10}0.280 & \cellcolor{steel!10}0.328 & \cellcolor[HTML]{FCE4D6}-0.015 & \cellcolor{steel!10}0.627 & \cellcolor[HTML]{FCE4D6}0.067 \\
W3                & \cellcolor[HTML]{FCE4D6}-0.037 & \cellcolor[HTML]{FCE4D6}-0.198 & \cellcolor{steel!10}0.545 & \cellcolor{steel!10}0.468  & \cellcolor{steel!10}0.520 & \cellcolor[HTML]{FCE4D6}0.123 & \cellcolor[HTML]{FCE4D6}-0.014 & \cellcolor{steel!10}0.641 & \cellcolor{steel!10}0.363 \\
W4                & \cellcolor[HTML]{FCE4D6}-0.078 & \cellcolor[HTML]{FCE4D6}-0.099 & \cellcolor{steel!10}0.492 & \cellcolor{steel!10}0.511  & \cellcolor{steel!10}0.550 & \cellcolor{steel!10}0.181 & \cellcolor[HTML]{FCE4D6}0.065  & \cellcolor{steel!10}0.257 & \cellcolor{steel!10}0.267 \\
W5                & \cellcolor[HTML]{FCE4D6}-0.050 & \cellcolor{steel!10}0.566  & \cellcolor[HTML]{FCE4D6}0.098 & \cellcolor{steel!10}0.523  & \cellcolor{steel!10}0.360 & \cellcolor[HTML]{FCE4D6}0.123 & \cellcolor[HTML]{FCE4D6}0.031  & \cellcolor{steel!10}0.637 & \cellcolor{steel!10}0.443 \\
W6                & \cellcolor[HTML]{FCE4D6}0.019  & \cellcolor{steel!10}0.612  & \cellcolor[HTML]{FCE4D6}0.018 & \cellcolor{steel!10}0.524  & \cellcolor{steel!10}0.438 & \cellcolor{steel!10}0.192 & \cellcolor{steel!10}0.155  & \cellcolor{steel!10}0.653 & \cellcolor[HTML]{FCE4D6}0.117 \\
W7                & \cellcolor{steel!10}0.196  & \cellcolor{steel!10}0.262  & \cellcolor{steel!10}0.318 & \cellcolor{steel!10}0.512  & \cellcolor{steel!10}0.341 & \cellcolor{steel!10}0.334 & \cellcolor[HTML]{FCE4D6}-0.045 & \cellcolor{steel!10}0.405 &                             \\
W8                & \cellcolor[HTML]{FCE4D6}-0.075 & \cellcolor{steel!10}0.383  & \cellcolor{steel!10}0.488 & \cellcolor{steel!10}0.232  & \cellcolor{steel!10}0.252 & \cellcolor{steel!10}0.150 & \cellcolor[HTML]{FCE4D6}0.015  & \cellcolor{steel!10}0.336 &                             \\
W9                & \cellcolor[HTML]{FCE4D6}-0.025 & \cellcolor{steel!10}0.504  & \cellcolor{steel!10}0.538 & \cellcolor{steel!10}0.471  & \cellcolor{steel!10}0.428 & \cellcolor[HTML]{FCE4D6}0.138 & \cellcolor[HTML]{FCE4D6}0.148  &                             &                             \\
W10               & \cellcolor[HTML]{FCE4D6}0.016  & \cellcolor{steel!10}0.456  & \cellcolor{steel!10}0.372 & \cellcolor[HTML]{FCE4D6}-0.005 & \cellcolor{steel!10}0.583 &                             &                              &                             &                             \\
W11               & \cellcolor[HTML]{FCE4D6}-0.072 & \cellcolor{steel!10}0.450  & \cellcolor{steel!10}0.225 & \cellcolor{steel!10}0.292  & \cellcolor{steel!10}0.188 &                             &                              &                             &                             \\
W12               & \cellcolor[HTML]{FCE4D6}0.026  & \cellcolor{steel!10}0.479  & \cellcolor{steel!10}0.245 & \cellcolor{steel!10}0.345  &                             &                             &                              &                             &                             \\
W13               & \cellcolor[HTML]{FCE4D6}0.073  & \cellcolor{steel!10}0.565  &                             &                              &                             &                             &                              &                             &                             \\ \midrule
\textbf{Median}   & \textbf{0.016}                 & \textbf{0.456}                 & \textbf{0.430}                & \textbf{0.469}                 & \textbf{0.360}                & \textbf{0.181}                & \textbf{0.015}                 & \textbf{0.632}                & \textbf{0.315}                \\ \midrule
\textbf{IQR}      & \textbf{0.076}                 & \textbf{0.123}                 & \textbf{0.263}                & \textbf{0.183}                 & \textbf{0.180}                & \textbf{0.070}                & \textbf{0.080}                 & \textbf{0.250}                & \textbf{0.265}                \\ \bottomrule
\end{tabular}

    \label{tb:FDC}
    \vspace{-5pt}
\end{table}


\begin{table}[t]
        \centering
	\caption{The proportion of configurations in the dataset that falls within the global optimum's basin of attraction. For each cases, the \colorbox{steel!10}{blue cells} indicate basin size of the global optimum is larger than 0.21; or \colorbox{red-}{red cells} otherwise.}
	\scriptsize
    \setlength{\tabcolsep}{0.7mm}
        \label{tb:basin_of_attraction_for_global_optimum}
	
        \setlength{\tabcolsep}{9pt}
\begin{tabular}{c|c|c|c|c|c|c|c|c|c}
\toprule
\textbf{Wrokload} & \textbf{\textsc{Lrzip}} & \textbf{\textsc{Xz}} & \textbf{\textsc{Z3}} & \textbf{\textsc{Dconvert}} & \textbf{\textsc{Batik}} & \textbf{\textsc{Kanzi}} & \textbf{\textsc{X264}} & \textbf{\textsc{H2}} & \textbf{\textsc{Jump3r}} \\ \midrule

W1              & \cellcolor[HTML]{FCE4D6}0.037 & \cellcolor[HTML]{FCE4D6}0.017 & \cellcolor{steel!10}0.643 & \cellcolor[HTML]{FCE4D6}0.079 & \cellcolor[HTML]{FCE4D6}0.162 & \cellcolor{steel!10}0.35  & \cellcolor[HTML]{FCE4D6}0.052 & \cellcolor[HTML]{FCE4D6}0.079 & \cellcolor[HTML]{FCE4D6}0.037 \\
W2              & \cellcolor{steel!10}0.565 & \cellcolor{steel!10}0.449 & \cellcolor{steel!10}0.735 & \cellcolor[HTML]{FCE4D6}0.107 & \cellcolor[HTML]{FCE4D6}0.11  & \cellcolor{steel!10}0.239 & \cellcolor[HTML]{FCE4D6}0.004 & \cellcolor[HTML]{FCE4D6}0.06  & \cellcolor[HTML]{FCE4D6}0.081 \\
W3              & \cellcolor[HTML]{FCE4D6}0.063 & \cellcolor[HTML]{FCE4D6}0.04  & \cellcolor{steel!10}0.753 & \cellcolor[HTML]{FCE4D6}0.051 & \cellcolor[HTML]{FCE4D6}0.077 & \cellcolor[HTML]{FCE4D6}0.081                         & \cellcolor[HTML]{FCE4D6}0.004 & \cellcolor[HTML]{FCE4D6}0.033 & \cellcolor[HTML]{FCE4D6}0.074 \\
W4              & \cellcolor[HTML]{FCE4D6}0.052 & \cellcolor[HTML]{FCE4D6}0.121 & \cellcolor[HTML]{FCE4D6}0.142 & \cellcolor[HTML]{FCE4D6}0.107 & \cellcolor[HTML]{FCE4D6}0.09  & \cellcolor{steel!10}0.301 & \cellcolor[HTML]{FCE4D6}0.013 & \cellcolor[HTML]{FCE4D6}0.045 & \cellcolor[HTML]{FCE4D6}0.109 \\
W5              & \cellcolor[HTML]{FCE4D6}0.131 & \cellcolor[HTML]{FCE4D6}0.116 & \cellcolor{steel!10}0.542 & \cellcolor[HTML]{FCE4D6}0.146 & \cellcolor[HTML]{FCE4D6}0.176 & \cellcolor{steel!10}0.331 & \cellcolor[HTML]{FCE4D6}0.02  & \cellcolor[HTML]{FCE4D6}0.061 & \cellcolor[HTML]{FCE4D6}0.101 \\
W6              & \cellcolor{steel!10}0.304 & \cellcolor[HTML]{FCE4D6}0.102 & \cellcolor{steel!10}0.642 & \cellcolor[HTML]{FCE4D6}0.085 & \cellcolor[HTML]{FCE4D6}0.109 & \cellcolor[HTML]{FCE4D6}0.11                          & \cellcolor[HTML]{FCE4D6}0.003 & \cellcolor[HTML]{FCE4D6}0.079 & \cellcolor[HTML]{FCE4D6}0.032 \\
W7              & \cellcolor[HTML]{FCE4D6}0.079 & \cellcolor{steel!10}0.338 & \cellcolor{steel!10}0.606 & \cellcolor{steel!10}0.408 & \cellcolor{steel!10}0.217 & \cellcolor{steel!10}0.226 & \cellcolor[HTML]{FCE4D6}0.034 & \cellcolor[HTML]{FCE4D6}0.054 &                             \\
W8              & \cellcolor[HTML]{FCE4D6}0.141 & \cellcolor[HTML]{FCE4D6}0.041 & \cellcolor{steel!10}0.26  & \cellcolor[HTML]{FCE4D6}0.137 & \cellcolor{steel!10}0.241 & \cellcolor{steel!10}0.266 & \cellcolor[HTML]{FCE4D6}0.206 & \cellcolor[HTML]{FCE4D6}0.05  &                             \\
W9              & \cellcolor[HTML]{FCE4D6}0.073 & \cellcolor[HTML]{FCE4D6}0.021 & \cellcolor{steel!10}0.375 & \cellcolor[HTML]{FCE4D6}0.066 & \cellcolor[HTML]{FCE4D6}0.078 & \cellcolor[HTML]{FCE4D6}0.17                          & \cellcolor[HTML]{FCE4D6}0.005 &                             &                             \\
W10             & \cellcolor[HTML]{FCE4D6}0.141 & \cellcolor[HTML]{FCE4D6}0.039 & \cellcolor[HTML]{FCE4D6}0.159 & \cellcolor[HTML]{FCE4D6}0.032 & \cellcolor{steel!10}0.384 &                             &                             &                             &                             \\
W11             & \cellcolor[HTML]{FCE4D6}0.12  & \cellcolor[HTML]{FCE4D6}0.03  & \cellcolor{steel!10}0.359 & \cellcolor[HTML]{FCE4D6}0.036 & \cellcolor[HTML]{FCE4D6}0.179 &                             &                             &                             &                             \\
W12             & \cellcolor[HTML]{FCE4D6}0.068 & \cellcolor[HTML]{FCE4D6}0.015 & \cellcolor{steel!10}0.514 & \cellcolor[HTML]{FCE4D6}0.115 &                             &                             &                             &                             &                             \\
W13             & \cellcolor[HTML]{FCE4D6}0.126 & \cellcolor[HTML]{FCE4D6}0.036 &                             &                             &                             &                             &                             &                             &                             \\ \midrule
\textbf{Median} & \textbf{0.12}                 & \textbf{0.04}                 & \textbf{0.528}                & \textbf{0.096}                & \textbf{0.162}                & \textbf{0.239}                & \textbf{0.013}                & \textbf{0.057}                & \textbf{0.078}                \\ \midrule
\textbf{IQR}    & \textbf{0.073}                & \textbf{0.086}                & \textbf{0.307}                & \textbf{0.058}                & \textbf{0.098}                & \textbf{0.131}                & \textbf{0.03}                 & \textbf{0.017}                & \textbf{0.05}                 \\ \bottomrule
\end{tabular}

\end{table}

\begin{table*}[t]
        \centering
	\caption{The proportion and quality of local optima across workloads for each system. The \colorbox{steel!10}{blue cells}, \colorbox{red-}{red cells}, and \colorbox{gold}{gold cells} indicate local optima with high ($\ell_q \geq 0.67$), medium (0.33 $\leq \ell_q \leq$ 0.67), and low quality ($\ell_q \leq 0.33$).}
        \scriptsize
        \label{tb:number_and_quality_of_local_optima}
	\setlength{\tabcolsep}{0.6 mm}
        
\begin{tabular}{c|cc|cc|cc|cc|cc|cc|cc|cc|cc}
\toprule
                                    & \multicolumn{2}{c|}{\textbf{\textsc{Lrzip}}}                                 & \multicolumn{2}{c|}{\textbf{\textsc{Xz}}}                                    & \multicolumn{2}{c|}{\textbf{\textsc{Z3}}}                                    & \multicolumn{2}{c|}{\textbf{\textsc{Dconvert}}}                              & \multicolumn{2}{c|}{\textbf{\textsc{Batik}}}                                & \multicolumn{2}{c|}{\textbf{\textsc{Kanzi}}}                                  & \multicolumn{2}{c|}{\textbf{\textsc{X264}}}                                  & \multicolumn{2}{c|}{\textbf{\textsc{H2}}}                                    & \multicolumn{2}{c}{\textbf{\textsc{Jump3r}}}                                 \\ \cline{2-19} 
\multirow{-2}{*}{\textbf{Workload}} & \multicolumn{1}{c|}{$\ell_p$}       & $\ell_q$                      & \multicolumn{1}{c|}{$\ell_p$}       & $\ell_q$                      & \multicolumn{1}{c|}{$\ell_p$}       & $\ell_q$                      & \multicolumn{1}{c|}{$\ell_p$}       & $\ell_q$                      & \multicolumn{1}{c|}{$\ell_p$}       & $\ell_q$                      & \multicolumn{1}{c|}{$\ell_p$}       & $\ell_q$                       & \multicolumn{1}{c|}{$\ell_p$}       & $\ell_q$                      & \multicolumn{1}{c|}{$\ell_p$}       & $\ell_q$                      & \multicolumn{1}{c|}{$\ell_p$}       & $\ell_q$                      \\ \midrule
W1              & 0.120          & \cellcolor{steel!10}0.972 & 0.025          & \cellcolor{steel!10}0.767 & 0.015          & \cellcolor[HTML]{FCE4D6}0.628 & 0.006          & \cellcolor{steel!10}0.787 & 0.017          & \cellcolor{steel!10}0.754 & 0.086          & \cellcolor[HTML]{FFF2CC}-0.333 & 0.180          & \cellcolor[HTML]{FFF2CC}0.096 & 0.026          & \cellcolor[HTML]{FFF2CC}0.244 & 0.041          & \cellcolor[HTML]{FCE4D6}0.376 \\
W2              & 0.194          & \cellcolor{steel!10}0.992 & 0.203          & \cellcolor[HTML]{FCE4D6}0.592 & 0.125          & \cellcolor{steel!10}0.784 & 0.005          & \cellcolor{steel!10}0.753 & 0.009          & \cellcolor{steel!10}0.906 & 0.105          & \cellcolor[HTML]{FFF2CC}-0.234 & 0.164          & \cellcolor[HTML]{FFF2CC}0.072 & 0.025          & \cellcolor[HTML]{FFF2CC}0.071 & 0.042          & \cellcolor[HTML]{FCE4D6}0.534 \\
W3              & 0.099          & \cellcolor{steel!10}0.987 & 0.025          & \cellcolor{steel!10}0.749 & 0.024          & \cellcolor{steel!10}0.967 & 0.026          & \cellcolor[HTML]{FCE4D6}0.569 & 0.010          & \cellcolor{steel!10}0.788 & 0.147          & \cellcolor[HTML]{FCE4D6}0.421  & 0.176          & \cellcolor[HTML]{FFF2CC}0.098 & 0.024          & \cellcolor[HTML]{FFF2CC}0.232 & 0.044          & \cellcolor[HTML]{FCE4D6}0.510 \\
W4              & 0.105          & \cellcolor{steel!10}0.945 & 0.021          & \cellcolor{steel!10}0.818 & 0.016          & \cellcolor{steel!10}0.966 & 0.005          & \cellcolor{steel!10}0.739 & 0.011          & \cellcolor{steel!10}0.890 & 0.084          & \cellcolor[HTML]{FFF2CC}-0.338 & 0.173          & \cellcolor[HTML]{FFF2CC}0.075 & 0.028          & \cellcolor[HTML]{FFF2CC}0.316 & 0.048          & \cellcolor[HTML]{FCE4D6}0.450 \\
W5              & 0.099          & \cellcolor{steel!10}0.830 & 0.025          & \cellcolor[HTML]{FCE4D6}0.633 & 0.012          & \cellcolor{steel!10}0.992 & 0.008          & \cellcolor{steel!10}0.718 & 0.011          & \cellcolor{steel!10}0.879 & 0.086          & \cellcolor[HTML]{FFF2CC}-0.247 & 0.176          & \cellcolor[HTML]{FFF2CC}0.092 & 0.030          & \cellcolor[HTML]{FFF2CC}0.101 & 0.042          & \cellcolor[HTML]{FCE4D6}0.396 \\
W6              & 0.099          & \cellcolor{steel!10}0.959 & 0.029          & \cellcolor{steel!10}0.673 & 0.022          & \cellcolor{steel!10}0.996 & 0.018          & \cellcolor[HTML]{FCE4D6}0.540 & 0.016          & \cellcolor{steel!10}0.826 & 0.089          & \cellcolor[HTML]{FFF2CC}-0.296 & 0.264          & \cellcolor[HTML]{FFF2CC}0.146 & 0.027          & \cellcolor[HTML]{FFF2CC}0.060 & 0.056          & \cellcolor[HTML]{FCE4D6}0.467 \\
W7              & 0.141          & \cellcolor{steel!10}0.904 & 0.045          & \cellcolor{steel!10}0.732 & 0.057          & \cellcolor{steel!10}0.829 & 0.004          & \cellcolor[HTML]{FCE4D6}0.610 & 0.008          & \cellcolor{steel!10}0.964 & 0.092          & \cellcolor[HTML]{FFF2CC}-0.282 & 0.194          & \cellcolor[HTML]{FFF2CC}0.138 & 0.028          & \cellcolor[HTML]{FFF2CC}0.283 &              &                             \\
W8              & 0.089          & \cellcolor{steel!10}0.974 & 0.021          & \cellcolor{steel!10}0.680 & 0.070          & \cellcolor{steel!10}0.973 & 0.009          & \cellcolor[HTML]{FCE4D6}0.652 & 0.012          & \cellcolor{steel!10}0.857 & 0.087          & \cellcolor[HTML]{FFF2CC}-0.293 & 0.246          & \cellcolor[HTML]{FFF2CC}0.143 & 0.028          & \cellcolor[HTML]{FCE4D6}0.376 &              &                             \\
W9              & 0.089          & \cellcolor{steel!10}0.972 & 0.021          & \cellcolor{steel!10}0.694 & 0.147          & \cellcolor{steel!10}0.938 & 0.016          & \cellcolor[HTML]{FCE4D6}0.578 & 0.022          & \cellcolor{steel!10}0.767 & 0.089          & \cellcolor[HTML]{FFF2CC}0.266  & 0.228          & \cellcolor[HTML]{FFF2CC}0.129 &              &                             &              &                             \\
W10             & 0.115          & \cellcolor{steel!10}0.945 & 0.020          & \cellcolor{steel!10}0.774 & 0.020          & \cellcolor{steel!10}0.803 & 0.033          & \cellcolor{steel!10}0.671 & 0.011          & \cellcolor{steel!10}0.900 &              &                              &              &                             &              &                             &              &                             \\
W11             & 0.120          & \cellcolor{steel!10}0.935 & 0.019          & \cellcolor{steel!10}0.698 & 0.013          & \cellcolor{steel!10}0.695 & 0.020          & \cellcolor[HTML]{FCE4D6}0.510 & 0.007          & \cellcolor{steel!10}0.964 &              &                              &              &                             &              &                             &              &                             \\
W12             & 0.105          & \cellcolor{steel!10}0.951 & 0.023          & \cellcolor{steel!10}0.792 & 0.010          & \cellcolor{steel!10}0.995 & 0.019          & \cellcolor{steel!10}0.775 &              &                             &              &                              &              &                             &              &                             &              &                             \\
W13             & 0.110          & \cellcolor{steel!10}0.980 & 0.025          & 0.637                         &              &                             &              &                             &              &                             &              &                              &              &                             &              &                             &              &                             \\ \midrule
\textbf{Median} & \textbf{0.105} & \textbf{0.959}                & \textbf{0.025} & \textbf{0.698}                & \textbf{0.021} & \textbf{0.952}                & \textbf{0.012} & \textbf{0.661}                & \textbf{0.011} & \textbf{0.879}                & \textbf{0.089} & \textbf{-0.282}                & \textbf{0.180} & \textbf{0.098}                & \textbf{0.028} & \textbf{0.238}                & \textbf{0.043} & \textbf{0.458}                \\ \midrule
\textbf{IQR}    & \textbf{0.021} & \textbf{0.029}                & \textbf{0.005} & \textbf{0.094}                & \textbf{0.046} & \textbf{0.180}                & \textbf{0.013} & \textbf{0.167}                & \textbf{0.004} & \textbf{0.096}                & \textbf{0.006} & \textbf{0.062}                 & \textbf{0.053} & \textbf{0.045}                & \textbf{0.003} & \textbf{0.198}                & \textbf{0.005} & \textbf{0.090}                \\ \bottomrule
\end{tabular}

\end{table*}

\subsubsection{Basin of Attraction \& Local Optima} For this, the results are summarized in Tables~\ref{tb:basin_of_attraction_for_global_optimum} and~\ref{tb:number_and_quality_of_local_optima}. 
(1) First, the proportion of configurations (against the whole search space) falling to the global optimum's basin are reported in Table~\ref{tb:basin_of_attraction_for_global_optimum}. Overall, based on the median values across workloads for each system, most systems exhibit a proportion below 0.21, suggesting that the global optimum is not effortlessly attainable. To further investigate the relationship between local optima superiority and their basin sizes, we assessed Spearman correlation ($\rho$) between them, as shown in Figure~\ref{fig:spearman_correlation_for_local_optima}. Overall, most systems show a moderate ($0.39 \textless \rho \leq 0.69$) and strong ($0.69 \textless \rho \leq 1$) correlations~\cite{wattanakriengkrai2022giving, chen2025accuracy}, especially for complex systems with larger LOC like \textsc{Z3}, \textsc{Batik}, and \textsc{H2}, indicating that higher-quality local optima generally attract more configurations, forming larger basins; (2)~Second, the proportion ($\ell_p$) and quality ($\ell_q$) of local optima are depicted in Table~\ref{tb:number_and_quality_of_local_optima}. The median proportion of local optima across workloads ranges from 1.1\% to 18\%, highlighting the multimodal nature of configuration landscapes. However, these local optima exhibit promising quality---\textsc{Lrzip}, \textsc{Xz}, \textsc{Z3}, \textsc{Dconvert}, and \textsc{Batik} contain mainly high-quality local optima ($\ell_q \geq 0.67$), while \textsc{Jump3r} mainly falls within the moderate-quality range ($0.33 \textless \ell_q \textless 0.67$); In contrast, \textsc{Kanzi}, \textsc{X264}, and \textsc{H2} tend to possess lower-quality local optima ($\ell_q \leq 0.33$).

\begin{figure}
    \centering
    \includegraphics[width=0.8\linewidth]{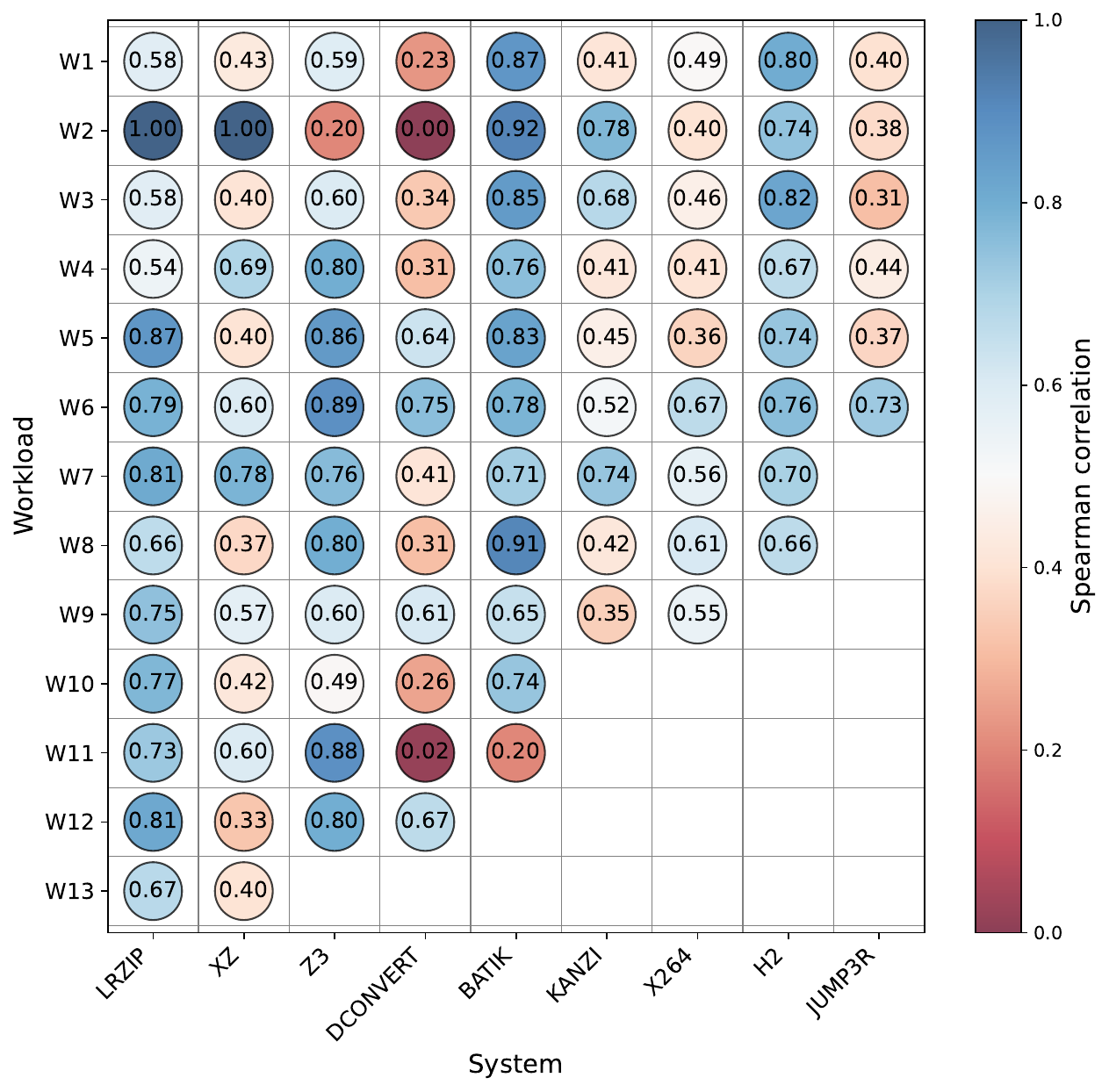}
    \caption{The Spearman correlation between  local optima superiority and basin size across workloads per system.}
    \label{fig:spearman_correlation_for_local_optima}
    \vspace{-15pt}
\end{figure}


\begin{quotebox}
   \noindent
   \textit{\textbf{Finding 4:} The global optimum's basin of attraction is not dominant and configuration landscapes have many superior local optima (especially for complex systems), often with larger basin size. 
   }
\end{quotebox}



\subsubsection{Autocorrelation}

The autocorrelation results are summarized in Table~\ref{tb:autocorrelation}. Based on median values across workloads, different systems exhibit varying ruggedness---\textsc{Xz}, \textsc{Z3}, \textsc{Dconvert}, \textsc{Batik}, and \textsc{H2} demonstrate a relatively smooth landscapes ($r \geq 0.5$); \textsc{Kanzi}, \textsc{X264}, and \textsc{Jump3r} fall into the moderately rugged category ($0.2 \textless r \textless 0.5$), while \textsc{Lrzip} illustrates a highly rugged landscape ($r \leq 0.2$).

\begin{quotebox}
   \noindent
   \textit{\textbf{Finding 5:} Software configuration landscapes exhibit diverse ruggedness patterns. Unlike prior studies~\cite{huang2024rethinking} that predominantly observe highly rugged landscapes, our investigation reveals that 5 out of the 9 studied systems display relatively smooth landscapes. On the other hand, \textsc{Lrzip}, \textsc{Kanzi}, \textsc{X264}, and \textsc{Jump3r}, despite spanning different areas, programming languages, and resource intensity, consistently exhibit higher degree of ruggedness. 
   }
\end{quotebox}

To understand what domain factors shape the ruggedness patterns, we examine the relations between the proportion of core options in each system and landscape characteristics. The results show that \textsc{Kanzi} (83\%), \textsc{X264} (72\%), and \textsc{Jump3r} (54\%) have a higher proportion of core options compared to \textsc{Xz} (50\%), \textsc{Z3} (33\%), \textsc{Dconvert} (50\%), \textsc{Batik} (13\%), and \textsc{H2} (20\%). This exhibits a clear trend with respect to the ruggedness of configuration landscape:

\begin{quotebox}
   \noindent
   \textit{\textbf{Finding 6:} Systems with a higher proportion of core options tend to exhibit greater landscape ruggedness, more deceptive structures, and a higher prevalence of local optima. 
   }
   
\end{quotebox}

\subsection{RQ3: Spatial Workload-level Synergy}
\label{subsecton:RQ3}



\begin{table}[t]
        \centering
	\scriptsize
	\setlength{\tabcolsep}{0.85mm}
	\caption{The autocorrelation across different workloads for nine software configurable systems. The \colorbox{steel!10}{blue cells}, \colorbox{red-}{red cells}, and \colorbox{gold}{gold cells} indicate smooth landscapes ($r \geq 0.5$), moderately rugged landscapes ($0.2 \textless r \textless 0.5$), and highly rugged landscapes ($r \leq 0.2$), respectively.}
    \label{tb:autocorrelation}
        \setlength{\tabcolsep}{9pt}
\begin{tabular}{c|c|c|c|c|c|c|c|c|c}
    \toprule
    \textbf{Workload} & \textbf{\textsc{Lrzip}} & \textbf{\textsc{Xz}} & \textbf{\textsc{Z3}} & \textbf{\textsc{Dconvert}} & \textbf{\textsc{Batik}} & \textbf{\textsc{Kanzi}} & \textbf{\textsc{X264}} & \textbf{\textsc{H2}} & \textbf{\textsc{Jump3r}} \\ \midrule
W1              & \cellcolor[HTML]{FFF2CC}-0.012 & \cellcolor{steel!10}0.630 & \cellcolor{steel!10}0.659 & \cellcolor{steel!10}0.781 & \cellcolor{steel!10}0.688 & \cellcolor[HTML]{FCE4D6}0.339 & \cellcolor[HTML]{FCE4D6}0.254 & \cellcolor{steel!10}0.779 & \cellcolor[HTML]{FCE4D6}0.411 \\
W2              & \cellcolor[HTML]{FFF2CC}0.167  & \cellcolor[HTML]{FFF2CC}0.196 & \cellcolor{steel!10}0.697 & \cellcolor{steel!10}0.779 & \cellcolor{steel!10}0.734 & \cellcolor[HTML]{FCE4D6}0.365 & \cellcolor[HTML]{FCE4D6}0.263 & \cellcolor{steel!10}0.828 & \cellcolor[HTML]{FCE4D6}0.320 \\
W3              & \cellcolor[HTML]{FFF2CC}0.017  & \cellcolor{steel!10}0.653 & \cellcolor{steel!10}0.604 & \cellcolor[HTML]{FCE4D6}0.441 & \cellcolor{steel!10}0.796 & \cellcolor[HTML]{FCE4D6}0.329 & \cellcolor[HTML]{FCE4D6}0.258 & \cellcolor{steel!10}0.770 & \cellcolor[HTML]{FCE4D6}0.400 \\
W4              & \cellcolor[HTML]{FFF2CC}-0.034 & \cellcolor[HTML]{FCE4D6}0.491 & \cellcolor{steel!10}0.729 & \cellcolor{steel!10}0.772 & \cellcolor{steel!10}0.825 & \cellcolor[HTML]{FCE4D6}0.316 & \cellcolor[HTML]{FCE4D6}0.244 & \cellcolor[HTML]{FFF2CC}0.129 & \cellcolor[HTML]{FCE4D6}0.420 \\
W5              & \cellcolor[HTML]{FFF2CC}-0.007 & \cellcolor{steel!10}0.626 & \cellcolor{steel!10}0.605 & \cellcolor{steel!10}0.754 & \cellcolor{steel!10}0.736 & \cellcolor[HTML]{FCE4D6}0.316 & \cellcolor[HTML]{FCE4D6}0.211 & \cellcolor{steel!10}0.807 & \cellcolor[HTML]{FCE4D6}0.442 \\
W6              & \cellcolor[HTML]{FFF2CC}0.026  & \cellcolor{steel!10}0.597 & \cellcolor{steel!10}0.617 & \cellcolor{steel!10}0.591 & \cellcolor{steel!10}0.725 & \cellcolor[HTML]{FCE4D6}0.336 & \cellcolor[HTML]{FCE4D6}0.282 & \cellcolor{steel!10}0.803 & \cellcolor[HTML]{FCE4D6}0.438 \\
W7              & \cellcolor[HTML]{FFF2CC}0.180  & \cellcolor{steel!10}0.534 & \cellcolor{steel!10}0.659 & \cellcolor{steel!10}0.892 & \cellcolor{steel!10}0.670 & \cellcolor[HTML]{FCE4D6}0.377 & \cellcolor[HTML]{FCE4D6}0.222 & \cellcolor{steel!10}0.608 &                             \\
W8              & \cellcolor[HTML]{FFF2CC}-0.023 & \cellcolor{steel!10}0.550 & \cellcolor{steel!10}0.680 & \cellcolor{steel!10}0.767 & \cellcolor{steel!10}0.723 & \cellcolor[HTML]{FCE4D6}0.329 & \cellcolor[HTML]{FCE4D6}0.246 & \cellcolor{steel!10}0.687 &                             \\
W9              & \cellcolor[HTML]{FFF2CC}-0.024 & \cellcolor{steel!10}0.581 & \cellcolor{steel!10}0.741 & \cellcolor{steel!10}0.557 & \cellcolor{steel!10}0.694 & \cellcolor[HTML]{FCE4D6}0.440 & \cellcolor[HTML]{FCE4D6}0.316 &                             &                             \\
W10             & \cellcolor[HTML]{FFF2CC}-0.009 & \cellcolor{steel!10}0.550 & \cellcolor{steel!10}0.678 & \cellcolor[HTML]{FFF2CC}0.005 & \cellcolor{steel!10}0.723 &                             &                             &                             &                             \\
W11             & \cellcolor[HTML]{FFF2CC}-0.035 & \cellcolor{steel!10}0.568 & \cellcolor{steel!10}0.703 & \cellcolor[HTML]{FCE4D6}0.322 & \cellcolor{steel!10}0.636 &                             &                             &                             &                             \\
W12             & \cellcolor[HTML]{FFF2CC}0.024  & \cellcolor{steel!10}0.612 & \cellcolor{steel!10}0.578 & \cellcolor{steel!10}0.672 &                             &                             &                             &                             &                             \\
W13             & \cellcolor[HTML]{FFF2CC}0.035  & \cellcolor{steel!10}0.714 &                             &                             &                             &                             &                             &                             &                             \\ \midrule
\textbf{Median} & \textbf{-0.007}                & \textbf{0.581}                & \textbf{0.669}                & \textbf{0.713}                & \textbf{0.723}                & \textbf{0.336}                & \textbf{0.254}                & \textbf{0.774}                & \textbf{0.415}                \\ \midrule
\textbf{IQR}    & \textbf{0.049}                 & \textbf{0.076}                & \textbf{0.085}                & \textbf{0.246}                & \textbf{0.044}                & \textbf{0.037}                & \textbf{0.019}                & \textbf{0.137}                & \textbf{0.031}                \\ \bottomrule
\end{tabular}

        \vspace{-5pt}
\end{table}

At workload level, following \approach, we focus on findings related to \textit{all workloads of a system} and \textit{all workloads across all systems}.

FDC values (see Table~\ref{tb:FDC}) and autocorrelation (see Table~\ref{tb:autocorrelation}) show that some systems (e.g., \textsc{Dconvert}, \textsc{H2}) exhibit high variability, while others remain relatively stable. Similarly, the proportion of configurations in the global optimum’s basin (see Table~\ref{tb:basin_of_attraction_for_global_optimum}) remains stable across workloads for most systems, except \textsc{Z3}, which shows notable fluctuations. The proportion and quality of local optima (see Table~\ref{tb:number_and_quality_of_local_optima}) remain largely consistent across workloads, with minor deviations in \textsc{Z3} and \textsc{Dconvert}.

\begin{quotebox}
   \noindent
   \textit{\textbf{Finding 7:} Workload-induced landscape variations are generally stable with only a minority of systems pronounce shifts.
  }
\end{quotebox}  




To uncover the factors driving workload-induced variations, we investigate whether workload properties consistently correlate with landscape shifts. \revision{Specifically, we focus our analysis on systems where landscape ruggedness exhibits significant fluctuations across workloads, assessing whether workload type and scale consistently impact landscape characteristics. Among all systems, \textsc{Dconvert}, \textsc{H2}, and \textsc{Xz} display the highest inter-workload variability in autocorrelation values (see Table \ref{tb:autocorrelation}). However, \textsc{Xz} is excluded from further analysis because all of its workloads belong to the same type and lack comprehensive workload scale information, limiting its suitability for correlating workload characteristics (e.g., type and scale) with landscape shifts. In contrast, \textsc{Dconvert} and \textsc{H2} contain diverse workload types and scale variations, enabling a more meaningful investigation. Therefore, we conduct a detailed analysis of \textsc{Dconvert} and \textsc{H2}, examining their workload characteristics (including type and scale) alongside their autocorrelation values.}



As shown in Table~\ref{tb:workload-factors}, even within the same type (e.g., \texttt{tpcc}), variations in scale cause substantial shifts in landscape structure. Likewise, workloads of the similar scale but different types also induce notable changes (e.g., comparing \texttt{png-large} to \texttt{svg-large}).

\begin{quotebox}
   \noindent
   \textit{\textbf{Finding 8:} Workload effects on landscape structure are not uniformly tied to type or scale. Both contribute to variations, but their impact is system-dependent, highlighting complex interactions with system characteristics. 
   }
\end{quotebox} 

\begin{table}[t]
\setlength{\tabcolsep}{3.5 mm}
\caption{Autocorrelation of workloads in \textsc{Dconvert} and \textsc{H2}.}
\label{tb:workload-factors}
\small
\begin{tabular}{lc|lc}
\toprule
\multicolumn{2}{c|}{\textbf{\textsc{Dconvert}}} & \multicolumn{2}{c}{\textbf{\textsc{H2}}} \\ \midrule
\textbf{Workload (Type-Scale)}         & \textbf{Autocorrelation}     & \textbf{Workload (Type-Scale)}      & \textbf{Autocorrelation} \\ \midrule

\rowcolor{steel!10} 
jpeg-large                                 & 0.781                                  & smallbank-1                             & 0.779                                  \\
\rowcolor{steel!10} 
jpeg-medium                                & 0.779                                  & smallbank-10                            & 0.828                                  \\
\cellcolor[HTML]{FCE4D6}jpeg-small                                 & \cellcolor[HTML]{FCE4D6}0.441                                  & \cellcolor{steel!10}tpcc-2                                  & \cellcolor{steel!10}0.770                                  \\
\rowcolor{steel!10} 
png-large                                  & 0.772                                  & \cellcolor[HTML]{FFF2CC}\textbf{tpcc-8} & \cellcolor[HTML]{FFF2CC}\textbf{0.129} \\
\rowcolor{steel!10} 
png-medium                                 & 0.754                                  & voter-16                                & 0.807                                  \\
\rowcolor{steel!10} 
png-small                                  & 0.591                                  & voter-2                                 & 0.803                                  \\
\rowcolor{steel!10} 
psd-large                                  & 0.892                                  & ycsb-2400                               & 0.608                                  \\
\rowcolor{steel!10} 
psd-medium                                 & 0.767                                  & ycsb-600                                & 0.687                                  \\
\cellcolor{steel!10}psd-small          & \cellcolor{steel!10}0.557          &                                       &                                      \\
\cellcolor[HTML]{FFF2CC}\textbf{svg-large} & \cellcolor[HTML]{FFF2CC}\textbf{0.005} &                                       &                                      \\
\cellcolor[HTML]{FCE4D6}svg-medium         & \cellcolor[HTML]{FCE4D6}0.322          &                                       &                                      \\
\cellcolor{steel!10}svg-small          & \cellcolor{steel!10}0.672          &                                       &                                      \\ \bottomrule
\end{tabular}

 \vspace{-5pt}
\end{table}

\subsection{Actionable Insights}\label{subsection:insights}
\quad\textbf{\textit{Findings 1 \& 2}} suggest that the proportion and magnitude of ruggedness-sensitive options are system-dependent, with core options ($F_1$) exhibiting a stronger impact on landscape ruggedness than the other types.

\begin{implicationbox}
   \noindent
   \textbf{Insight 1:} Tuner design should be system-specific, where prioritizing ruggedness-sensitive options may benefit exploitation-driven tuners, potentially leading to significant performance improvements. For system designers, selectively exposing core options helps strike a trade-off between configurability and stability.
\end{implicationbox}

\textbf{\textit{Finding 3}} suggests that the majority of the configuration landscapes exhibit clear trends toward the global optimum, providing rich guidance to a tuner.

\begin{implicationbox}
   \noindent
   \textbf{Insight 2:} 
   Given the observed global trends in many configuration landscapes, tuners that favor convergence/exploitation (e.g., iterated local search~\cite{lourencco2003iterated}) could often work well.
\end{implicationbox}

\textbf{\textit{Finding 4}} implies that tuners are prone to being trapped in suboptimal configurations, which often exhibit a generally acceptable quality.

\begin{implicationbox}
   \noindent
   \textbf{Insight 3:} While suboptimal configurations often yield acceptable results, global exploration is still crucial for performance-critical tasks. Strategies like multi-objectivization~\cite{chen2024mmo, chen2021multi} and landscape smoothing~\cite{DBLP:conf/iclr/KirjnerYSBJBF24}, have shown promise in counteracting premature convergence.
\end{implicationbox}

\textbf{\textit{Findings 5 \& 6}} indicate that software configuration landscapes exhibit diverse ruggedness patterns, with some systems displaying smooth structures while others, particularly those with a higher proportion of core options, exhibiting a higher degree of ruggedness.

\begin{implicationbox}
   \noindent
   \textbf{Insight 4:} The system-specific ruggedness highlights the need to analyze the landscape characteristics before configuration tuning. Smooth landscapes benefit performance modeling, making model-based tuners preferable. Conversely, rugged landscapes require strong exploration to handle abrupt performance fluctuations. For system designers, coding fewer core options in the systems to configure can greatly help to relieve the difficulty of tuning them.
\end{implicationbox}

\textbf{\textit{Findings 7 \& 8}} suggest that workload-induced variations are system-specific---most exhibit stable landscape properties across workloads, while only a few show dramatic shifts. Moreover, workload effects on landscape structure do not consistently align with type or scale, indicating complex interactions with system properties.

\begin{implicationbox}
   \noindent
   \textbf{Insight 5:} The tuning strategies should be workload-aware rather than one-size-fits-all. Systems with stable landscapes may benefit from static tuning~\cite{kinneer2021information, chen2022lifelong}, while those with significant shifts require self-adaptation~\cite{DBLP:journals/corr/abs-2501-00840}. 
\end{implicationbox}

\subsection{Experimental Validation of Actionable Insights}\label{subsection:validation}
We have systematically analyzed the configuration landscapes from multiple perspectives, including spatial option-level synergy, spatial system-level synergy, and spatial workload-level synergy. Based on these analyses, we draw several actionable insights that aim to guide tuner researchers and system designers in making more informed decisions. To verify whether the excavated insights hold in practice, we conduct a series of algorithmic experiments on configuration tuning and discuss how the results can be related to our insights above.

\subsubsection{Validating Insight 1: Prioritizing Ruggedness-Sensitive Options}


To examine the role of ruggedness-sensitive options, we compare tuners that either uniformly treat all configuration options or prioritize those identified as ruggedness-sensitive. \revminor{In addition, to further contextualize this design choice, we construct a variant where sensitive options are identified using ANOVA. This allows us to contrast landscape-derived sensitivity with a commonly used statistical sensitivity measure.} We consider two representative types of tuners with different exploration-exploitation characteristics: hill climbing (HC) and genetic algorithm (GA), each of which is evaluated in two variants, as follows:

\begin{itemize}
    \item \textbf{Vanilla HC:} This approach treats all configuration options equally. At each iteration, an option is selected uniformly at random and mutated to explore a neighboring configuration.
    \item \textbf{Priority HC:} This approach gives higher priority to sensitive options during mutation. Specifically, at each iteration, there is a 90\% probability that a mutation is applied to one of the sensitive options, and only a 10\% chance that a non-sensitive option is selected. \revminor{In our experiments, sensitive options are identified either by ruggedness analysis or by ANOVA.}
   
    \item \textbf{Vanilla GA:} This approach uses uniform crossover, where each configuration option has an equal 50\% probability of being exchanged between parent configurations, ensuring all options are treated equally.
    \item \textbf{Priority GA:} This approach biases the crossover toward sensitive options. At each crossover operation, there is a 90\% probability that a sensitive option is exchanged. \revminor{As with HC, sensitive options are determined either through ruggedness analysis or ANOVA.}
\end{itemize}

\begin{figure}
    \centering
    \includegraphics[width=1.0\linewidth]{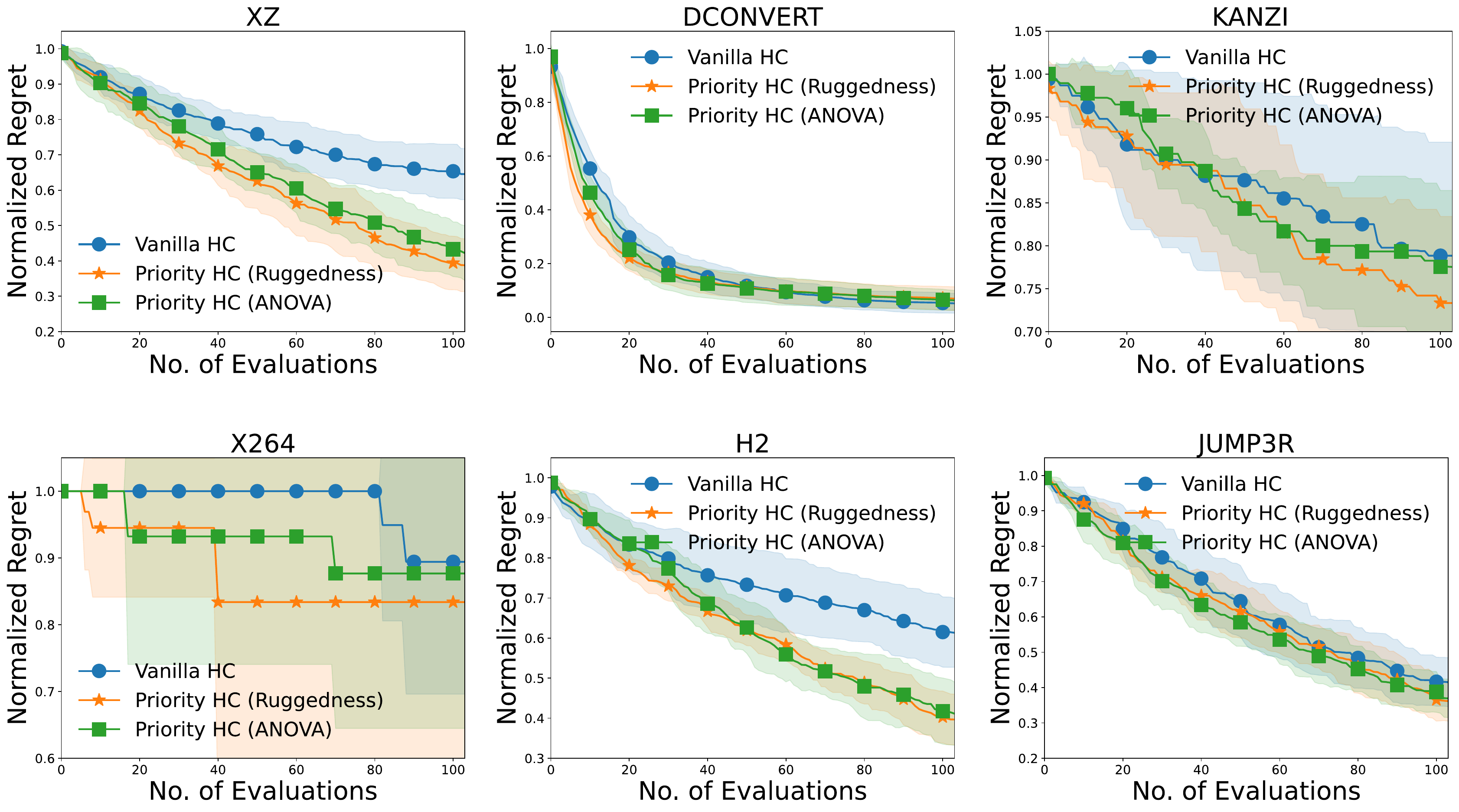}
    \caption{\revminor{The regret trajectories of Vanilla HC and Priority HC, aggregated across all workloads for each system.}}
    \label{fig:insight_1}
\end{figure}

In our experiment, we focus on systems that contain a mix of both ruggedness-sensitive and non-ruggedness-sensitive configuration options, i.e., \textsc{Xz, Dconvert, Kanzi, X264, H2}, and \textsc{Jump3r}. The remaining three systems were excluded because all their options either strongly or weakly affect landscape ruggedness, which makes the comparison between Vanilla HC and Priority HC (as well as Vanilla GA and Priority GA) uninformative for this validation. All experiments are repeated 30 times independently, and the tuning trajectories of the above tuners are illustrated in Figures~\ref{fig:insight_1} and~\ref{fig:insight_1_ga}, where the results are averaged over all workloads for each system to simplify the exposition. As shown in Figure \ref{fig:insight_1}, Priority HC achieves faster convergence and lower final regret than Vanilla HC in the majority of systems, suggesting that prioritizing ruggedness-sensitive options leads to more efficient tuning. \revminor{Notably, prioritization based on ruggedness-sensitive options tends to yield larger improvements than prioritization based on ANOVA-identified options. This indicates that ruggedness analysis more accurately captures the option characteristics that influence the search behavior of HC.} An exception is \textsc{Dconvert}, where both variants exhibit similar performance. This may be attributed to the limited number of ruggedness-sensitive options in this system---only two were identified---increasing the likelihood of premature convergence due to an overly narrow search focus. \revminor{In contrast, Figure~\ref{fig:insight_1_ga} presents the comparison between Vanilla GA and Priority GA. As expected, prioritizing ruggedness-sensitive options leads to nearly identical convergence trajectories across most studied systems. This suggests that tuners with strong exploration are inherently more robust to ruggedness and therefore gain limited advantage from prioritizing ruggedness-sensitive options. Surprisingly, contrary to the common intuition that prioritizing performance-sensitive options should generally improve tuning efficiency, prioritizing the options identified by ANOVA does not yield noticeable improvements. This suggests that relying solely on performance-sensitivity analysis, without considering the characteristics of the tuning algorithm, may lead to misleading guidance for search strategies. ANOVA measures how much an option contributes to overall performance variance, but it does not capture how options reshape the landscape structure or interact with the search dynamics of different tuners. Consequently, ANOVA-based prioritization fails to differentiate the behaviors of HC and GA, whereas ruggedness-based analysis provides algorithm-aware insights that better explain when option prioritization is beneficial.}

\rev{
Overall, these observations provide empirical evidence for \textbf{\textit{Insight 1}}, suggesting that tuner design should be system-specific. In systems where ruggedness-sensitive options are prevalent, giving them priority during the tuning process can enhance the efficiency, particularly for exploitation-driven tuners, whereas exploration-driven tuners tend to be inherently robust to such ruggedness effects and therefore benefit less from such prioritization.}

\begin{figure}
    \centering
    \includegraphics[width=1.0\linewidth]{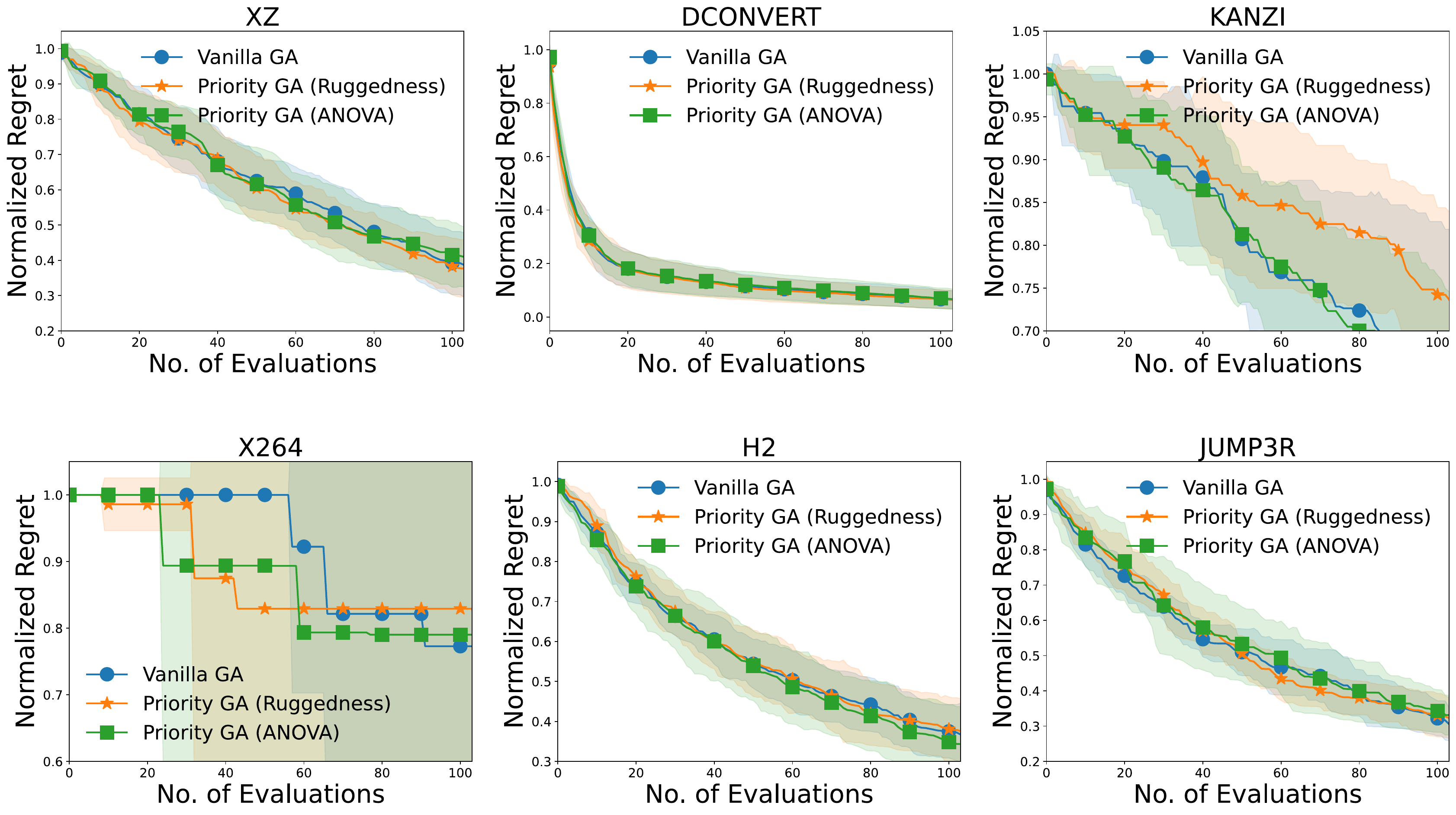}
    \caption{\revminor{The regret trajectories of Vanilla GA and Priority GA, aggregated across all workloads for each system.}}
    \label{fig:insight_1_ga}
\end{figure}

\begin{figure}
    \centering
    \includegraphics[width=1.0\linewidth]{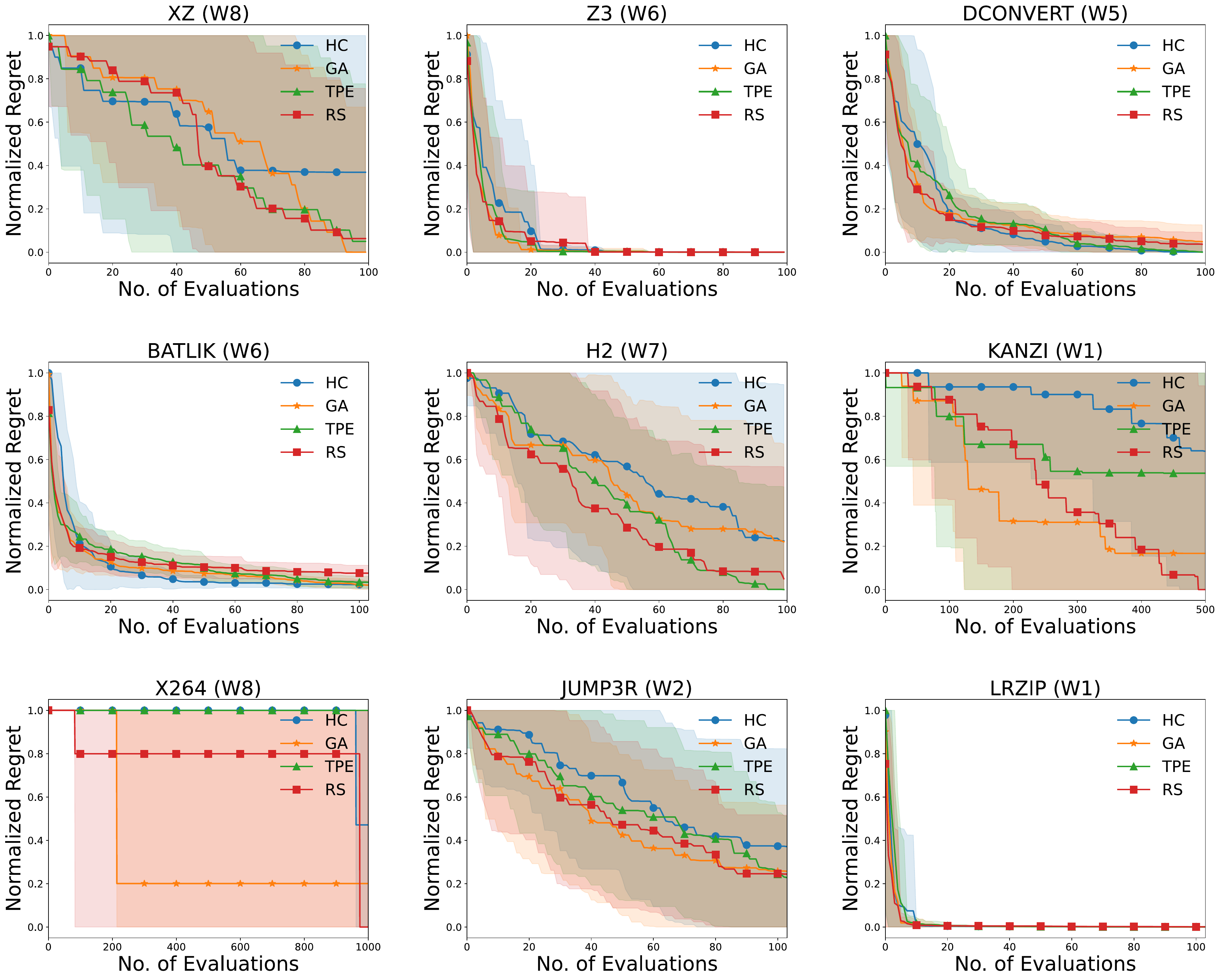}
    \caption{\rev{The regret trajectories using GA, RS, HC, and TPE to tune nine systems.}}
    \label{fig:insight_2_4}
\end{figure}

\subsubsection{Validating Insights 2 to 4: Tuner Performance across Landscape Characteristics.} 

Insights 2, 3, and 4 are derived from a detailed analysis of configuration landscapes from three aspects, including global structure, local structure, and ruggedness. These insights suggest that the performance of tuners may depend on the underlying characteristics of the configuration landscape. For instance, model-based tuners are expected to perform well on landscapes with smooth global structures, where accurate performance modeling is feasible. In contrast, landscapes that are rugged or highly multimodal may favor exploration-driven tuners that can better escape local optima. To evaluate whether these actionable insights hold in practice, we conduct a comparative experiment involving four tuners (i.e., GA, RS, HC, and TPE) with diverse properties across nine representative systems. 

Specifically, the characteristics and implementation details of the four tuners are summarized as follows:
\begin{itemize}
    \item \textbf{Genetic Algorithm (GA)~\cite{olaechea2014comparison,back1993overview}:} GA is a population-based tuner that explores configuration space via selection, crossover, and mutation, equipped with strong exploration capability. Specifically, we use binary tournament selection for mating, a population size of 20, and apply uniform crossover with a probability of 0.9 and boundary mutation with a probability of 0.1, respectively.
    \item \textbf{Random Search (RS)~\cite{oh2017finding}:} RS is a pure exploration tuner that samples configurations randomly from the configuration space.
    \item \textbf{Hill Climbing (HC)~\cite{li2014mronline,xi2004smart}:}  HC is a deterministic local search tuner that iteratively explores the neighborhood of the current configuration and moves to an improved one when available. In our implementation, we adopt the first-improvement strategy, where neighbors are evaluated at random and the search accepts the first configuration that yields better performance. 
    \item \textbf{TPE~\cite{bergstra2011algorithms}:} TPE is a model-based tuner that uses density estimation to model good and bad configurations separately and selects new configurations by maximizing expected improvement.
\end{itemize}

\rev{
Accordingly, we summarize the landscape characteristics of the nine studied systems into three representative types as follows:
\begin{itemize}
    \item \textbf{Type I (\textsc{Xz, Z3, Dconvert, Batik}, and \textsc{H2})}: These systems often exhibit relatively smooth landscapes ($r \geq 0.5$) with a small number of local optima, most of which yield high-quality performance.
    \item \textbf{Type II (\textsc{Kanzi, X264}, and \textsc{Jump3r})}: These system present moderately rugged landscapes ($0.2 \textless r \textless 0.5$), containing a certain number of local optima with medium quality.
    \item \textbf{Type III (\textsc{Lrzip})}: This system features a highly rugged landscape ($r \leq 0.2$) characterized by numerous local optima, many of which are of high-quality and close to the global optimum ($\ell_p$=10.5\% and $\ell_q$=0.959).
\end{itemize}
}

In this experiment, we select one representative workload for each system, chosen to closely reflect the system's overall landscape characteristics. Each tuner is independently executed 30 times to mitigate stochastic variance, and the aggregated regret trajectories are illustrated in Figure \ref{fig:insight_2_4}. The results reveal several notable patterns, which we outline as follows:


\begin{itemize}
    \item [1)] \textbf{Type I}: These systems exhibit smooth landscapes with a clear global structure, HC converges slowly but steadily to high-quality configurations. Meanwhile, the model-based tuner TPE performs comparably well, probably due to the reliability of surrogate modeling in such a smooth landscape. These results align with \textbf{\textit{Insights 2 \& 4}}, which suggest that exploitation-driven and model-based methods could be effective when the landscape exhibits clear global guidance and smooth structure.
    \item [2)] \textbf{Type II}: These systems are characterized by moderate ruggedness and multimodality; both HC and TPE exhibit degraded performance. For HC, the relatively rugged landscape with abundant local optima may cause the search to get trapped in suboptimal regions and struggle to escape. For TPE, the performance drop may stem from the difficulty of learning an accurate surrogate model under such an irregular and multimodal landscape structure. In contrast, RS and GA, which offer stronger global exploration capabilities, achieve superior performance. These observations lend support to \textbf{\textit{Insights 3 \& 4}}, underscoring the importance of exploration in rugged and multimodal landscapes.
    \item [3)] \textbf{Type III}: On workload W1 of \textsc{Lrzip}, which is characterized by a highly rugged landscape with numerous high-quality local optima, all tuners, including those with limited exploration capabilities, unexpectedly achieve good performance and rapid convergence. This counterintuitive result can be attributed to the dense distribution of high-quality local optima ($\ell_p$=10.5\% and $\ell_q$=0.959), which allows search processes to quickly settle into high-performing regions (\textbf{\textit{Insights~3~\&~4}}). \rev{To further investigate this seemingly counterintuitive phenomenon, we visualize the configuration landscape and distribution of local optima for \textsc{Lrzip}, as shown in Figure~\ref{fig:lrzip_analysis}. From Figure~\ref{fig:lrzip_analysis}(a), it can be observed that while a few configurations exhibit notably poor performance (corresponding to deep valleys), the majority of configurations achieve similar and relatively high performance, resulting in a flat upper region where the performance differences among good configurations are marginal. Furthermore, Figure~\ref{fig:lrzip_analysis}(b) shows that the identified local optima are densely distributed ($\ell_p$=10.5\%) and exhibit small performance variations, which enables different tuners, even those with limited exploration, to quickly converge to high-performing regions. Moreover, the configuration space of \textsc{Lrzip} involves only seven configuration options, further reducing the search difficulty and explaining why all tuners achieve similar performance despite the high ruggedness.} 
    
\end{itemize}

\begin{figure}[t]
  \centering
  \begin{subfigure}{0.48\textwidth}
    \centering
    \includegraphics[width=\linewidth]{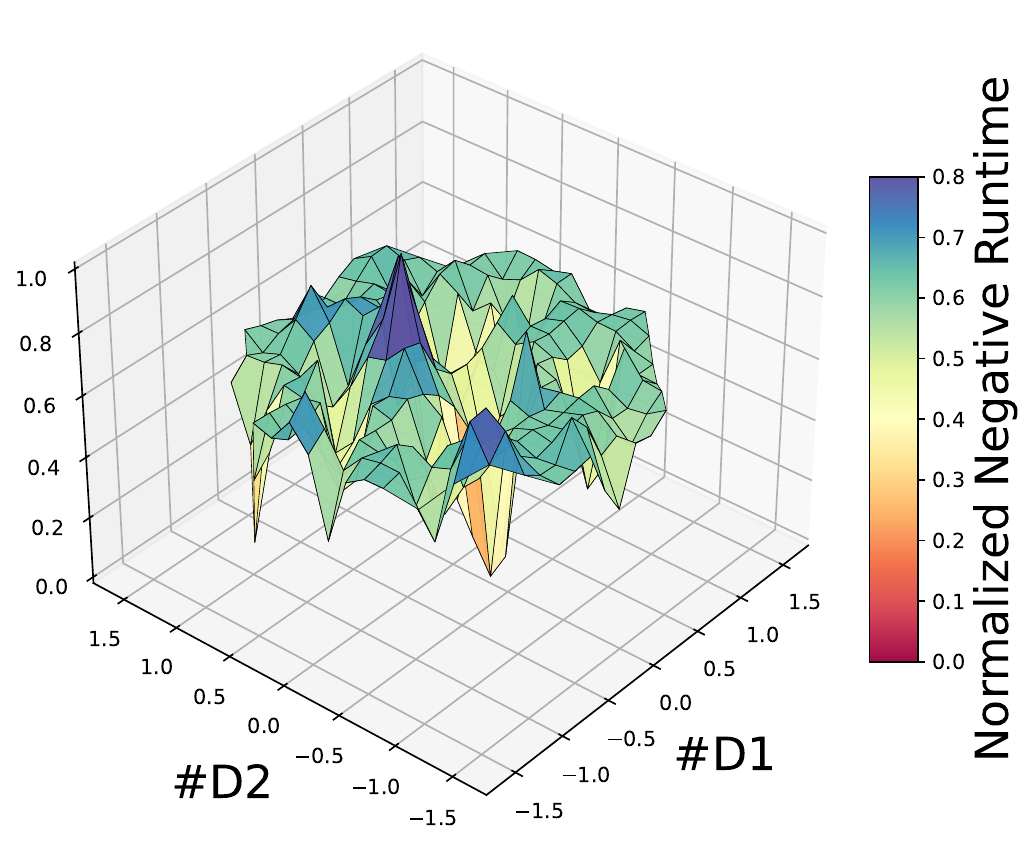}
    \caption{Landscape}
    \label{fig:a}
  \end{subfigure}\hfill
  \begin{subfigure}{0.48\textwidth}
    \centering
    \includegraphics[width=\linewidth]{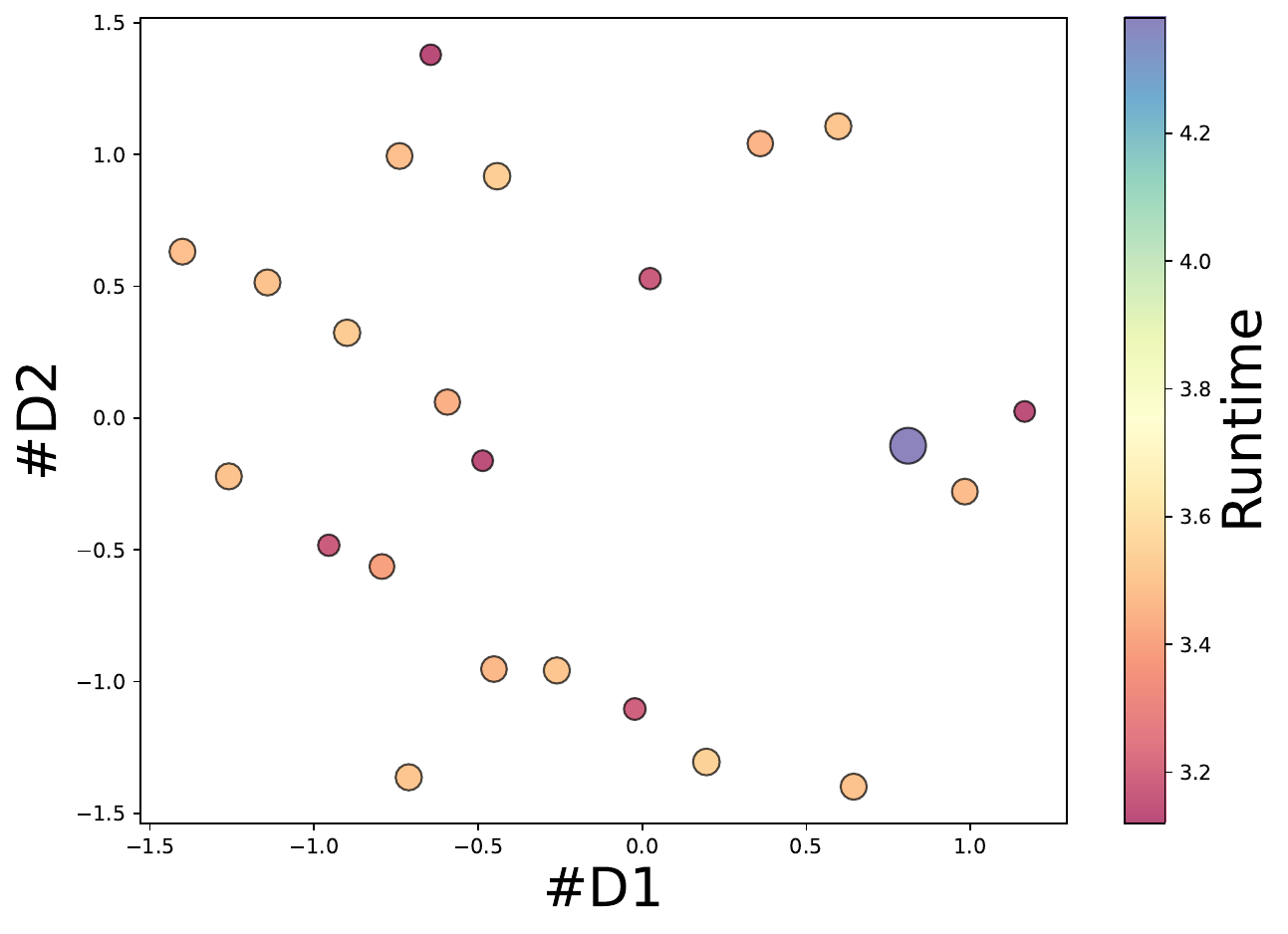}
    \caption{Local Optima}
    \label{fig:b}
  \end{subfigure}
  \caption{\rev{Illustration of the configuration landscape and local optima for workload W1 of \textsc{Lrzip}. (a) Normalized negative runtime landscape (higher values indicate better performance). (b) Spatial distribution of identified local optima and their corresponding performance values.}}
  \label{fig:lrzip_analysis}
\end{figure}

Overall, the above analyses suggest that tuner performance often relies on the underlying landscape characteristics of the optimization problem. By systematically analyzing multiple aspects of the configuration landscape, such as ruggedness, local structure, and global structure, practitioners can better assess the problem's inherent difficulty and interpret the behavior of different tuning strategies. Such understanding not only facilitates informed tuner selection and design but also enhances the explainability of why a tuner succeeds or fails under specific scenarios. 

\subsubsection{Validating Insight 5: Transferability Across Workloads}

To investigate whether configuration transfer across workloads can improve tuning performance, we conduct an experiment under dynamic workload settings, comparing tuning with and without configuration transfer. That is, we examine whether configuration transfer is beneficial under similar configuration landscapes, and conversely, whether it loses effectiveness---thus calling for self-adaptation strategies~\cite{DBLP:journals/corr/abs-2501-00840}---when landscape structures vary significantly across workloads, as suggested by \textbf{\textit{Insight 5}}. Specifically, we compare the following two variants of the genetic algorithm (GA)~\cite{olaechea2014comparison,back1993overview}:

\begin{itemize}
    \item \textbf{Restarted GA:} This approach responds to workload changes by triggering a new search from scratch with randomly initialized configurations.
    \item \textbf{Transfer GA:} This approach responds to workload changes by transferring all the configurations optimized from the most recent past workload as the initial population. \revision{For example, given a random sequence of workloads (e.g., W1 $\rightarrow$ W3 $\rightarrow$ W4 $\rightarrow$ W2 $\rightarrow$ W6 $\rightarrow$ W5), when tuning starts on W3, the configurations optimized under W1 are used as initialization; for W4, the ones from W3 are transferred, and so on.}
\end{itemize}

In our experiment, each run begins with a randomly shuffled order of workloads, mimicking a dynamic operational environment. For each workload arrival, the tuner is triggered to search for an optimal configuration under a fixed budget of 80 measurements, a setting aligned with the prior study on configuration tuning~\cite{DBLP:journals/corr/abs-2501-00840}. To obtain a statistically sound comparison, all experiments are run 30 times independently. Then, we use the Wilcoxon rank-sum test ($p$)~\cite{arcuri2011practical} with 0.05 significance level and $\hat{A}_{12}$ effect size~\cite{vargha2000critique} to assess statistical differences between the above tuners. We say a comparison is statistically significant only if it has $\hat{A}_{12} \textgreater$ 0.56 (or $\hat{A}_{12} \textless 0.44$) and $p \textless 0.05$. 

The experimental results are summarized in Table~\ref{tb:insight5}. Clearly, Transfer GA significantly outperforms Restarted GA, achieving better performance in 63 out of 93 cases. This advantage may stem from the fact that, in most systems, the configuration landscapes across different workloads exhibit structural similarity. As a result, directly reusing configurations optimized under the previous workload can provide a strong starting point for tuning under the current workload, reducing search overhead and accelerating convergence. However, on some systems such as \textsc{Z3} and \textsc{H2}, the effectiveness of Transfer GA diminishes, yielding no clear advantage over Restarted GA, and in some cases even leading to negative transfer. This may be attributed to the relatively greater variability in landscape structure across workloads observed in these systems, as discussed in Section \ref{subsecton:RQ3}. Overall, this comparative experiment provides empirical support for \textbf{\textit{Insight 5}}, confirming that transferability of configurations is generally beneficial when landscape structures remain stable across workloads, while its effectiveness diminishes as landscape variability increases, requiring more sophisticated self-adaptation approaches to handle such challenges~\cite{DBLP:journals/corr/abs-2501-00840}.
\begin{table}[t]
        \centering
	\tiny
	\setlength{\tabcolsep}{0.5 mm}
        \renewcommand{\arraystretch}{1.1}
	\caption{The Mean and Standard deviation (Std) of performance objectives between Transfer GA and Restarted GA for all cases over 30 runs; The \colorbox{steel!10}{blue cells}, \colorbox{red-}{red cells}, and \colorbox{gold}{gold cells} respectively indicate Transfer GA performing significantly better than, similarly to, or worse than Restarted GA.}
    \label{tb:insight5}
        
\begin{tabular}{c|c|c|c|c|c|c|c|c|c|c}
\hline
                                    &                                  & \textbf{Lrzip}                         & \textbf{Xz}                            & \textbf{Z3}                           & \textbf{Dconvert}                      & \textbf{Batik}                        & \textbf{Kanzi}                        & \textbf{X264}                         & \textbf{H2}                                  & \textbf{Jump3r}                       \\ \cline{3-11} 
\multirow{-2}{*}{\textbf{Workload}} & \multirow{-2}{*}{\textbf{Tuner}} & \textbf{Mean (Std)}                          & \textbf{Mean (Std)}                          & \textbf{Mean (Std)}                         & \textbf{Mean (Std)}                          & \textbf{Mean (Std)}                         & \textbf{Mean (Std)}                         & \textbf{Mean (Std)}                         & \textbf{Mean (Std)}                                & \textbf{Mean (Std)}                         \\ \hline
                                    & Transfer GA                      & \cellcolor[HTML]{FCE4D6}3.131 (0.014)  & \cellcolor{steel!10}3.773 (0.955)  & \cellcolor[HTML]{FCE4D6}5.860 (0.016) & \cellcolor{steel!10}1.797 (0.009)  & \cellcolor{steel!10}0.903 (0.009) & \cellcolor{steel!10}1.129 (1.529) & \cellcolor{steel!10}0.929 (0.258) & \cellcolor{steel!10}26336.284 (1348.381) & \cellcolor{steel!10}2.376 (0.478) \\
\multirow{-2}{*}{\textbf{W1}}       & Restarted GA                       & \cellcolor[HTML]{FCE4D6}3.133 (0.017)  & \cellcolor{steel!10}4.732 (1.299)  & \cellcolor[HTML]{FCE4D6}5.907 (0.226) & \cellcolor{steel!10}1.860 (0.092)  & \cellcolor{steel!10}0.924 (0.019) & \cellcolor{steel!10}1.832 (1.368) & \cellcolor{steel!10}1.097 (0.348) & \cellcolor{steel!10}25721.608 (816.896)  & \cellcolor{steel!10}2.768 (0.658) \\ \hline
                                    & Transfer GA                      & \cellcolor[HTML]{FCE4D6}0.030 (0.000)  & \cellcolor{steel!10}0.011 (0.003)  & \cellcolor[HTML]{FFF2CC}2.333 (0.656) & \cellcolor{steel!10}1.078 (0.040)  & \cellcolor{steel!10}1.334 (0.006) & \cellcolor{steel!10}0.133 (0.048) & \cellcolor[HTML]{FCE4D6}3.778 (1.110) & \cellcolor[HTML]{FFF2CC}18266.502 (1838.219) & \cellcolor{steel!10}0.753 (0.131) \\
\multirow{-2}{*}{\textbf{W2}}       & Restarted GA                       & \cellcolor[HTML]{FCE4D6}0.030 (0.000)  & \cellcolor{steel!10}0.014 (0.005)  & \cellcolor[HTML]{FFF2CC}1.761 (0.004) & \cellcolor{steel!10}1.110 (0.040)  & \cellcolor{steel!10}1.360 (0.021) & \cellcolor{steel!10}0.160 (0.044) & \cellcolor[HTML]{FCE4D6}4.005 (1.024) & \cellcolor[HTML]{FFF2CC}18378.785 (822.054)  & \cellcolor{steel!10}0.934 (0.272) \\ \hline
                                    & Transfer GA                      & \cellcolor{steel!10}3.301 (0.003)  & \cellcolor{steel!10}4.014 (1.188)  & \cellcolor[HTML]{FCE4D6}0.402 (0.695) & \cellcolor{steel!10}0.371 (0.003)  & \cellcolor{steel!10}4.177 (0.020) & \cellcolor[HTML]{FFF2CC}0.427 (0.311) & \cellcolor[HTML]{FCE4D6}1.322 (0.300) & \cellcolor{steel!10}932.575 (51.553)     & \cellcolor{steel!10}1.270 (0.357) \\
\multirow{-2}{*}{\textbf{W3}}       & Restarted GA                       & \cellcolor{steel!10}3.314 (0.024)  & \cellcolor{steel!10}4.857 (1.347)  & \cellcolor[HTML]{FCE4D6}0.573 (0.905) & \cellcolor{steel!10}0.377 (0.006)  & \cellcolor{steel!10}4.254 (0.089) & \cellcolor[HTML]{FFF2CC}0.281 (0.066) & \cellcolor[HTML]{FCE4D6}1.449 (0.375) & \cellcolor{steel!10}878.662 (63.803)     & \cellcolor{steel!10}1.548 (0.487) \\ \hline
                                    & Transfer GA                      & \cellcolor[HTML]{FCE4D6}7.170 (0.022)  & \cellcolor{steel!10}11.096 (3.137) & \cellcolor[HTML]{FCE4D6}2.428 (0.247) & \cellcolor{steel!10}1.561 (0.003)  & \cellcolor{steel!10}1.186 (0.020) & \cellcolor{steel!10}2.020 (4.084) & \cellcolor{steel!10}1.603 (0.412) & \cellcolor{steel!10}984.067 (141.645)    & \cellcolor{steel!10}0.619 (0.063) \\
\multirow{-2}{*}{\textbf{W4}}       & Restarted GA                       & \cellcolor[HTML]{FCE4D6}7.158 (0.032)  & \cellcolor{steel!10}14.333 (4.261) & \cellcolor[HTML]{FCE4D6}2.376 (0.332) & \cellcolor{steel!10}1.603 (0.048)  & \cellcolor{steel!10}1.214 (0.019) & \cellcolor{steel!10}2.207 (1.423) & \cellcolor{steel!10}1.886 (0.617) & \cellcolor{steel!10}934.248 (131.108)    & \cellcolor{steel!10}0.698 (0.121) \\ \hline
                                    & Transfer GA                      & \cellcolor{steel!10}33.431 (0.195) & \cellcolor[HTML]{FCE4D6}11.596 (3.487) & \cellcolor[HTML]{FFF2CC}3.378 (0.876) & \cellcolor{steel!10}0.489 (0.005)  & \cellcolor{steel!10}2.393 (0.008) & \cellcolor{steel!10}1.116 (1.692) & \cellcolor[HTML]{FCE4D6}3.414 (0.834) & \cellcolor[HTML]{FFF2CC}46953.445 (4509.310) & \cellcolor{steel!10}1.103 (0.282) \\
\multirow{-2}{*}{\textbf{W5}}       & Restarted GA                       & \cellcolor{steel!10}33.443 (0.194) & \cellcolor[HTML]{FCE4D6}12.438 (3.627) & \cellcolor[HTML]{FFF2CC}3.299 (0.275) & \cellcolor{steel!10}0.504 (0.014)  & \cellcolor{steel!10}2.460 (0.061) & \cellcolor{steel!10}1.539 (0.982) & \cellcolor[HTML]{FCE4D6}3.561 (0.774) & \cellcolor[HTML]{FFF2CC}47123.373 (2904.705) & \cellcolor{steel!10}1.363 (0.503) \\ \hline
                                    & Transfer GA                      & \cellcolor{steel!10}0.970 (0.000)  & \cellcolor{steel!10}1.610 (0.458)  & \cellcolor{steel!10}1.324 (0.131) & \cellcolor{steel!10}0.365 (0.008)  & \cellcolor{steel!10}3.145 (0.042) & \cellcolor{steel!10}0.446 (0.396) & \cellcolor[HTML]{FCE4D6}0.102 (0.014) & \cellcolor[HTML]{FFF2CC}47071.542 (4573.392) & \cellcolor{steel!10}0.297 (0.021) \\
\multirow{-2}{*}{\textbf{W6}}       & Restarted GA                       & \cellcolor{steel!10}0.972 (0.005)  & \cellcolor{steel!10}1.959 (0.585)  & \cellcolor{steel!10}1.370 (0.190) & \cellcolor{steel!10}0.382 (0.009)  & \cellcolor{steel!10}3.242 (0.076) & \cellcolor{steel!10}0.548 (0.328) & \cellcolor[HTML]{FCE4D6}0.110 (0.022) & \cellcolor[HTML]{FFF2CC}47652.393 (822.434)  & \cellcolor{steel!10}0.319 (0.036) \\ \hline
                                    & Transfer GA                      & \cellcolor{steel!10}0.191 (0.003)  & \cellcolor[HTML]{FCE4D6}0.199 (0.014)  & \cellcolor[HTML]{FCE4D6}0.319 (0.401) & \cellcolor{steel!10}16.148 (0.022) & \cellcolor{steel!10}1.135 (0.014) & \cellcolor{steel!10}0.173 (0.101) & \cellcolor{steel!10}0.595 (0.150) & \cellcolor{steel!10}19787.726 (1770.222) &                                     \\
\multirow{-2}{*}{\textbf{W7}}       & Restarted GA                       & \cellcolor{steel!10}0.195 (0.005)  & \cellcolor[HTML]{FCE4D6}0.206 (0.017)  & \cellcolor[HTML]{FCE4D6}0.343 (0.306) & \cellcolor{steel!10}16.475 (1.411) & \cellcolor{steel!10}1.149 (0.016) & \cellcolor{steel!10}0.241 (0.131) & \cellcolor{steel!10}0.651 (0.134) & \cellcolor{steel!10}18676.683 (2001.725) &                                     \\ \hline
                                    & Transfer GA                      & \cellcolor{steel!10}10.902 (0.009) & \cellcolor{steel!10}23.830 (6.797) & \cellcolor[HTML]{FCE4D6}8.747 (0.005) & \cellcolor{steel!10}1.013 (0.007)  & \cellcolor{steel!10}7.065 (0.041) & \cellcolor{steel!10}3.448 (7.086) & \cellcolor{steel!10}0.141 (0.033) & \cellcolor{steel!10}27809.413 (1992.050) &                                     \\
\multirow{-2}{*}{\textbf{W8}}       & Restarted GA                       & \cellcolor{steel!10}10.912 (0.023) & \cellcolor{steel!10}30.210 (8.771) & \cellcolor[HTML]{FCE4D6}8.748 (0.004) & \cellcolor{steel!10}1.034 (0.025)  & \cellcolor{steel!10}7.105 (0.062) & \cellcolor{steel!10}4.937 (4.133) & \cellcolor{steel!10}0.170 (0.043) & \cellcolor{steel!10}25719.406 (2141.486) &                                     \\ \hline
                                    & Transfer GA                      & \cellcolor[HTML]{FCE4D6}9.127 (0.316)  & \cellcolor[HTML]{FCE4D6}21.255 (5.355) & \cellcolor{steel!10}3.182 (0.004) & \cellcolor{steel!10}0.462 (0.013)  & \cellcolor{steel!10}1.046 (0.009) & \cellcolor{steel!10}0.780 (1.248) & \cellcolor[HTML]{FCE4D6}0.251 (0.041) &                                            &                                     \\
\multirow{-2}{*}{\textbf{W9}}       & Restarted GA                       & \cellcolor[HTML]{FCE4D6}9.192 (0.413)  & \cellcolor[HTML]{FCE4D6}22.277 (6.795) & \cellcolor{steel!10}3.185 (0.005) & \cellcolor{steel!10}0.477 (0.016)  & \cellcolor{steel!10}1.062 (0.017) & \cellcolor{steel!10}1.205 (0.951) & \cellcolor[HTML]{FCE4D6}0.263 (0.053) &                                            &                                     \\ \hline
                                    & Transfer GA                      & \cellcolor{steel!10}5.291 (0.035)  & \cellcolor{steel!10}10.306 (2.266) & \cellcolor[HTML]{FCE4D6}6.838 (0.246) & \cellcolor[HTML]{FCE4D6}1.435 (0.008)  & \cellcolor{steel!10}1.112 (0.006) &                                     &                                     &                                            &                                     \\
\multirow{-2}{*}{\textbf{W10}}      & Restarted GA                       & \cellcolor{steel!10}5.407 (0.270)  & \cellcolor{steel!10}13.123 (4.756) & \cellcolor[HTML]{FCE4D6}6.803 (0.251) & \cellcolor[HTML]{FCE4D6}1.437 (0.008)  & \cellcolor{steel!10}1.127 (0.019) &                                     &                                     &                                            &                                     \\ \hline
                                    & Transfer GA                      & \cellcolor{steel!10}2.081 (0.003)  & \cellcolor[HTML]{FCE4D6}2.825 (0.687)  & \cellcolor[HTML]{FCE4D6}8.240 (0.870) & \cellcolor{steel!10}1.429 (0.014)  & \cellcolor{steel!10}1.619 (0.025) &                                     &                                     &                                            &                                     \\
\multirow{-2}{*}{\textbf{W11}}      & Restarted GA                       & \cellcolor{steel!10}2.095 (0.028)  & \cellcolor[HTML]{FCE4D6}3.202 (0.880)  & \cellcolor[HTML]{FCE4D6}8.121 (0.730) & \cellcolor{steel!10}1.448 (0.020)  & \cellcolor{steel!10}1.668 (0.035) &                                     &                                     &                                            &                                     \\ \hline
                                    & Transfer GA                      & \cellcolor[HTML]{FCE4D6}3.483 (0.076)  & \cellcolor{steel!10}5.653 (1.647)  & \cellcolor[HTML]{FCE4D6}3.899 (0.088) & \cellcolor[HTML]{FCE4D6}0.480 (0.004)  &                                     &                                     &                                     &                                            &                                     \\
\multirow{-2}{*}{\textbf{W12}}      & Restarted GA                       & \cellcolor[HTML]{FCE4D6}3.510 (0.097)  & \cellcolor{steel!10}7.278 (1.908)  & \cellcolor[HTML]{FCE4D6}3.936 (0.262) & \cellcolor[HTML]{FCE4D6}0.481 (0.007)  &                                     &                                     &                                     &                                            &                                     \\ \hline
                                    & Transfer GA                      & \cellcolor{steel!10}2.525 (0.013)  & \cellcolor{steel!10}2.947 (0.822)  &                                     &                                      &                                     &                                     &                                     &                                            &                                     \\
\multirow{-2}{*}{\textbf{W13}}      & Restarted GA                       & \cellcolor{steel!10}2.534 (0.020)  & \cellcolor{steel!10}3.432 (0.927)  &                                     &                                      &                                     &                                     &                                     &                                            &                                     \\ \hline
                                    & \cellcolor{steel!10}                        & \textbf{8}                             & \textbf{9}                             & \textbf{2}                            & \textbf{10}                            & \textbf{11}                           & \textbf{8}                            & \textbf{4}                            & \textbf{5}                                   & \textbf{6}                            \\
                                    & \cellcolor[HTML]{FCE4D6}                        & \textbf{5}                             & \textbf{4}                             & \textbf{8}                            & \textbf{2}                             & \textbf{0}                            & \textbf{0}                            & \textbf{5}                            & \textbf{0}                                   & \textbf{0}                            \\
\multirow{-3}{*}{\textbf{Summary}}  & \cellcolor[HTML]{FFF2CC}                       & \textbf{0}                             & \textbf{0}                             & \textbf{2}                            & \textbf{0}                             & \textbf{0}                            & \textbf{1}                            & \textbf{0}                            & \textbf{3}                                   & \textbf{0}                            \\ \hline
\end{tabular}

        \vspace{-5pt}
\end{table}


\section{Discussion}
\label{section:implication}



The results of our study demonstrate that tuning effectiveness is closely tied to the structural properties of the configuration landscape, which are shaped by the system's domain characteristics. These observations reinforce the importance of understanding not only what a tuner does, but why it succeeds or fails under specific tuning tasks.

Our proposed methodology, \approach, offers a systematic way to close this gap. By synergizing domain knowledge and spatial information of the configuration landscape mined via FLA, \approach~ enables researchers and practitioners to reason about tuning difficulty, explain tuner behavior, and inform the design of more robust tuning strategies. Therefore, our methodology has important implications for both tuner researchers and system designers, together with several relevant practices in software engineering, which we outline below.

\subsection{Implications to Tuner Researchers}
\approach~ provides actionable insights for tuner researchers, guiding decisions across three key aspects:

\begin{enumerate}
    \item \textbf{Tuner selection:} \approach~offers a comprehensive analysis on configuration landscape (e.g., ruggedness, local structure, global structure), shedding light on tuner selection. For instance, smooth landscapes with high FDC (\textbf{\textit{Finding 3}}) favor exploitation-driven methods such as iterated local search~\cite{lourencco2003iterated}, whereas highly rugged or with rich local optima structures (\textbf{\textit{Finding 5}}) require global exploration strategies (e.g., multi-objectivization~\cite{chen2024mmo, chen2021multi}) to escape local optima. In practical optimization scenarios, this process can be carried out as follows:
    \begin{itemize}
    \item \textbf{Step 1:} Collect a small set of configuration–performance measurements, for example, from lightweight sampling or historical observations.
    \item \textbf{Step 2:} Apply analysis workflow described in Section~\ref{section:methodology} to characterize key structural properties of the landscape (e.g., ruggedness, local structure, and global structure). 
    \item \textbf{Step 3:} Use the resulting structural patterns to guide tuner selection/design. For instance, landscapes exhibiting high ruggedness may call for stronger exploration, whereas smoother landscapes may favor exploitation-oriented or model-based approaches.
    \end{itemize}


    \item \textbf{Operator design:} Beyond tuner selection, \approach~can also guide operator design. For example, in landscape with structured basins of attraction and weak FDC correlation (e.g., \textsc{Batik} in our case study), integrating global exploration with local search operators (e.g., variable neighborhood search~\cite{mladenovic1997variable}) might improve search efficiency while mitigating premature convergence.

    \item \textbf{Budget allocation:} Researchers can also utilize \approach~ to assess the hardness of the tuning tasks and inform budget allocations. For instance, deceptive landscapes with numerous local optima (\textbf{\textit{Finding 4}}) may demand increased evaluation budgets for global exploration, whereas smoother landscape allow faster convergence with fewer evaluations (\textbf{\textit{Finding 5}}). By aligning tuner effort with system complexity, \approach~helps the designs of smarter budget allocation.
\end{enumerate}





\subsection{Implications to System Designers}

For system designers, \approach, for the first time, reveals how the options and the workloads contribute to the characteristics of tuning configurable systems, thereby making the (re-)designs thereof with more informed decisions:

\begin{enumerate}
    \item \approach~helps to understand how domain-specific factors (e.g., option types, system attributes, and workload characteristics) shape the configuration landscape structures. For example, by synergizing the spatiality and domain analysis in \approach, we found that core options exert stronger impact on landscape ruggedness (\textbf{\textit{Findings 2 and 6}}).
    
    \item \approach~can help enhance configuration manuals by refining option descriptions. For example, highlighting the option's impacts on ruggedness and implication to tuner design in API documentation (\textbf{\textit{Findings 1 and 2}}).
    
    \item \approach~assists system designers in refining system configurability, such as selectively refactoring tunable options based on their impact on ruggedness (\textbf{\textit{Findings 2 and 6}}).

\end{enumerate}

In practical system (re-)design, practitioners can follow the procedures below:
\begin{itemize}
    \item \textbf{Step 1:} Apply the workflow described in Section~\ref{section:methodology} to analyze the domain-spatiality patterns of the case and identify options (or option types) that are strongly associated with structural difficulty (e.g., sensitive to landscape ruggedness).
    \item \textbf{Step 2:} Based on these insights, implement system-level design adjustments that constrain or regulate such options, for instance, introducing new code that bounds their permissible value ranges or reduces harmful interactions.
    \item \textbf{Step 3:} Directly apply a tuner with predefined ranges/bounds. By limiting configurations that induce highly rugged landscapes, the resulting configuration space becomes smoother and more amenable to efficient tuning (e.g., by local search tuners).
\end{itemize}

\subsection{Implications to Practitioners}

\rev{In addition to the above, the domain-spatiality patterns revealed by \approach~also offers a broader foundation for understanding and explaining diverse aspects of ``difficulty'' for tuning configurable systems. This would benefit other relevant practices of software engineering, such as exploratory analysis and performance analysis:}


%
\rev{
\begin{itemize}
\item [(1)] \revminor{\textbf{Exploratory analysis for understanding tuning cases:} \approach~can be used for exploratory analysis to understand the structural characteristics of a tuning case and to help explain why a tuner behaves as observed (e.g., fast convergence, stagnation, or high variance). By analyzing the domain–spatiality patterns, practitioners can obtain a structured view of the tuning difficulty and the likely behaviors of different tuners. For example, when the goal is to explain why different tuners exhibit distinct search behaviors on a given configurable system, the following analysis can be carried out:
\begin{itemize}
    \item \textbf{Step 1:} Apply the workflow described in Section~\ref{section:methodology} to reveal the domain–spatiality patterns of the case, including structural properties such as ruggedness, local structure, and global structure.
    \item \textbf{Step 2:} Summarize the core search mechanism of the tuners under comparison, such as their balance between exploration and exploitation, use of restart or diversity maintenance, or dependence on surrogate modeling.
    \item \textbf{Step 3:} Interpret the observed behavioral differences by aligning the structural properties of the tuning case with the search characteristics of the tuners. For instance, highly rugged and multi-modal landscapes may hinder strongly exploitative strategies, whereas smoother and more coherent landscapes may favor model-based approaches.
\end{itemize}
}
\item [(2)] \textbf{Assisted performance analysis prioritization:} \approach~helps to reveal the workloads (or options), including their types, that should be studied first, in order to analyze the system. \revminor{In practical performance analysis scenarios, practitioners can follow the procedures below:
\begin{itemize}
    \item \textbf{Step 1:} Apply the workflow described in Section~\ref{section:methodology} to characterize the spatiality patterns of the target system across workloads or option settings, with particular attention to indicators such as autocorrelation and FDC.
    \item \textbf{Step 2:} Identify workloads or options that are associated with structurally difficult landscapes, such those exhibiting low autocorrelation or weak FDC, since these patterns indicate irregular search spaces and potentially unstable performance behavior.
    \item \textbf{Step 3:} Rank those workloads/options according to the landscape metrics. For example, testing configurations with prioritized workloads or options, as they are more likely to expose performance anomalies, hidden sensitivities, or bugs. This helps to reveal configuration-related performance bugs at earlier stage in configuration bug testing~\cite{DBLP:conf/icse/MaCL25}.
\end{itemize}
}
\end{itemize}
}

\section{Threats to Validity}
\label{section:threats_to_validity}

Threats to \textbf{internal validity} is related to the definition of neighborhood structures in \approach, since we do not have a complete dataset of all configurations. To mitigate this, we adopt an adaptive neighborhood strategy, attempting to ensure connectivity while preserving landscape integrity. However, this remains an approximation that may impact the characterization of the landscape. Besides, the exact landscape characteristics might vary depending on specific FLA metrics used in \approach, which, although widely adopted~\cite{huang2024rethinking,jones1995fitness,malan2021survey,weinberger1990correlated}, may not capture all structural aspects of complex configuration spaces. For example, alternative metrics may be more suitable for quantifying landscape ruggedness if the software system satisfies certain implicit assumptions required by those measures~\cite{de2014empirical,DBLP:conf/iclr/KeskarMNST17}. The inclusion of alternative or complementary metrics could provide a more comprehensive understanding of the spatial characteristics of a software system. \rev{Another potential threat arises from our use of the best-performing configuration within the sampled data as a proxy for the global optimum in computing landscape metrics. While this practice follows the setting in existing literature~\cite{huang2018self,tanabe2020analyzing}, it may introduce bias if the sampled configurations do not sufficiently approximate the true global optimum. Nevertheless, the datasets we used were collected through diverse coverage-based and random sampling strategies, which help reduce this risk to a certain extent. A further limitation lies in our focus on option-specific ruggedness where each configuration option is analyzed independently. This design choice ensures interpretability and consistency with the option-level perspective adopted in earlier analyses. Nevertheless, {if resources permitted}, the methodology can be readily extended to capture interaction-level ruggedness by partitioning the configuration space jointly over multiple options (e.g., pairs or triplets) and computing the corresponding autocorrelation metric. Such an extension would enable systematic investigation of how option interactions, for example, between core and resource-related options, jointly reshape the configuration landscape. We consider this as an interesting direction for future work.}


Threats to \textbf{external validity} may arise from dataset selection. While our case study spans nine configurable systems and 93 workloads, a more exhaustive exploration of the configuration space—potentially covering a complete set of configurations—could further strengthen the findings. \rev{Our case study used sampled datasets rather than exhaustive configuration enumeration, potential biases may arise from incomplete coverage of the configuration space. Sampling-based datasets, while practical for large-scale configurable systems, may omit certain configurations or interactions, potentially distorting the true configuration landscape. Nonetheless, the datasets we use were collected using coverage-based and random sampling strategies, which aims to approximate the landscape's overall structure with minimal bias. This trade-off between feasibility and completeness is consistent with realistic tuning scenarios, where exhaustive evaluation is infeasible.} \revminor{To the best of our knowledge, the adopted dataset is one of the largest publicly available configuration–performance datasets supporting multi-system and multi-workload landscape analysis. While no dataset can be fully representative of all modern configurable systems, the breadth of this dataset provides a strong empirical foundation for the domain–spatiality analyses conducted in this study.} In addition, while our analysis reveals some common patterns across systems or workloads, different types of configurable software systems may exhibit fundamentally distinct landscape features. Therefore, the findings reported here might not directly or fully generalize to all configurable systems. Nonetheless, the methodology is designed to be general-purpose and can be applied to a wide range of black-box configurable systems. We believe the FLA perspective presented in this work offers a systematic lens through which diverse configuration-performance spatial correlations can be analyzed, and it can be extended to support system-specific insights in other application domains.

On domain analysis in \approach, the human analysis might pose threats to \textbf{construct validity}. To mitigate this, we followed a systematic classification protocol, involving multiple rounds of independent reviews. Yet, unintentional subjective biases are still possible. Future work could integrate automated program analysis techniques to enhance consistency and reduce potential biases.

\section{Related Work}
\label{section:related_work}

\textbf{Static Domain Analysis:} Over the past decade, there exist methodologies that seek to pry open the black box of systems for understanding their internal characteristics~\cite{DBLP:conf/icse/LiangChen25}. These, typically, leverage \textit{domain-specific knowledge} (e.g., system manual)~\cite{sullivan2004using,he2022multi, DBLP:journals/pvldb/LaoWLWZCCTW24} or \textit{code-level analysis} (e.g., static code inspection) to excavate how parameters influence system behaviors~\cite{chen2023diagconfig}. \revision{For instance, Li et al. \cite{li2020statically} propose a static analysis approach to infer the performance properties of configuration options by analyzing control and data dependencies between option values and performance-critical operations. Chen et al. \cite{chen2023diagconfig} propose DiagConfig, a configuration-oriented diagnosis method that explains performance violations by tracing how configuration options propagate through code. DiagConfig combines static code analysis with performance violation detection to identify performance-sensitive options and construct cause-effect chains linking misconfigurations to observed performance issues.} These analyses can help identify and prioritize critical options; reveal causal relationships between configuration options and system performance; and even extract implicit constraints~\cite{ chen2020understanding}. Nevertheless, existing domain analysis methods heavily rely on expert assumptions, which alone cannot fully reveal when a tuner might be successful or fail to tune a system. \revision{In contrast, our methodology, \approach~, synergizes the spatial information of the configuration space, mined using FLA, with domain knowledge, allowing us to better profile the characteristics of the systems and understand the behaviors of the tuners.}


\textbf{Dynamic Data Analysis:} A common data-driven method to study configurable systems is to apply statistical methods to extract the characteristics of performance distribution~\cite{jamshidi2017transfer,muhlbauer2023analyzing}. \revision{For example, Valov et al. \cite{valov2017transferring} analyze performance distribution across platforms and find that they exhibit stable trends, enabling low-cost performance prediction via model transfer. Jamshidi et al. \cite{jamshidi2017transfer} study how performance distributions drift across environments and leverage this knowledge to guide transfer learning for performance modeling. Xiang et al.~\cite{DBLP:conf/icse/XiangChen26} model online configuration performance drifts.} While statistical methods can mine useful information regarding the systems, such as trends and skewness, they typically consider performance values as isolated numerical observations, neglecting the inherent spatial information within the configuration space. In fact, each performance value is intrinsically tied to a specific configuration, and configurations themselves exhibit spatial relationships defined by their parameter values. As a result, the associated performance values are naturally distributed across the configuration space, forming spatially structured patterns rather than isolated data points. The lack of spatial information makes it challenging to answer fundamental questions such as: \textit{“where is the global optimum located?”, “what does the local structure look like?” and “how rugged is the search space?”}---the key factors that shape the tuner's behaviors. 
Recently, FLA, which explicitly incorporates spatial information, has been applied in software testing~\cite{neelofar2022instance,aleti2021effectiveness,10.1145/3699596}, configuration tuning~\cite{huang2024rethinking, chen2025accuracy}. The most relevant work is perhaps \texttt{GraphFLA}~\cite{huang2024rethinking}, a graph-based method to extract spatial relations within configuration space, aiming to uncover key system characteristics. Compared with \texttt{GraphFLA}, \approach~of this study differs in several key aspects:

\begin{itemize}
    \item While \texttt{GraphFLA} analyzes configuration  landscape purely from a data-driven spatial perspective, \approach~extends this by incorporating a structured system-specific domain analysis at three levels of abstractions.
    
    \item Lacking domain-specific analysis, \texttt{GraphFLA} is unable to provide a fundamental explanation for spatial insights, while \approach~synergizes configuration spatiality and domain analysis, aligning landscape with system-specific attributes. This synergy enables a more context-aware explanation of how configurations influence performance across systems.
    
    \item Our case study with \approach~involves a more comprehensive empirical evaluation across nine software systems and 93 workloads, covering a wider range of configurable systems and workloads compared to the more focused case studies in \texttt{GraphFLA}, which examines only three systems.
\end{itemize}


\textbf{Empirical Studies of Configuration:} There exist a few empirical studies for configurable systems~\cite{DBLP:journals/tosem/ChenL23a,DBLP:journals/tosem/ChenL23,DBLP:conf/icse/ZhangHLL0X21,chen2025accuracy,muhlbauer2023analyzing,weber2023twins}. Among others, Zhang et al.~\cite{DBLP:conf/icse/ZhangHLL0X21} present a source-code level study on how configuration design and implementation evolve in cloud systems. By analyzing over a thousand configuration-related commits, they summarize developer practices and call for new techniques to proactively reduce misconfigurations. Chen et al.~\cite{chen2025accuracy} conduct a large-scale empirical study to examine whether higher surrogate model accuracy truly leads to better tuning results. Their findings challenge the common assumption that model accuracy determines tuning quality and call for rethinking the role of accuracy in model-based configuration tuning. Muhlbauer et al.~\cite{muhlbauer2023analyzing} systematically study how workload variation impacts the accuracy and generalizability of performance models. Their findings show that workload variation can lead to significant and non-monotonic performance shifts, revealing the limitations of single-workload modeling and the necessity to account for workload-configuration interactions. Weber et al.~\cite{weber2023twins} investigate how runtime performance correlates with energy consumption across configurable software systems and observe that the relationship is highly system-dependent. However, these studies have not emphasized a systematic methodology that connects spatial information to domain understanding of the configurable systems and their tuning.

\section{Conclusion}
\label{section:conclusion}

In this paper, we propose \approach, a methodology that bridges spatial fitness landscape analysis with domain knowledge to understand configuration tuning. \revision{\approach~enables a structured interpretation of configuration-performance correlation, helping to assess tuning difficulty and inform algorithm design.} Through extensive experiments across nine configurable systems and 93 workloads, we show how to follow \approach~and reveal several key insights:

\begin{itemize}
    \item \textbf{System-specific landscapes:} Configuration landscape vary across systems; no single domain (e.g., area, language, or resource intensity) consistently shapes their structure.
    \item \textbf{Impact of core option:} Core options have a stronger influence on landscape ruggedness than resource options.
    \item \textbf{Workload-induced variations:} Workload effects on landscape structure are not uniformly tied to type or scale, but are more system-dependent.
\end{itemize}

We hope that domain-spatiality synergies for understanding configuration tuning in this work pave the way for better adaptive tuner designs and the evolution of configurable systems. By integrating \approach~into the tuner selection and design process, future work can move toward landscape-aware autotuning, where optimization strategies are dynamically matched to system and workload characteristics for better performance and adaptability. \rev{Additionally, integrating causal reasoning into \approach~ may represent a promising future direction to move beyond correlation analysis and toward a deeper understanding of how domain factors shape configuration landscapes.}



\begin{acks}
This work was supported by an NSFC Grant (62372084) and a UKRI Grant (10054084).
\end{acks}

\bibliographystyle{ACM-Reference-Format}
\bibliography{references.bib}

\end{document}